\documentclass[reprint,aps,prb,showpacs,superscriptaddress,english]{revtex4-2}

\usepackage[utf8]{inputenc}
\usepackage{hyperref}

\usepackage{natbib}

\usepackage{graphicx,color}
\usepackage{amssymb,amsmath,amsfonts,mathrsfs,fancyhdr}

\usepackage[caption=false,subrefformat=parens,labelformat=empty]{subfig}

\newcommand{\veps}{\varepsilon}
\renewcommand{\Re}{\mathrm{Re \,}}

\newcommand{\Tr}{\mathrm{Tr}}

\newcommand{\PV}{\mathrm{PV}}

\newcommand{\bra}[1]{\left \langle #1 \right |}
\newcommand{\ket}[1]{\left| #1 \right \rangle}

\begin{document}

\title{Signatures of Many-Body Localization in the Dynamics of Two-Level Systems in Glasses}
\author{Claudia Artiaco}
\email{cartiaco@sissa.it}
\author{Federico Balducci}
\email{fbalducc@sissa.it}
\affiliation{The Abdus Salam ICTP, Strada Costiera 11, 34151, Trieste, Italy}
\affiliation{INFN Sezione di Trieste, Via Valerio 2, 34127 Trieste, Italy}
\affiliation{SISSA, via Bonomea 265, 34136, Trieste, Italy}
\author{Antonello Scardicchio}
\affiliation{The Abdus Salam ICTP, Strada Costiera 11, 34151, Trieste, Italy}
\affiliation{INFN Sezione di Trieste, Via Valerio 2, 34127 Trieste, Italy}
\date{\today}

\begin{abstract}
    We investigate the quantum dynamics of Two-Level Systems (TLS) in glasses at low temperatures (1 K and below). We study an ensemble of TLSs coupled to phonons. By integrating out the phonons within the framework of the Gorini-Kossakowski-Sudarshan-Lindblad (GKSL) master equation, we derive analytically the explicit form of the interactions among TLSs, and of the dissipation terms. We find that the unitary dynamics of the system shows clear signatures of Many-Body Localization physics. We study numerically the time behavior of the concurrence, which measures pairwise entanglement also in non-isolated systems, and show that it presents a power-law decay both in the absence and in the presence of dissipation, if the latter is not too large. These features can be ascribed to the strong, long-tailed disorder characterizing the distributions of the model parameters. Our findings show that assuming ergodicity when discussing TLS physics might not be justified for all kinds of experiments on low-temperature glasses.
\end{abstract}

\maketitle

\section{Introduction}

Recent years have witnessed several advances in our understanding of the dynamics of many-body quantum systems. On one hand, the mechanism by which thermal equilibrium appears in isolated quantum systems has been explained via the Eigenstate Thermalization Hypothesis (ETH) \cite{deutsch1991quantum,srednicki1994chaos,d2016quantum}, and its connection to the classic von Neumann ergodic theorem has been made clear \cite{rigol2012alternatives}. On the other hand, a generic mechanism by which quantum systems can \emph{avoid} going to thermal equilibrium has been identified in Many-Body Localization (MBL) \cite{gornyi2005interacting,Basko:2006hh,deLuca2013,huse2014phenomenology,luitz2015many,huse2015review,abanin2017recent}. Analogous phenomena take place in driven periodic systems (time crystals) \cite{Abanin2016Theory,zhang2017observation,sacha2017time}, and in systems without disorder \cite{schiulaz2014glass,Papic2015,pino2015metallic,yao2016quasi,Nandkishore2017LR,brenes2018many,Schulz2019Stark,vanNieuwenburg2019From,giudici2020breakdown}. These progresses give now a more or less complete picture of the various ways of thermalization in quantum systems, under different conditions.

One of the places in which one routinely finds disorder \emph{and} quantum effects at the same time is in the study of low-temperature properties of glasses. A series of classic experiments \cite{ZellerPohl,stephens1973low} has made manifest that the properties of glasses at temperatures of 1K and below show a surprising degree of universality, and deviate significantly from Debye theory. Several theoretical ideas aimed at explaining these results, mostly on the lines of two seminal works \cite{anderson1972anomalous,phillips1972tunneling}. There, the authors introduced the idea of bi-stable tunnelling systems (or Two-Level Systems, TLSs), whose parameters (energy difference and tunnelling rates) are very broadly distributed. With an appropriate choice of such distributions, one can reproduce quantitatively the values of several equilibrium quantities, including specific heat, conductivity, and sound attenuation. The range of TLS models has been expanded considerably beyond the original works to account for various experimental facts \cite{Phillips87,Esquinazi2013}, and even criticized as a glorified curve-fitting procedure \cite{YuLeggett,Leggett91,LeggettVural}.

In a glass, TLSs interact with phonons and, if they have an electric dipole moment, also with photons. The consequence of the interaction between TLSs and the phonon (and photon) bath is twofold: it generates TLS-TLS interactions, which have been observed in several experiments \cite{Arnold75,Enss97,Strehlow98,Boiron99,Classen00,Lisenfeld2015}, and it is responsible for the equilibration of the TLSs at the bath temperature. While the former have been subject of extensive studies \cite{JoffrinLevelut,black1977spectral,Kassner-89,Burin96,Schechter-PRB17}, and also have been used to explain the discrepancies between experiments and the original TLS model, the issue of thermalization has been overlooked so far. Namely, TLSs are always assumed thermal on all experimentally accessible timescales, and standard thermodynamic ensembles are applied.

The purpose of this paper is to investigate the quantum dynamics of TLSs coupled to phonons, and to discuss in particular how they reach thermal equilibrium. We idealize the system TLSs + phonons as an isolated system, and we analytically derive the Gorini–Kossakowski–Sudarshan–Lindblad (GKSL) master equation for the reduced density matrix of the TLSs, tracing out the phonons. We find that the TLS unitary evolution (the so-called Liouvillian) is governed by a Hamiltonian with an extensive number of local conserved quantities, as the effective Hamiltonian of MBL systems; the dissipative term (the so-called Lindbladian) destroys localization and drives the system to a thermal state. We show that, considering the typical values of the TLS disorder parameters, dissipation is much slower than any other time scale of the problem, and TLS relaxation dynamics shows the fingerprint of localization for a long time window.

Recently, a growing body of literature has investigated the impact of dissipation and dephasing on MBL systems \cite{nandkishore2014spectral,levi2016robustness,fischer2016dynamics,medvedyeva2016influence,everest2017role,nandkishore2017many,vakulchyk2018signatures,Gopalakrishnan20,wybo2020entanglement}. The research question underlying these works concerns how the imperfect isolation from the environment enters the experimental measurements on MBL systems \cite{schreiber2015observation,bordia2016coupling,smith2016many,luschen2017signatures}. Even though dissipative baths necessarily lead to delocalization, it has been found that at intermediate and long time scales the relaxation dynamics of MBL systems coupled to heat baths shows clear signatures of the localized phase, and differs from the one of ergodic systems. These findings are in agreement with the results of the present study, as will become evident in the following.

We focus in particular on the creation and spreading of entanglement as measured by the concurrence \cite{hill1997entanglement,Wootters98,Amico08} and the entanglement entropy. The former measures the amount of entanglement between two TLSs; under time evolution it grows to a maximum, and then decays and vanishes. The latter instead increases monotonically with time to reach a thermodynamic value. We simulate both the artificially isolated TLS system (i.e. dissipation is set to zero), and the open system. For the artificially isolated system, we can confidently investigate the thermodynamic limit (our numerics goes up to $N=60$ TLSs). We observe that, for long time scales, the concurrence decays as a power-law $C \sim t^{-\beta_{i}}$, down to a plateau value which is exponentially small in the number of TLSs. This slow power-law decay is the signature of localization, and contrasts with the exponentially fast decay one would observe for an ergodic system. In the open system we find that the concurrence always vanishes, never reaching the plateau observed in the unitary case. This is not surprising, since the phonon (and photon) bath to which TLSs are coupled is effectively infinite, and entanglement can spread indefinitely. Moreover, for not too large dissipation, we find that the concurrence decays as a power-law $C \sim t^{-\beta_{o}}$, as in the artificially isolated system, indicating that the signatures of TLS localization are observable even in this case for long time windows. The exponents $\beta_{i,o}$ in the two scenarios are of the same order of magnitude. Their comparison shows that, within the statistical errors and finite-size corrections, $\beta$ increases in the presence of dissipation.

The structure of the paper is as follows. In Sec.~\ref{sec:TLSmodel}, we introduce the Hamiltonian of the full system (TLSs + phonons), and comment on the various parameters that are needed to describe TLSs in amorphous media. In Sec.~\ref{sec:Lindblad_derivation}, we introduce the GKSL equation for the TLS density matrix which is obtained integrating out the phonons. We present the explicit form of the TLS-TLS interactions and dissipation terms, and discuss their competition. Thus, we sketch the dynamical phase diagram of the system. In Sec.~\ref{sec:numerical_simulations}, we present the numerical results on the real-time evolution of the TLSs. In particular, in Sec.~\ref{sec:unitary_dynamics} we consider the system as artificially isolated, studying the unitary part of the GKSL dynamics, and we analyze the signatures of MBL on the entanglement quantifiers. In Sec.~\ref{sec:dissipation}, instead, we reintroduce the dissipative terms of the GKSL dynamics, and show how they affect the results of Sec.~\ref{sec:unitary_dynamics}. Finally, in Sec.~\ref{sec:conclusions} we summarize our findings and indicate possible future research directions.

\section{The TLS model}

\label{sec:TLSmodel}

We define the total Hamiltonian of the TLSs system and the thermal bath as \cite{JoffrinLevelut,Kassner-89,Carruzzo-Yu}
\begin{equation}
    \label{eq:total_hamiltonian}
    H = H_{\mathit{TLS}} + H_B + H_{int}.
\end{equation}
The phonon bath is described by
\begin{equation}
    H_B = \sum_{k} \hbar \omega_{k} \psi^\dagger_k \psi^{\phantom{\dagger}}_k,
\end{equation}
$\psi_k$ (resp.\ $\psi_k^\dagger$) being the annihilation (resp.\ creation) operator of a phonon with wavevector and polarization $k = (\mathbf{q}, \alpha)$. The dispersion relation in amorphous solids is, to a good approximation at low temperatures \cite{Ruzicka-04}, $\omega_{\mathbf{q} \alpha} \simeq v_\alpha q$ with $v=v_L$ for longitudinal modes and $v=v_T$ for transverse modes \footnote{Isotropy is due to structural disorder and holds up to short scales. A different dispersion is encountered in ultra-stable vapor-deposited glasses \cite{Ediger17}, which are essentially two-dimensional.}. Typically $v_L \simeq 1.6 \, v_T$ (see Table \ref{tab:parameters}).

The TLS Hamiltonian is
\begin{equation}
    \label{eq:H_TLS}
    H_{\mathit{TLS}} = \sum_i (\Delta_i \sigma_i^x + \veps_i \sigma_i^z).
\end{equation}
We employ Pauli spins to represent the two states of a TLS; $\veps_i$ is the energy splitting and $\Delta_i$ the tunnelling amplitude in the $i$-th double well. According to the original works \cite{anderson1972anomalous,phillips1972tunneling}, we consider $\veps$ as drawn from a uniform distribution of width $W\simeq 0.1$ eV:
\begin{equation}
    \label{eq:pdf_veps}
    p_{\veps}(\veps)=\frac{1}{W}\Theta(W-\veps)\Theta(\veps)
\end{equation}
($\Theta$ is the Heaviside step function). In \cite{anderson1972anomalous,phillips1972tunneling} it is also argued that the tunnelling amplitudes $\Delta_i$ are broadly distributed, and that the most reasonable distribution, from a simplicity standpoint, is log-uniform:
\begin{equation}
    \label{eq:pdf_Delta}
    p_\Delta(\Delta) = \frac{\Theta(\Delta-\Delta_{\mathit{min}})\Theta(\Delta_{\mathit{max}}-\Delta)}{\log(\Delta_{\mathit{max}}/\Delta_{\mathit{min}})\Delta}
\end{equation}
where
\begin{equation}
    \Delta_{\mathit{min}} = \overline \Delta \cdot 10^{-n_\Delta/2}, \qquad
    \Delta_{\mathit{max}} = \overline \Delta \cdot 10^{n_\Delta/2}.
\end{equation}
The parameter $n_\Delta$ defines the span of the distribution: $\Delta_{\mathit{max}} / \Delta_{\mathit{min}} = 10^{n_\Delta}$. Since $\langle \log \Delta\rangle = \log \overline{\Delta}$, we note that $\overline{\Delta}$ is the typical value. Usually in the literature, $n_{\Delta} \simeq 8$ and $\overline{\Delta}/W \approx 10^{-5}$, making $p_{\Delta}(\Delta)$ very wide. 

The interaction Hamiltonian of the localized degrees of freedom with the strain field is, to lowest order \cite{JoffrinLevelut,Kassner-89,Carruzzo-Yu},
\begin{equation}
    \label{eq:interaction_hamiltonian}
    H_{int} = \sum_{ik} \sigma_i^z \left( \xi_{ik} \psi_k + {\rm h.c}. \right),
\end{equation}
with
\begin{equation}
    \label{eq:def_xi_vectors}
    \xi_{i k} = - i \sqrt{\frac{\hbar}{2 V \rho \, \omega_k}} \gamma_i D_i^{ab} e^{ab}_k e^{i \mathbf{q} \cdot \mathbf{r}_i}.
\end{equation}
Above, $\rho$ is the material density, $V$ the volume, $\gamma_i D_i^{ab}$ the elastic dipole tensor of the $i$-th TLS (the strength $\gamma_i$ has the dimension of an energy and $D_i^{ab}$ is dimensionless), and 
$e^{ab}_k := \frac{1}{2} \big(q^a \hat e_{\mathbf{q}\alpha}^b + q^b \hat e_{\mathbf{q}\alpha}^a \big)$ ($\mathbf{q}$ is the wavevector and $\mathbf{\hat e}_{\mathbf{q}\alpha}$ the unit $(\mathbf{q},\alpha)$-polarization vector). $\gamma_i$ and $D_i^{ab}$ are random variables; their probability distributions are induced by the distributions of the shapes and directions of the TLSs in space. In the literature \cite{Phillips87,Esquinazi2013} it is argued that $\gamma_i$ should be of the same order of magnitude of $W$, since the former is related to the energy shift induced in a TLS by a phonon, and it must be comparable with the energy imbalance of the two minima in the double well. Therefore, for simplicity, we set $\gamma_i \equiv W$ and absorb in the dipole entries $D_i^{ab}$ all the disorder fluctuations: we consider $D_i^{ab}$ to be random variables of order 1. We will not specify the full distribution of their entries, since in Sec.~\ref{sec:coupling_phonons} we will show that only some combinations are needed. We refer to those Sections for more details.

We report in Table \ref{tab:parameters} the experimental values of the TLS model parameters for three well-known structural glasses.

\begin{table}[t]
    \centering
    \begin{tabular}{|c|c|c|c|}
        \hline
        & SiO$_2$ & BK7 & PMMA \\ \hline 
        
        $W$ [meV]  & 130 & 70 & 30 \\ 
        $\Delta_\mathit{max}$ [meV] & 13 & 7 & 3 \\
        $\overline{\Delta}$ [meV] & $10^{-3}$ & $10^{-3}$ & $10^{-4}$ \\
        $\Delta_\mathit{min}$ [meV] & $10^{-7}$ & $10^{-7}$ & $10^{-8}$ \\
        $\gamma$ [eV] & 0.8 & 0.7 & 0.3 \\ 
        
        $\rho$ [g/cm$^3$] & 2.2 & 2.5 & 1.2 \\ 
        $v_L$ [km/s] & 5.8 & 6.2 & 3.2  \\ 
        $v_T$ [km/s] & 3.8 & 3.8 & 1.6 \\ 
        
        $k_B T_D$ [meV] & 30 & 30 & 10 \\ 
        $\rho_{\mathit{TLS}}$ [nm$^{-3}$] & 0.3 & 0.2 & 0.05 \\ 
        $\hbar \tau^{-1}$ [meV] & 1.8 & 1.7 & 0.45 \\
         \hline
    \end{tabular}
    \caption{Summary of the TLS model parameters for fused quartz (SiO$_2$), borosilicate glass (BK7), and plexiglass (PMMA). The parameters $v_L,v_T,\rho$ and the Debye temperature $T_D$ are independent; their values are derived from experimental measurements \cite{Berret1988How,Carruzzo-Yu}. The (average) TLS-phonon coupling $\gamma$ is experimentally accessible too \cite{Berret1988How}. One can reasonably assume $W \approx k_B T_{\mathit{glass}}$: indeed the TLSs are formed at the glass transition \cite{Phillips87}. As a consequence, one should also set $\Delta_\mathit{max} \approx 10^{-1} \,W $ in order to have a density of states that goes to zero above $W$ \cite{anderson1972anomalous}, and $\Delta_\mathit{min} \approx 10^{-9}\, W$ to reproduce instead a flat DOS at low temperatures \cite{Hunklinger1986Thermal}. The precise value of $\Delta_{max}$ and $\Delta_{min}$ is not crucial, since they enter only logarithmically in the quantities of interest. One can obtain the numerical density of the TLSs, $\rho_{\mathit{TLS}}$, from the experimentally measurable parameter $\bar P=\rho_{\mathit{TLS}}/W \log(\Delta_{max}/\Delta_{min})$ \cite{Phillips87,Berret1988How}.}
    \label{tab:parameters}
\end{table}

\section{The GKSL master equation}
\label{sec:Lindblad_derivation}

To study the dynamics of the TLSs, we need to integrate out the phonons. We choose to work in the GKSL framework \cite{BreuerPetruccione,Manzano2020Lindblad}, obtaining a master equation for the (reduced) density matrix of the TLSs $\rho$, that reads:
\begin{equation}
    \label{eq:Lindblad}
    \partial_t \rho(t) = -\frac{i}{\hbar} [H_{\mathit{TLS}} + H_{LS},\rho(t)] +
    \sum_{\kappa} \mathcal{L}_\kappa \rho(t).
\end{equation}
The first term on the r.h.s.\ describes the unitary evolution of the system, and it is called the Liouvillian. It is governed by $H_{\mathit{TLS}}$, which is the TLS Hamiltonian of Eq.~\eqref{eq:H_TLS}, and $H_{\mathit{LS}}$, which is the Lamb-Stark shift Hamiltonian (it will be specified below in Eq.~\eqref{eq:Lamb-Stark}). The second term on the r.h.s.,\ the so-called Lindbladian, describes instead dissipation and decoherence. $\mathcal{L}_\kappa$ are the Lindblad super-operators; in general, the label $\kappa$ can assume $O(N^2)$ values but, as we will show in the following, in our system the dominant terms are on-site, reducing $\kappa \equiv i = 1,2,\dots,N$.

The GKSL master equation \eqref{eq:Lindblad} relies on some approximations \cite{BreuerPetruccione,Manzano2020Lindblad}. First, one assumes weak coupling between TLSs and phonons. This assumption is usually taken in literature \cite{Phillips87}; its validity has to be checked a posteriori, verifying that the energy scales of decoherence and dissipation induced by phonons are smaller than the TLS energy set by $W$. The GKSL framework consists in three further approximations: the Born, the Markov, and the rotating wave approximation. In the Born approximation, one assumes that at all times the influence of the TLSs on the phonon thermal population is negligible. This is a consequence of weak coupling, and of the TLSs being a dilute system in the (amorphous) lattice. Therefore, we expect the Born approximation to be valid to a good extent in our systems. The Markov approximation instead entails that all the bath excitations decay on very fast timescales with respect to those of the TLSs. This is not guaranteed when working at ultra-low temperatures, but it is still a good starting point. Finally,the rotating wave approximation assumes that, when considering two TLSs, the \emph{resonant} processes are dominant, or equivalently that the relaxation time of TLSs in the open-system, $\tau_R$, is long with respect to the time scale of the intrinsic evolution of the system \cite{BreuerPetruccione}; in formulas: $\tau_R \gg |\nu_i - \nu_j|^{-1}$. We will validate a posteriori this assumption in Sec.~\ref{sec:dynamical_phases}.

Within these assumptions, the TLS-TLS interactions in $H_{\mathit{LS}}$ commute with the isolated TLS Hamiltonian: $[H_{\mathit{TLS}},H_{\mathit{LS}}] = 0$, ultimately leading to the MBL character of the unitary dynamics. In further studies, it might be interesting to go beyond the GKSL master equation, and relax its assumptions.

\subsection{The free TLS eigenoperators}

In order to compute the Lamb-Stark shift $H_{\mathit{LS}}$ and the Lindblad super-operators $\mathcal{L}_\kappa$, it is convenient to diagonalize the TLS Hamiltonian $H_{\mathit{TLS}}$ \cite{BreuerPetruccione,Manzano2020Lindblad}. We look for single-site operators $S_i$ such that 
\begin{equation}
    [H_{\it TLS}, S_i] = -\hbar \nu S_i.
\end{equation}
The linear problem is easily solved, finding eigenvalues
\begin{equation}
    \label{eq:freeTLS_eigenvalues}
    \hbar \nu_{i,0} = 0, \qquad
    \hbar \nu_{i,\pm} = \pm \hbar \nu_{i} = \pm 2 \sqrt{\veps^2_i + \Delta_i^2},
\end{equation}
with corresponding eigenoperators
\begin{equation}
    \label{eq:S_basis}
    S_i^z = \vec{v}_{i,0} \cdot \vec{\sigma}_i, \qquad
    S_i^\pm = \vec{v}_{i,\pm} \cdot \vec{\sigma}_i,
\end{equation}
where 
\begin{equation}
    \vec{v}_{i,0} =- \frac{2}{\hbar \nu_i} (\Delta_i, 0, \veps_i), \quad
    \vec{v}_{i,\pm} = \frac{2}{\hbar \nu_i} (-\veps_i,\pm i \hbar \nu_i/2, \Delta_i).
\end{equation}
Notice that, since typically $\Delta_i \ll \veps_i \sim W$, $\hbar \nu_i$ will be of order $W$. Also, defining $S_i^x = (S_i^+ + S_i^-)/2$ and $S_i^y = (S_i^+ - S_i^-)/2i$, the operators $S_i^x, S_i^y, S_i^z$ form a Pauli basis.

At this point, it is easy to verify that the TLS Hamiltonian reads
\begin{equation}
    H_{\mathit{TLS}} = - \frac{1}{2} \sum_i \hbar \nu_i S_i^z.
\end{equation}

\subsection{Coupling to phonons}
\label{sec:coupling_phonons}

The coupling with phonons induces both dissipation and TLS--TLS interactions. Under the assumptions discussed above, they can be be modelled via the GKSL master equation. Its final form for TLSs in glasses is given by:
\begin{widetext}
\begin{multline}
    \label{eq:our_Lindblad}
    \partial_t \rho(t) = -\frac{i}{\hbar} \bigg[ -\frac{1}{2} \sum_i \hbar \nu_i S_i^z + \sum_{ij} J_{ij} S_i^z S_j^z,\rho(t) \bigg] 
    + \sum_{i} Y_i f_T(\hbar \nu_i) \biggl( S_i^+ \rho(t) S_i^- +  S_i^- \rho(t) S_i^+  - 4 \rho(t) \biggr) \\
    + \sum_{i} Y_i \biggl( S_i^+ \rho(t) S_i^- + \left\{ \rho(t),S^z_i \right\} - 2 \rho(t) \biggr).
\end{multline}
\end{widetext}
In the previous equation, the first term on the r.h.s.\ corresponds to the commutator $- \frac{i}{\hbar}[ H_{\mathit{TLS}} + H_{\mathit{LS}} , \rho(t) ]$, where
\begin{equation}
    \label{eq:Lamb-Stark}
    H_{\it LS} = \sum_{ij} J_{ij} S_i^z S_j^z
\end{equation}
is the Lamb-Stark shift Hamiltonian. The second term on the r.h.s.\ contains the dissipative terms; it is written separating explicitly the temperature dependent and independent contributions: $f_T(\epsilon) := (e^{\epsilon/k_B T} - 1)^{-1} $ is, indeed, the Bose-Einstein distribution function  at temperature $T$. Considering that $\hbar\nu_i \sim W \sim 0.1$ eV, however, at ultra-low temperature ($T \sim 1$ K and below) $f_T \simeq 0$, and our system is effectively at \emph{zero temperature}. Thus, in the following we will keep only the temperature-independent contributions.

\begin{figure}[t]
    \centering
    \includegraphics[width=\columnwidth]{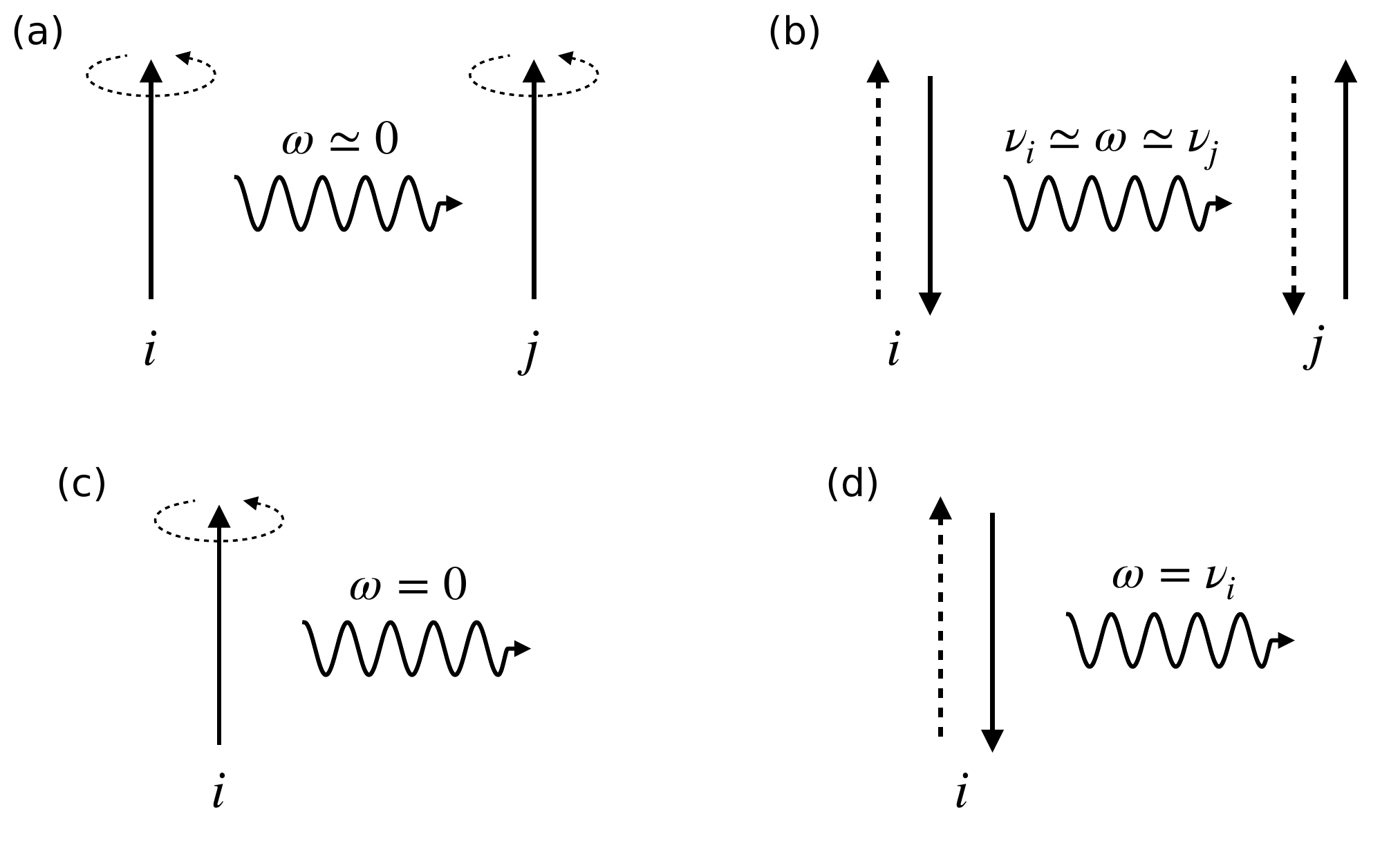}
    \caption{Virtual (a,b) and real (c,d) phonon processes that lead to TLS-TLS interactions and dissipation, respectively. (a) Interactions in the $S^z$--$S^z$ channel are mediated by phonons of vanishing frequency $\omega$, and take place among each couple $ij$, yielding $H_{\mathit{LS}}$ of Eq.~\eqref{eq:Lamb-Stark}. (b) Flip-flop interactions can take place only if the two TLSs resonate: this is a very rare event, because the $\nu_i$'s are widely distributed random variables. We disregard this possibility altogether throughout this study. (c) Dephasing of a single TLS under the action of the phonon bath. This process is negligible because there are no real phonons at $\omega=0$. (d) Decay of a TLS into a phonon. Considering that resonating TLSs are very rare and the phonon density of states vanishes at $\omega=0$, as noted above, it is easy to see that non-unitary processes involving two TLSs can be neglected.}
    \label{fig:processes}
\end{figure}

Before introducing the expressions for $Y_i$ and $J_{ij}$, a few comments are in order. As depicted in Fig.~\ref{fig:processes}, in general interactions can take place either in the $S^z$--$S^z$ channel (panel (a)), or by flipping two spins with the emission and absorption of a virtual phonon (panel (b)). This latter case, for our system, can be neglected: since $\nu_i$ and $\nu_j$ are random variables, the matching condition $\omega = \nu_i = \nu_j$ ($\omega$ is the phonon frequency), entailed by the rotating wave approximation, is a rare event. Thus, the Lamb-Stark shifts are always diagonal in $S^z$ \footnote{Even accounting for rare interactions in the $S^x$--$S^x$ channel, the picture is not modified. Indeed, terms of the form $K_{ij} S_i^x S_j^x$ will still decay with the distance $r_{ij}$: the probability of having a resonant $ij$ couple that is \emph{also close in real space} is vanishingly small. Therefore, the MBL-breaking effect of weak $S^x_i S^x_j$ terms \cite{Yao14dipolar,Burin15MBL,deng2020anisotropymediated} is negligible in comparison to the Lindblad dissipator.}.

Moreover, the Lindblad superoperators of Eq.~\eqref{eq:Lindblad} correspond only to the decay processes in Fig.~\ref{fig:processes}d, since purely dephasing processes (panel (c)) are absent. This is simply because there is no density of states of the phonons at zero frequency. 

Having understood what are the physical processes behind the GKSL evolution, we can compute explicitly the dissipation rates $Y_i$ and the interaction strengths $J_{ij}$. As stated above, they both come from phonon processes; therefore, one can treat them in a unified way. We start by rewriting the interaction Hamiltonian, Eq.~\eqref{eq:interaction_hamiltonian}, as
\begin{equation}
    \label{eq:def_Ei}
    H_{int} = \sum_{ik} \sigma_i^z \left( \xi_{ik} \psi_k + {\rm h.c}. \right)
    =: \sum_{i} \sigma_i^z E_i:
\end{equation}
$E_i$ are the environment operators that need to be traced out. Then, following \cite{BreuerPetruccione,Manzano2020Lindblad}, we define 
\begin{equation}
    \label{eq:def_Gamma}
    \Gamma_{ij}^\omega := \frac{1}{\hbar^2} \int_0^\infty ds \, e^{i \omega s} \, \Tr_B \left[ \rho_B^T \, \hat E_i^\dagger(t) \, \hat E_j(t-s) \right]
\end{equation}
with the hat on $\hat E_i(t)$ indicating the interaction picture. It then holds
\begin{align}
    \label{eq:def_Yi}
    Y_i &= \left(\frac{\Delta_i}{\hbar \nu_i} \right)^2 \big[ \Gamma_{ii}^{\nu_i} + \big(\Gamma_{ii}^{\nu_i} \big)^* \big] \Big|_{T=0}, \\
    \label{eq:def_Jij}
    J_{ij} &= \frac{2 \veps_i}{\hbar \nu_i} \frac{2 \veps_j}{\hbar \nu_j} \,\frac{ \hbar}{2i} \big[ \Gamma_{ij}^0 - \big( \Gamma_{ji}^0 \big)^* \big].
\end{align}
The prefactors $\Delta_i / \hbar \nu_i$ and $2 \veps_i / \hbar \nu_i$ come from the basis rotation in Eq.~\eqref{eq:S_basis}. 

We leave to Appendix~\ref{app:sec:computation_Lindblad} all the details of the computation of $\Gamma_{ij}^\omega$, which is rather straightforward, while we present here the results obtained:
\begin{align}
    \label{eq:Y_i}
    Y_i &= \frac{\Delta_i^2 \gamma_i^2 \nu_i \Tr (D^2_i) }{12 \pi\rho \hbar^3 v^5}, \\
    \label{eq:J_ij}
    J_{ij} &= \frac{\gamma_i \veps_i}{\hbar \nu_i} \frac{\gamma_j \veps_j}{\hbar \nu_j} \, \frac{\mathbb{D}_{ij} }{4 \pi \rho v^2 r_{ij}^3 }.
\end{align}
Above, $\Tr (D^2_i)=\sum_{ab}D^{ab}_iD^{ba}_i$, and $\mathbb{D}_{ij}$ is a specific contraction of the dipoles $D_i^{ab}$ and $D_j^{cd}$, defined in Eq.~\eqref{eq:def_Dij}.

At this point, we can check a posteriori whether the weak coupling and the rotating wave approximations are valid. Plugging in Eqs.~\eqref{eq:Y_i} and \eqref{eq:J_ij} the typical values of the parameters, we find $\hbar Y_i/W \sim 10^{-8}$ and $J_{ij}/W \sim 10^{-3}$. Therefore, even if the coupling constant is comparable to the on-site energies $\gamma \sim W$, we see that assuming weak coupling is perfectly justified a posteriori. Moreover, as anticipated at the beginning of this Section, the rotating wave approximation is amply valid too. Indeed, the relaxation time in the open system is much longer than the intrinsic time scale of TLSs: $Y_i ^{-1} \gg |\nu_i - \nu_j|^{-1} \sim \hbar W^{-1}$.

\subsection{Dynamical phases from the GKSL equation}
\label{sec:dynamical_phases}

The GKSL equation \eqref{eq:Lindblad} constitutes the starting point for exploring the quantum dynamics of the TLSs. As a first thing, we notice that in the absence of dissipation the evolution would be unitary, governed by the Hamiltonian 
\begin{equation}
    \label{eq:lbit_hamiltonian}
    H_{\it TLS} + H_{\it LS} = - \frac{1}{2} \sum_i \hbar \nu_i S_i^z + \sum_{ij} J_{ij} S_i^z S_j^z.
\end{equation}
$H_{\it TLS} + H_{\it LS}$ is completely expressed in terms of the extensive set of local conserved quantities $S^z_i$. This is the same property of the effective Hamiltonian of MBL systems, known as the \emph{l-bit Hamiltonian} \cite{huse2014phenomenology,Abanin2016Explicit,ros2015integrals,imbrie2014many,imbrie2016review}. Borrowing the terminology from MBL, we can refer to the $S^z_i$'s as the l-bits, or Local Integrals Of Motion (LIOMs); indeed, they are on-site operators whose values are conserved during time evolution. However, $H_{\it TLS} + H_{\it LS}$ presents two main differences with respect to the l-bit Hamiltonian of standard MBL systems. First, in the TLS Hamiltonian the l-bits are formed by \emph{single} spins, not exponentially localized groups of them. Second, the interaction between the TLS decays with distance as a power law, $J_{ij} \propto r_{ij}^{-3}$, rather than exponentially. We will comment more on this point later, in Sec.~\ref{sec:unitary_dynamics}.

The diagonal interactions in $H_{\mathit{LS}}$ are responsible for the dephasing of the spins. That is to say, if one artificially turns off the jump operators, i.e.\ if one sets the dissipation rates $Y_i \equiv 0$, diffusive transport is suppressed but the entanglement spreading persists. We will present numerical results on this artificial situation in Sec.~\ref{sec:unitary_dynamics}, showing that the entanglement entropy grows slowly, but indefinitely in time, while the concurrence decays as a power law. 

\begin{figure}[t]
    \centering
    \includegraphics[width=\columnwidth]{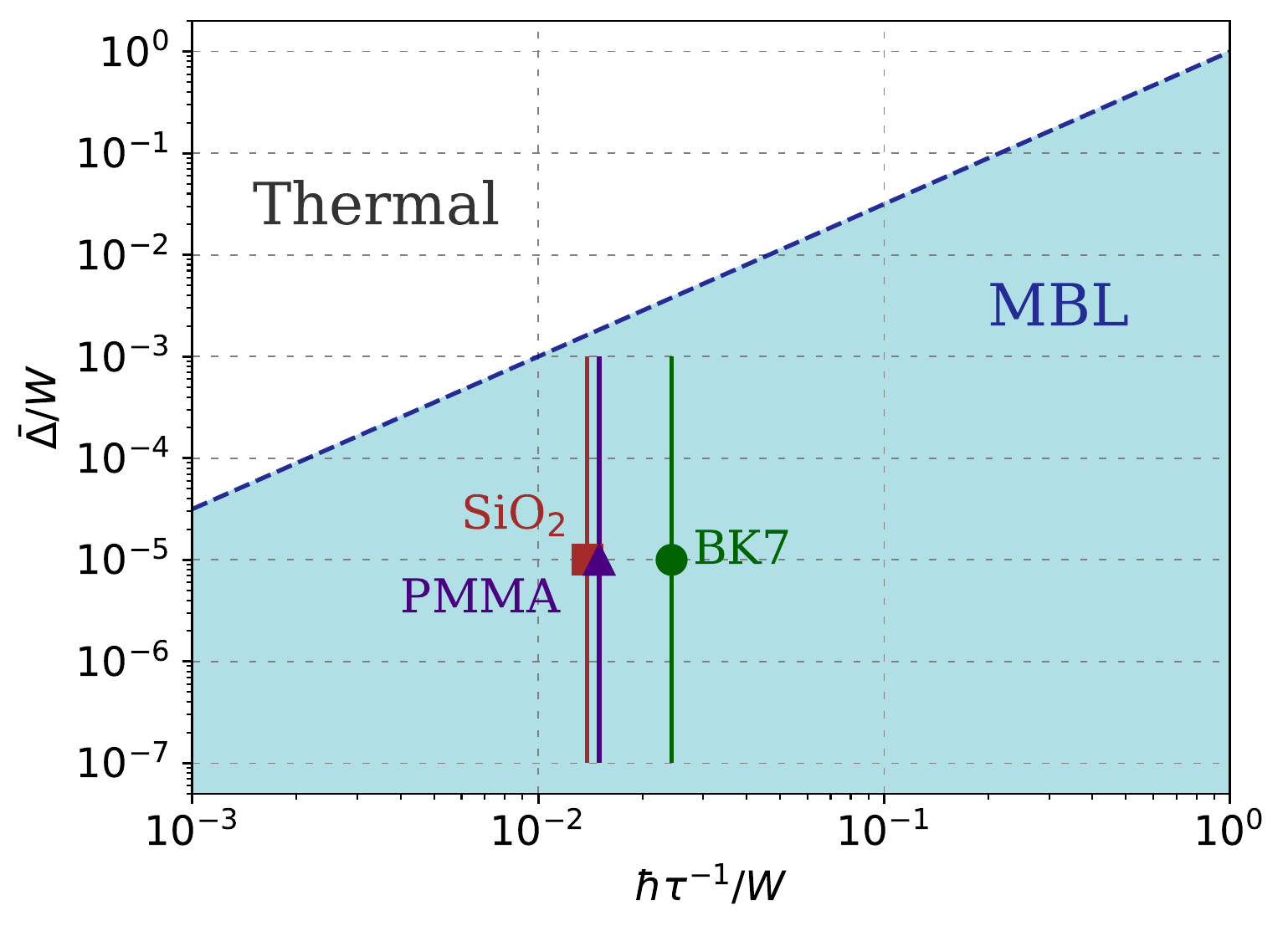}
    \caption{Sketch of the expected phase diagram for TLSs in glasses. From Eq.~\eqref{eq:interaction_dephasing_ratio} we see that an MBL transient regime can be observed before thermalization takes place, if the typical time scales of interaction are short with respect to the dissipation time scales (blue-shaded area). The three glassy materials reported in Table \ref{tab:parameters} lie well within the MBL region, even accounting for the large uncertainties in the parameter $\overline{\Delta}$ (the standard deviation of $\log( \Delta_i)$ is plotted as an errorbar). Thus, the localized regime should be experimentally observable.}
    \label{fig:phaseDiag}
\end{figure}

The picture described above is broken by the introduction of the jump operators: dissipative terms in the GKSL equation kill long-time coherence and drive the system to a thermal state. Nevertheless, one can observe an MBL transient regime in the relaxation dynamics, if the time scales of dissipation are appreciably longer than those of interactions. Such competition is quantified by the dimensionless ratio
\begin{equation}
    \label{eq:interaction_dephasing_ratio}
    \frac{\hbar Y_i}{J_{ij}}\sim \left(\frac{\overline{\Delta}}{W}\right)^2\left(\frac{W}{\hbar \tau^{-1}}\right)^3,
\end{equation}
where $\tau=r/v$, $r$ being the typical distance between TLSs and $v$ the speed of sound in the glass. If this ratio is sensibly smaller than 1, the signatures of the localized phase should be observed in the dynamics of the system, and in particular in the spreading of entanglement. In Fig.~\ref{fig:phaseDiag}, we show a tentative dynamical phase diagram for the TLS system.

Recalling that in experiments $\overline{\Delta} \sim 10^{-5}\, W$ while $W \sim 0.1$ eV and, considering $v \sim 5$ km/s and $r \sim 10$ nm, we have $\hbar \tau^{-1}\sim 1$ meV. Thus, the ratio is approximately $\hbar Y/J \sim 10^{-5}\div10^{-4}$, making dissipation much slower than the interaction part of the unitary dynamics. Even if one allows $\overline{\Delta}$ --- the most difficult parameter to infer from experiments --- to vary few orders of magnitude, the system will still present an observable MBL transient regime.

\section{Numerical Simulations}
\label{sec:numerical_simulations}

In this Section we present the results of our numerical simulations on the real-time evolution of the TLSs. The analysis will be divided into two parts. In Sec.~\ref{sec:unitary_dynamics}, we will consider the artificially isolated system (i.e. the one evolving only under the unitary dynamics given by the Liouvillian of the GKSL) governed by the Hamiltonian in Eq.~\eqref{eq:lbit_hamiltonian}. In Sec.~\ref{sec:dissipation}, we will re-introduce the dissipative terms and consider the full TLS evolution governed by the GKSL master equation \eqref{eq:our_Lindblad}.

Before going through that, in the next Sections \ref{sec:numerical_parameters}--\ref{sec:numerical_definitions} we will briefly discuss the assumptions involved in our numerical simulations, and define the dynamical observables.


\subsection{Disorder distributions of the parameters}
\label{sec:numerical_parameters}

As discussed in Sec.~\ref{sec:TLSmodel}, in the literature the parameters defining the TLS model are drawn from wide probability distributions (see $p_{\veps}$ and $p_{\Delta}$ in Eq.~\eqref{eq:pdf_veps} and \eqref{eq:pdf_Delta}, respectively). It follows that the competing time scales in the GKSL master equation (namely $\nu_i^{-1}$, $\hbar/J_{ij}$, and $Y_i^{-1}$) are distributed across several orders of magnitude and, even though their typical values are very different, they overlap one with another. In our numerical simulations we employ simplified and less broad distributions, arguing that this choice, if properly taken, does not qualitatively alter the physical content and predictions of the model.

\begin{figure}[t]
    \centering
    \includegraphics[width=0.65\columnwidth]{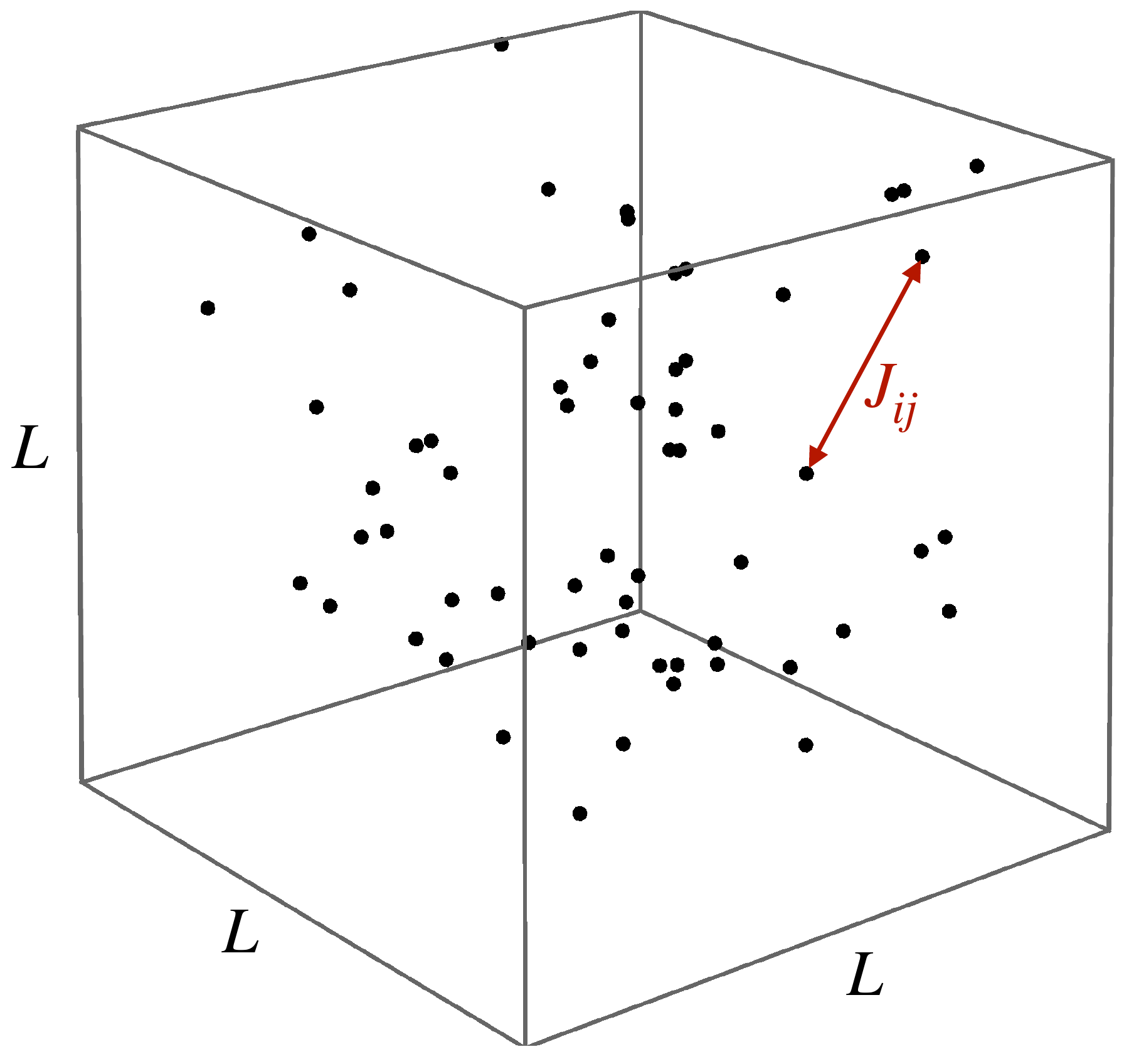}
    \caption{The TLSs are uniformly distributed in a cube of size $L$, at constant density. The pairwise interactions $J_{ij}$ in Eq.~\eqref{eq:J_ij} are mediated by phonons. These are also responsible for the dissipation in Eq.~\eqref{eq:Y_i}. We employ periodic boundary conditions to minimize finite-size effects. }
    \label{fig:TLS_points}
\end{figure}

We fix $W \equiv 1$, thus setting the (dimensionless) energy scale; $\overline{\Delta} / W=10^{-1}$, unless otherwise specified, and $n_\Delta = 2$. We also set $\gamma_i \equiv W$, the material density $\rho = 2$ g/cm$^3$, and the speed of sound $v_{L,T} = 5$ km/s, irrespective of polarization. We consider $\Tr(D_i^2)$ to be the square of a Gaussian random variable of zero average and variance 1, since it must be positive, and $\mathbb{D}_{ij}$ to be a Gaussian random variable of zero average and standard deviation 1, since it can take both signs (see also App.~\ref{app:sec:computation_Lindblad}). Finally, we consider the TLSs as uniformly distributed in a cube with side $L$, and compute their distances $r_{ij}$ using periodic boundary conditions. The cube side depends on the number of TLSs as $L = L_0 N^{1/3}$, with $L_0 \simeq \rho_{\mathit{TLS}}^{-1/3}$, so that we keep fixed the TLS number density $\rho_{\mathit{TLS}}$. For numerical purposes, we fix $L_0 = 1$ nm. See Table~\ref{tab:parameters} for a comparison with the experimental values, and Fig.~\ref{fig:TLS_points} for a sketch of the system.

In order to explore the phase diagram obtained in the GKSL framework, and shown in Fig.~\ref{fig:phaseDiag}, we introduce two further artificial parameters to tune interaction and dissipation strengths:
\begin{equation}
    \label{eq:rescalings}
    J_{ij} \to \eta J_{ij} , \qquad
    Y_i \to \epsilon Y_i .
\end{equation}
In Sec.~\ref{sec:unitary_dynamics} we study the artificially isolated system, setting $\eta=10^5$ and $\epsilon=0$. In Sec.~\ref{sec:dissipation} we re-introduce the dissipator in the GKSL master equation, and we set $\eta=10^5$ and $\epsilon=10^{-6},10^{-4},1$.

With these choices of the parameters, the on-site frequencies $\nu_i$, the TLS-TLS interactions $\eta J_{ij}/\hbar$ (with $\eta = 10^5$), and the dissipation rates $\epsilon Y_i$ (for $\epsilon=1$) are of comparable orders of magnitude and are much less widely distributed than originally. The latter feature is particularly useful for numerical purposes, since one can access only small system sizes and, hence, cannot sample well broad distributions. Our results will be discussed in view of these choices.

\subsection{Initial state and dynamical observables}
\label{sec:numerical_definitions}

We always take the initial state of the dynamics to be a product state, in which each TLS is represented by a random vector on the Bloch sphere:
\begin{equation}
    | \psi (0) \rangle = \bigotimes_{i=1}^{N}\bigl(\cos (\theta_i/2) | \uparrow \rangle_i + e^{i \phi_i} \sin (\theta_i/2) | \downarrow \rangle_i\bigr),
    \label{eq:lbit_initial_state}
\end{equation}
where $\theta_i \in [0,\pi]$ and $\phi_i \in [0,2 \pi)$. Thus, the system is initially at infinite temperature, and we can track precisely the entanglement growth and spreading. 

The choice of the appropriate entanglement measure is not obvious: since we are dealing with an open quantum system, we wish to discriminate between quantum entanglement and thermal entropy. A reliable measure of (pairwise) quantum entanglement in open systems is the \emph{concurrence} $C_{ij}$ \cite{hill1997entanglement,Wootters98,Amico08}, where $i$ and $j$ are TLS indices. The concurrence quantifies the distance of the two-site reduced density matrix $\rho_{ij}$ from the manifold of mixed, separable states whose reduced density matrix can be written as $\rho=\sum_a p_a \rho_a^{sep}$, where $\rho_a^{sep}$ are separable, $p_a\geq 0$, and $\sum_a p_a = 1$ . This implies that, if $C_{ij}> 0$, there is no mixture of separable states that can account for the correlations between sites $i$ and $j$. For two spins $1/2$, it can be shown \cite{Amico08} that 
\begin{equation}
    \label{eq:def_concurrence}
    C_{ij} = \max{ \{0, \, \lambda_1 - \lambda_2 - \lambda_3 - \lambda_4 \}},
\end{equation}
where $\lambda_a^2$ are the eigenvalues of the matrix $R_{ij} = \rho_{ij} (\sigma_y \otimes \sigma_y) \rho^*_{ij} (\sigma_y \otimes \sigma_y)$ sorted in descending order, and the complex conjugation is done in the standard computational basis.

We define the average concurrence as
\begin{equation}
    \label{eq:average_concurrence}
    C(t) := \frac{1}{N}\sum_{1\leq i<j\leq N}C_{ij}(t).
\end{equation}
The normalization factor $1/N$ (instead of the seemingly natural $1/N^2$) is due to the \emph{monogamy of entanglement}: each TLS can be highly entangled only with another TLS, so among the $N(N-1)/2$ terms in the sum, only $O(N)$ will be non-negligible.

Its particular definition allows the concurrence to quantify the entanglement between the two TLSs considered, irrespective of how they are entangled with other degrees of freedom. Thus, it spots entanglement between two TLSs even if they are thermal, i.e.\ also entangled with a heat bath. For this reason, we employ the concurrence as a well-defined entanglement measure both in the absence (Sec.~\ref{sec:unitary_dynamics}) and in the presence (Sec.~\ref{sec:dissipation}) of dissipation. 

It is interesting to compare the time behavior of the concurrence with the half-system entanglement entropy (HSEE)
\begin{equation}
    \label{eq:HSEE}
    S_E(t) = - \Tr ( \rho_A \log \rho_A),
\end{equation}
where $\rho_A$ is the reduced density matrix of the half system $A$ in the bipartition $A|B$. Since the system is three-dimensional, and the TLSs do not fall on a regular lattice, we bipartite the system in the following way. For each TLS, a bubble is constructed around it so that $N/2$ TLSs fall inside and $N/2$ outside the bubble. The entanglement entropy relative to the bipartition is computed as in Eq.~\eqref{eq:HSEE}, and then averaged over all such bipartitions. We measure $S_E(t)$ both with and without the dissipator (see Sec.~\ref{sec:unitary_dynamics} and \ref{sec:dissipation}, respectively).

\subsection{Unitary evolution of the TLSs}
\label{sec:unitary_dynamics}

This Section is entirely devoted to the study of the unitary time evolution of a system of $N$ TLSs governed by the Hamiltonian $H_{\it TLS} + H_{\it LS}$ of Eq.~\eqref{eq:lbit_hamiltonian}, that we reproduce here for clarity:
\begin{equation*}
    H_{\it TLS} + H_{\it LS} = - \frac{1}{2} \sum_i \hbar \nu_i S_i^z + \sum_{ij} J_{ij} S_i^z S_j^z.
\end{equation*}
As discussed in Sec.~\ref{sec:dynamical_phases}, this Hamiltonian is diagonal in the operators $S^z_i$, i.e.\ the values assumed by $S^z_i$ are conserved quantities; therefore, adopting the terminology of MBL systems, we say that $H_{\it TLS} + H_{\it LS}$ is in the l-bit form, and $S^z_i$ are LIOMs. 

Studying the dynamics induced only by the Hamiltonian term of the GKSL equation \eqref{eq:our_Lindblad} is equivalent to set $\epsilon=0$ (see Eq.~\eqref{eq:rescalings}), i.e.\ to assume that the time scales of dissipation are much longer than those of interactions: $1/Y_i \gg \hbar/J_{ij}$. In this limit, it is clear that a coherent many-body dynamics can take place before thermal equilibrium is reached. This situation corresponds to the bulk of the MBL phase depicted in the phase diagram of Fig.~\ref{fig:phaseDiag}.

Thanks to the diagonal nature of the Hamiltonian \eqref{eq:lbit_hamiltonian} and to the choice of initial product states, few-sites observables are efficient to compute, as was recognized in previous studies \cite{serbyn2014quantum,iemini2016signatures,Znidaric18}. We refer the interested reader to App.~\ref{app:sec:numerics_unitary_dynamics} for more details on the computation. Here, we just mention that to compute the concurrence, which is a two-site observable, within the diagonal Hamiltonian \eqref{eq:lbit_hamiltonian} it is not necessary to perform the time evolution of the whole $2^N\times 2^N$ density matrix, but only to carry out $O(N)$ operations. Therefore, we could easily simulate systems of $N=60$ TLSs. 

\begin{figure}[t]
    \centering
    \includegraphics[width=\columnwidth]{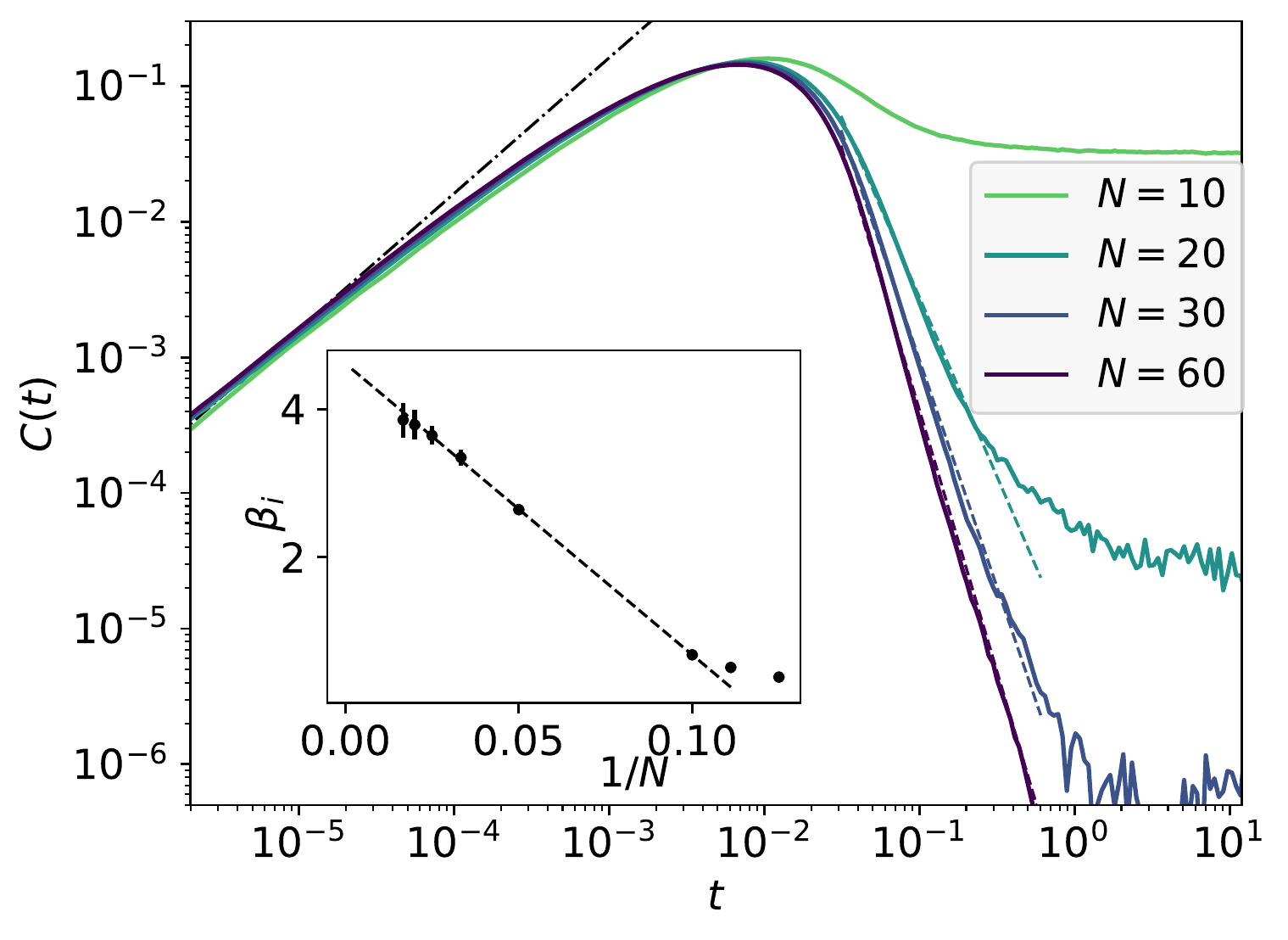}
    \caption{Average concurrence within the unitary dynamics, $\epsilon=0$ (solid lines). After a linear raise $C \sim t$ (black dashed-dotted line), the average concurrence decays with a power-law $C\sim t^{-\beta_i}$ (dashed lines), down to a value which is exponentially small in $N$. We set $\overline \Delta = 0.1$, $\eta = 10^5$; the results are averaged over 5000 disorder realizations.
    \emph{Inset:} The exponent $\beta_i$ depends on $N$ and reaches a finite value in the thermodynamic limit. The errors are computed by using the statistical uncertainties of the concurrence values. Not all datasets were shown in the main figure to improve readability.}
    \label{fig:Ct-Unit}
\end{figure}

\begin{figure}[t]
    \centering
    \includegraphics[width=\columnwidth]{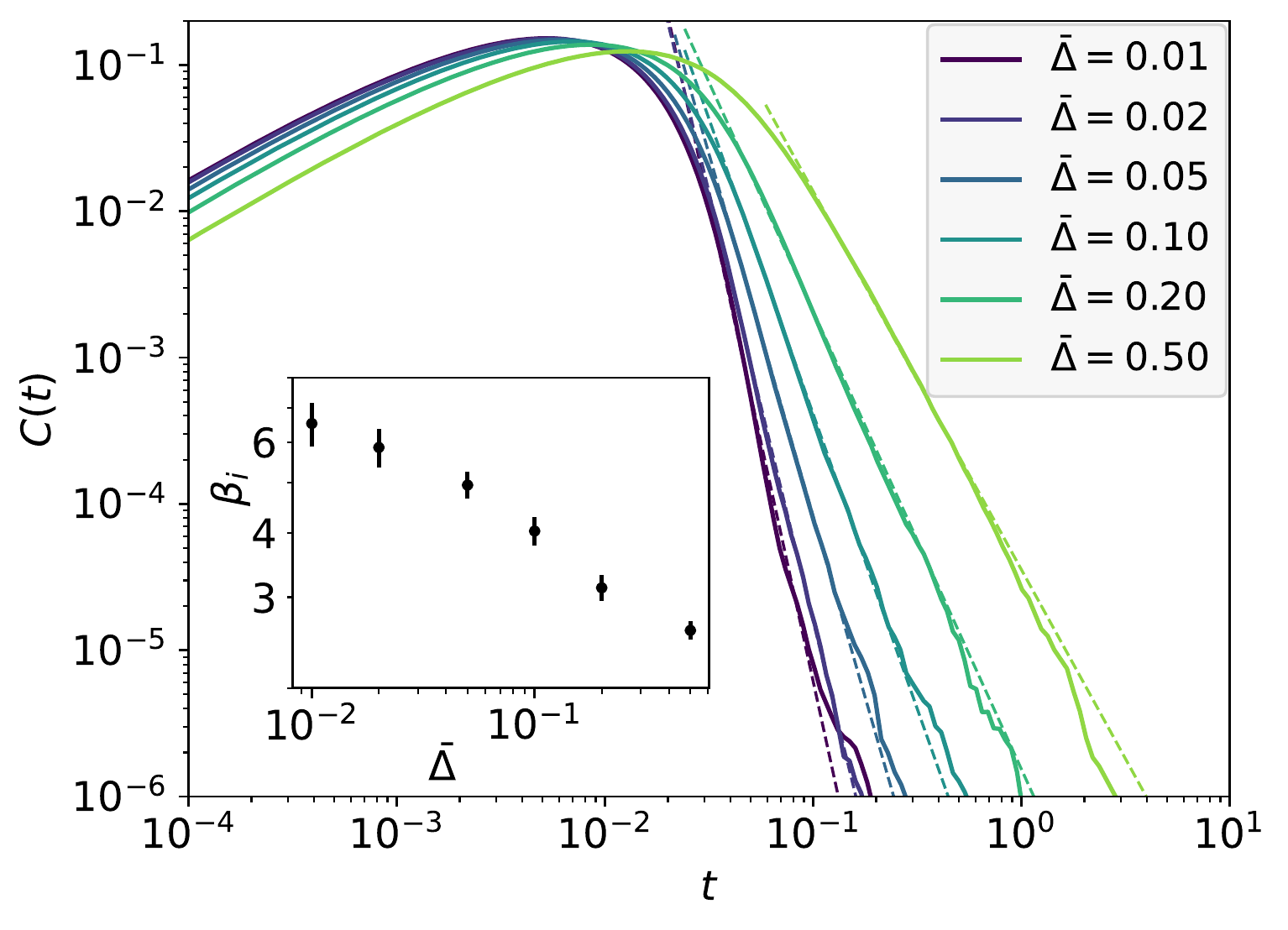}
    \caption{Dependence of the average concurrence decay exponent $\beta_i$ on $\overline{\Delta}$ in the case of unitary evolution ($\epsilon=0$). We set $N = 50$, $\eta = 10^5$ and averaged over 5000 disorder realization. We see that the smaller $\overline{\Delta}$, the faster the decay, which remains however compatible with a power-law $C(t)\sim t^{-\beta_i}$ (dashed lines).}
    \label{fig:Ct-Unit_Delta}
\end{figure}

\begin{figure}[t]
    \centering
    \subfloat[]{\includegraphics[width=\columnwidth]{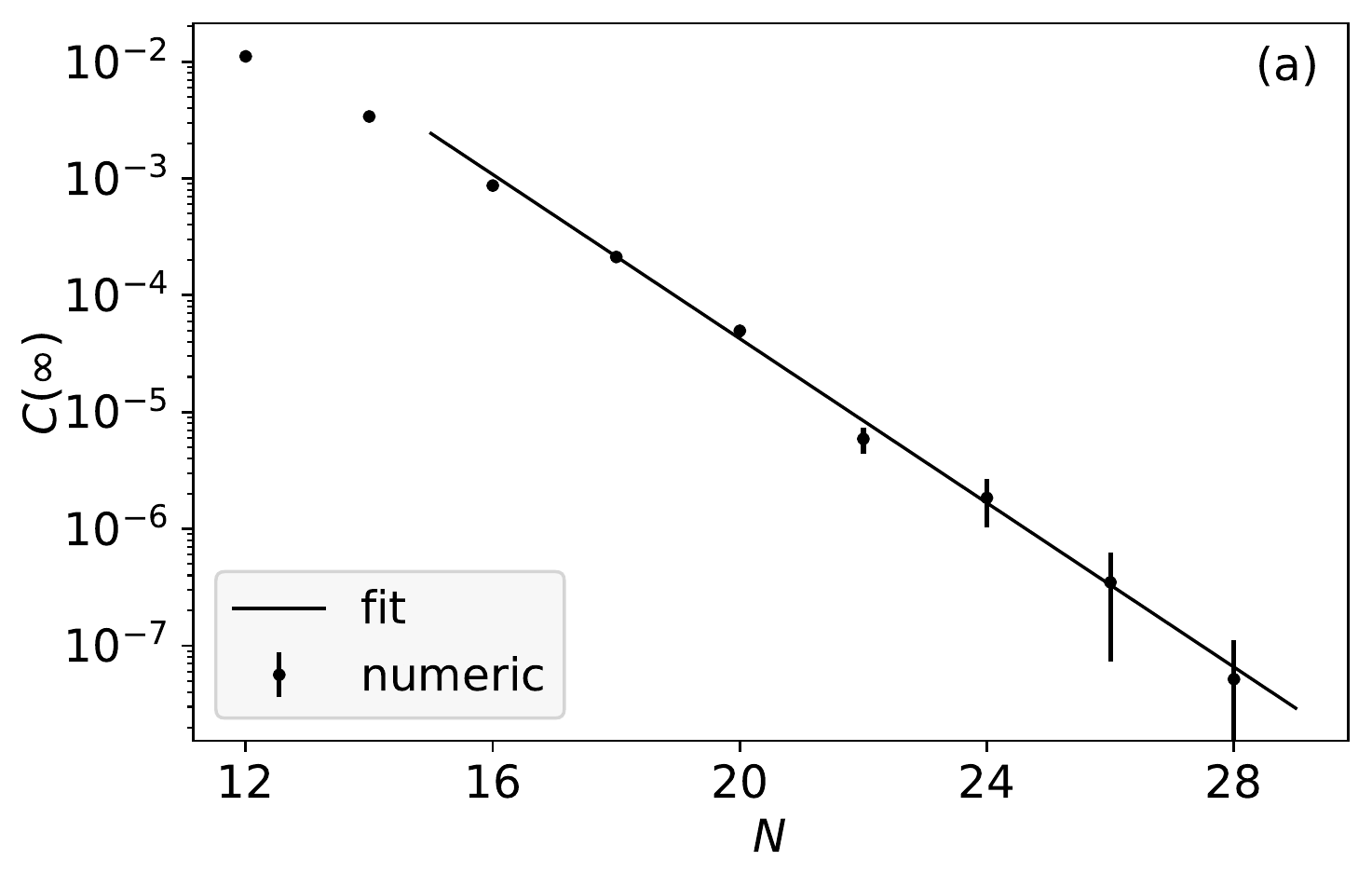}
    \label{fig:plateau-Unit}}
    \vspace{-6mm}
    \subfloat[]{\includegraphics[width=\columnwidth]{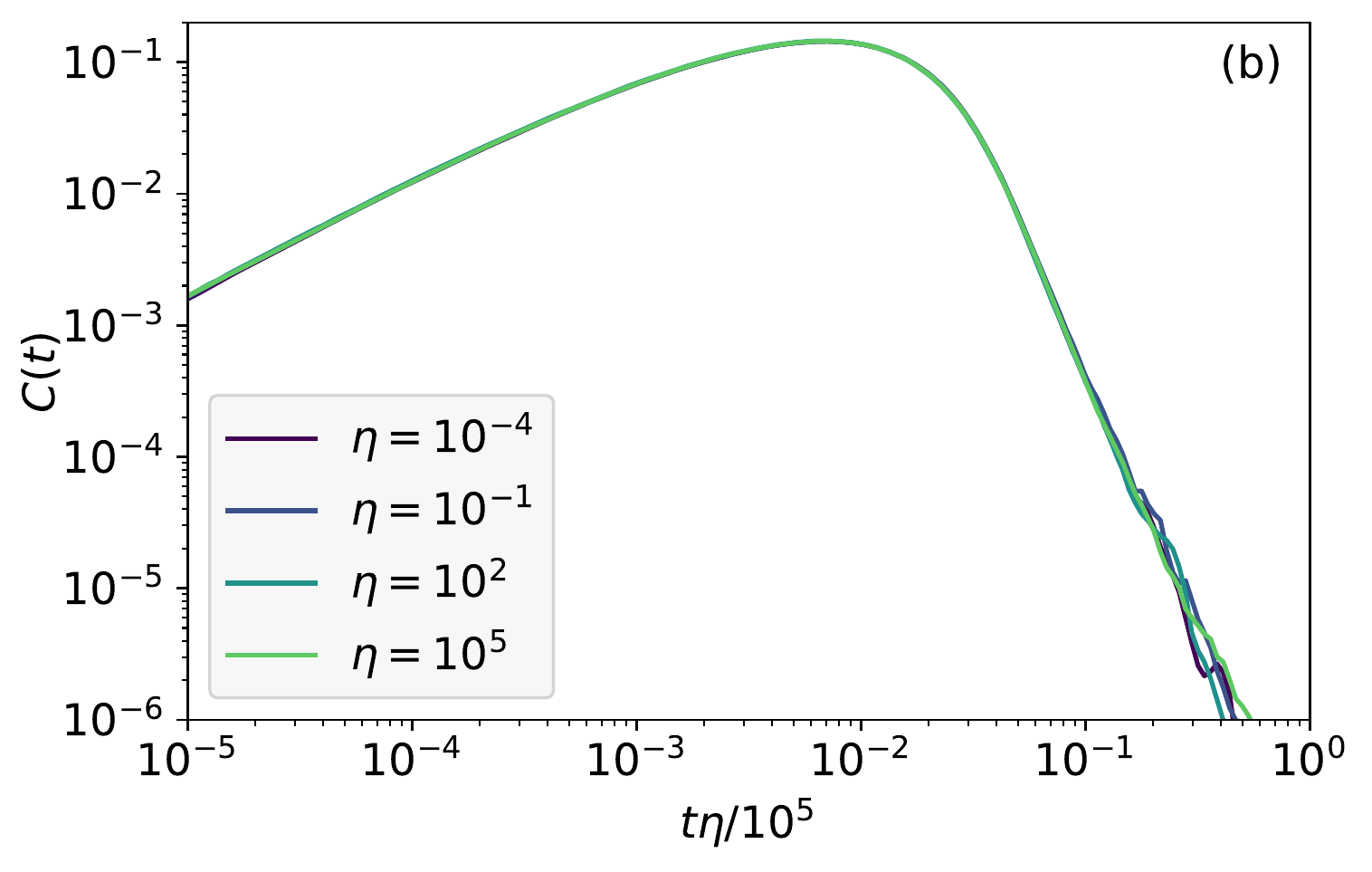}
    \label{fig:rescaled-R}}
    \caption{Results for the unitary dynamics, $\epsilon=0$. (a) Plateau value of the average concurrence at long times (dots), with errors coming from statistical fluctuations. From a fit (solid line) we find that $C(\infty) \propto e^{-\alpha N}$ with $\alpha \approx 0.8$. This is considerably larger than the value given by the ETH prediction, i.e.\ a random state, which obeys $C \propto e^{-a 2^{N}}$ with $a \approx 0.127$ (see App.~\ref{app:sec:random_state}). Here $\eta=10^5$, $\overline\Delta=0.1$, and an average over 10000 disorder realizations was performed.
    (b) Average concurrence for different interaction strengths $\eta$. Rescaling the time as $t \to t \eta/ 10^5$ (we normalize to $\eta=10^5$ to compare to the other plots) the curves collapse, showing that the value of $\eta$ only shifts the timescale but does not modify the shape of the curve $C(t)$. Here $N=50$, $\overline{\Delta}=0.1$, and an average over 1000 disorder realizations was performed.}
    \label{fig:plateauUnit&rescaledR}
\end{figure}

The results of the simulations for the unitary evolution are shown in Figs.~\ref{fig:Ct-Unit}, \ref{fig:Ct-Unit_Delta}, and \ref{fig:plateauUnit&rescaledR}. One can see that the concurrence $C(t)$, defined in Eq.~\eqref{eq:average_concurrence}, raises linearly from the initial value 0 (the initial state is factorized) to a value independent of $N$ (Fig.~\ref{fig:Ct-Unit}), but slightly dependent on $\overline{\Delta}$ (Fig.~\ref{fig:Ct-Unit_Delta}). It then falls off to a plateau via a power-law decay, whose exponent $\beta_i$ remains finite in the thermodynamic limit (inset of Fig.~\ref{fig:Ct-Unit}), and depends on $\overline{\Delta}$ (inset of Fig.~\ref{fig:Ct-Unit_Delta}). Fig.~\ref{fig:plateau-Unit} shows that the concurrence plateau decays exponentially with the system size: $C(\infty) \propto e^{-\alpha N}$. Finally, from Fig.~\ref{fig:rescaled-R} we see that the concurrence reaches its maximum on time scales of order $\hbar / J_{ij}$. In conclusion, the concurrence time behavior can be schematized as
\begin{equation}
    C(t)\sim
    \begin{cases}
        t                   &\mathrm{if}\ t<t_{1}\\
        t^{-\beta_i}        &\mathrm{if}\ t_1<t<t_2\\
        e^{-\alpha N}       &\mathrm{if}\ t>t_2,
    \end{cases}
\end{equation}
where $t_1$ does not depend significantly on $N, \overline{\Delta}$ but depends parametrically on $\hbar/J_{ij}$, while $t_2$ grows with $N$ and diverges in the thermodynamic limit.

\begin{figure}[t]
    \centering
    \includegraphics[width=\columnwidth]{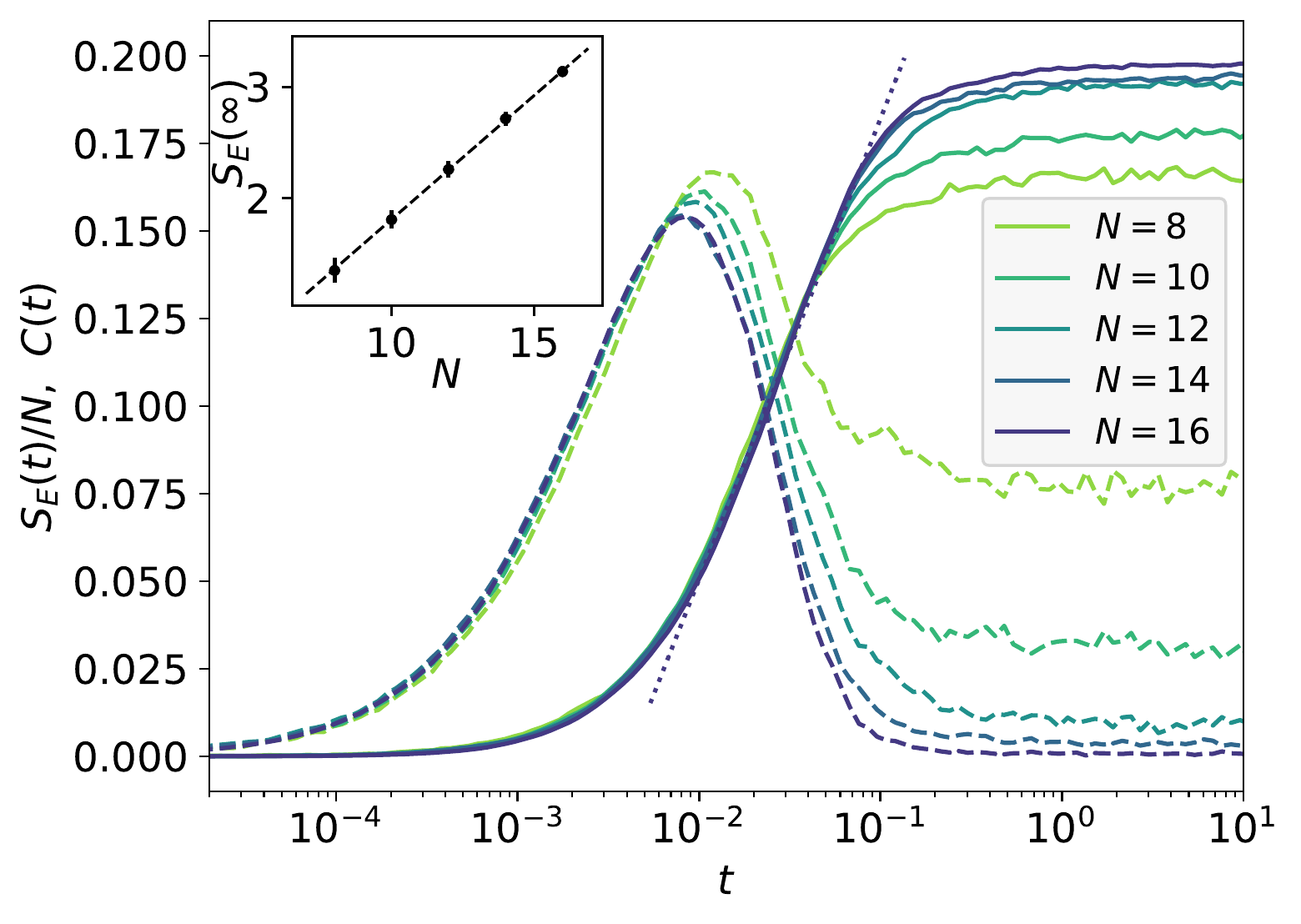}
    \caption{Unitary evolution: half-system entanglement entropy per unit volume $S_E(t) / N$, as defined in Eq.~\eqref{eq:HSEE}, for various system sizes (solid lines). We set $\overline \Delta = 0.1$, $\eta = 10^5$, and averaged over 1000 disorder realization. The average concurrences $C(t)$ (Eq.~\eqref{eq:average_concurrence}) are shown as dashed lines for comparison. We see that the concurrence reaches a maximum at short times, as nearby TLSs start to evolve coherently. Then, it starts to decay because the entanglement becomes many-body, as shown by the increase in the HSEE. In this regime, the growth of the HSEE is compatible both with a small power law $S_E(t) \sim t^{\alpha}$ with $\alpha \sim 1$, as well as $\log(t)$; the dotted line shows $\log(t)$ as a guide for the eye. \emph{Inset:} The HSEE saturates to a volume law, as expected for an MBL system: the phase of each spin depends on all the others. The error bars are computed from the statistical fluctuations of the plateau values.}
    \label{fig:EE-Unit}
\end{figure}

The decay of the concurrence from its maximum is due to the fact that the interactions $J_{ij}$ make the entanglement spread among many TLSs, as illustrated in Fig.~\ref{fig:EE-Unit}, while each TLS cannot be highly entangled with more than one other TLS because of the monogamy of the entanglement. The power-law decay of the concurrence from its maximum is in contrast to the behavior of ergodic systems, in which the concurrence vanishes exponentially fast \cite{iemini2016signatures}. The slowness of such decay is the fingerprint of the lack of thermalization and of the presence of many-body localization in the artificially isolated TLS system. In fact, slow decays of correlation functions are known \cite{serbyn2014quantum} to be a feature of MBL dynamics, and the concurrence (albeit not an operator nor a correlation function) follows the same behavior.

We stress again that $H_{TLS}+H_{LS}$, although completely expressed in terms of local integrals of motion, is different from the effective l-bit Hamiltonian of MBL systems, as already pointed out in Sec.~\ref{sec:dynamical_phases}. In particular, the TLS interactions in $H_{\it LS}$ scale as a power law with distance. Following general arguments \cite{serbyn2014quantum,pino2014entanglement}, one would expect that for long-range interactions the correlation functions decay in time as stretched exponentials. We cannot exclude that a stretched-exponential behavior would be observed in TLSs if one pushes the dynamics at larger times. In the present study, however, we are only interested in the TLS relaxation dynamics at intermediate time scales since, at long times, dissipation would always bring the system to a thermal state.

The results on the half-system entanglement entropy (HSEE) are shown in Fig.~\ref{fig:EE-Unit}, compared with the behavior of $C(t)$. This comparison confirms, as anticipated, that the concurrence starts to decrease when the entanglement spreads and, thus, $S_E(t)$ starts to increase. 

In addition, Fig.~\ref{fig:EE-Unit} shows that $S_E(t)$ grows slowly for a large time window. This slowness is known \cite{bardarson2012unbounded} to be the signature of localization, and shows that TLSs remain coherent and non-ergodic during the time-evolution. According to the arguments in \cite{serbyn2013universal,huse2014phenomenology,pino2014entanglement}, we expect that for a long-range, 3$d$ system as the TLS one entanglement would grow algebraically in time, $S_{E}(t) \sim t^\alpha$ with $\alpha\sim1$. From our data, the entanglement growth is compatible with both a power-law with small exponent ($\sim 1$), and a logarithmic growth. In the inset of Fig.~\ref{fig:EE-Unit}, we see that the asymptotic value of HSEE, $S_E(\infty)$, is proportional to $N$, indicating a volume law.

\subsection{Full evolution of the TLSs}
\label{sec:dissipation}

This Section is entirely devoted to the study of the time evolution of the TLSs governed by the GKSL master equation \eqref{eq:our_Lindblad}, that we reproduce here for clarity:
\begin{align*} 
\partial_t \rho(t) &= - \frac{i}{\hbar} \bigg[ - \sum_i \frac{\hbar \nu_i}{2} S_i^z + \sum_{ij} J_{ij} S_i^z S_j^z, \; \rho(t) \bigg] \\
    &+ \sum_{i} Y_i f_T(\hbar \nu_i) \left[ S_i^+ \rho(t) S_i^- +  S_i^- \rho(t) S_i^+  - 4 \rho(t) \right] \\
    &+ \sum_{i} Y_i \left[ S_i^+ \rho(t) S_i^- + \left\{ \rho(t),S^z_i \right\} - 2 \rho(t) \right].
\end{align*}
We set $T=0$, but $\epsilon \neq 0$ (see Eq.~\eqref{eq:rescalings}), i.e.\ the system is in the presence of dissipation and decoherence. Increasing $\epsilon$, we increase the typical dissipation rate. For our particular choice of parameters (Sec.~\ref{sec:numerical_parameters}), when $\epsilon=1$, dissipation ultimately becomes comparable with the timescale of the interactions $J_{ij}$.

To investigate the time evolution of the system, one has to integrate numerically the GKSL master equation for the TLS density matrix (see App.~\ref{app:sec:full_lindblad} for more details). Because of the doubling of the Hilbert space dimension, we are forced to small system sizes, up to $N=9$. In the following analysis, we varied both $N$ (to perform a finite-size scaling) and $\epsilon$.

\begin{figure}
    \centering
    \includegraphics[width=\columnwidth]{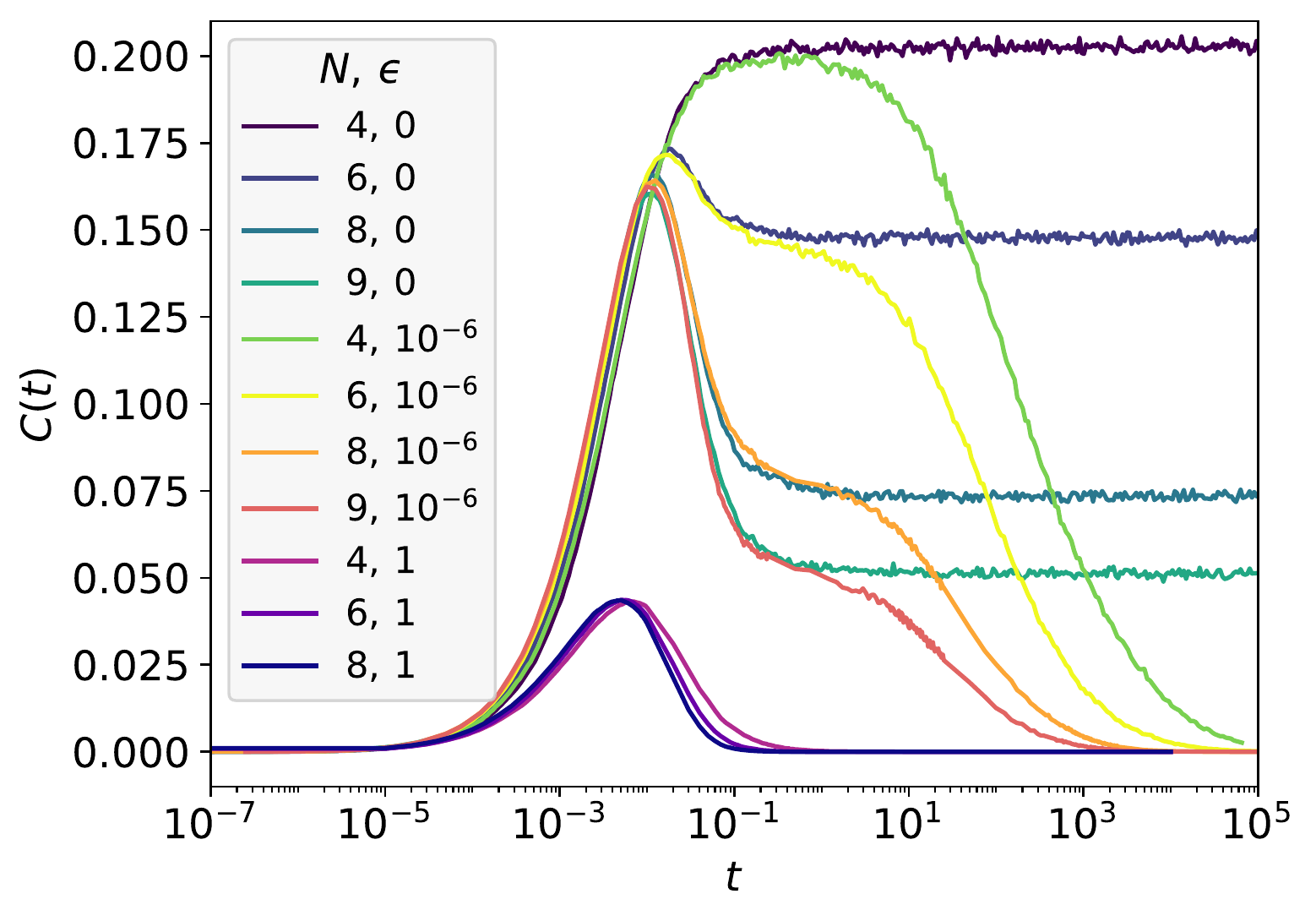}
    \caption{$C(t)$, as defined in Eq.~\eqref{eq:average_concurrence}, for $\epsilon=0,10^{-6},1$, and different values of $N$. We see that the presence of dissipation in the GKSL master equation \eqref{eq:our_Lindblad} decreases the concurrence maximum and moves it at earlier times. We set $\overline{\Delta}=0.1$, $\eta=10^5$, and averaged over at least 1000 disorder realizations.}
    \label{fig:concAllCurves}
\end{figure}

\begin{figure}
    \centering
    \includegraphics[width=\columnwidth]{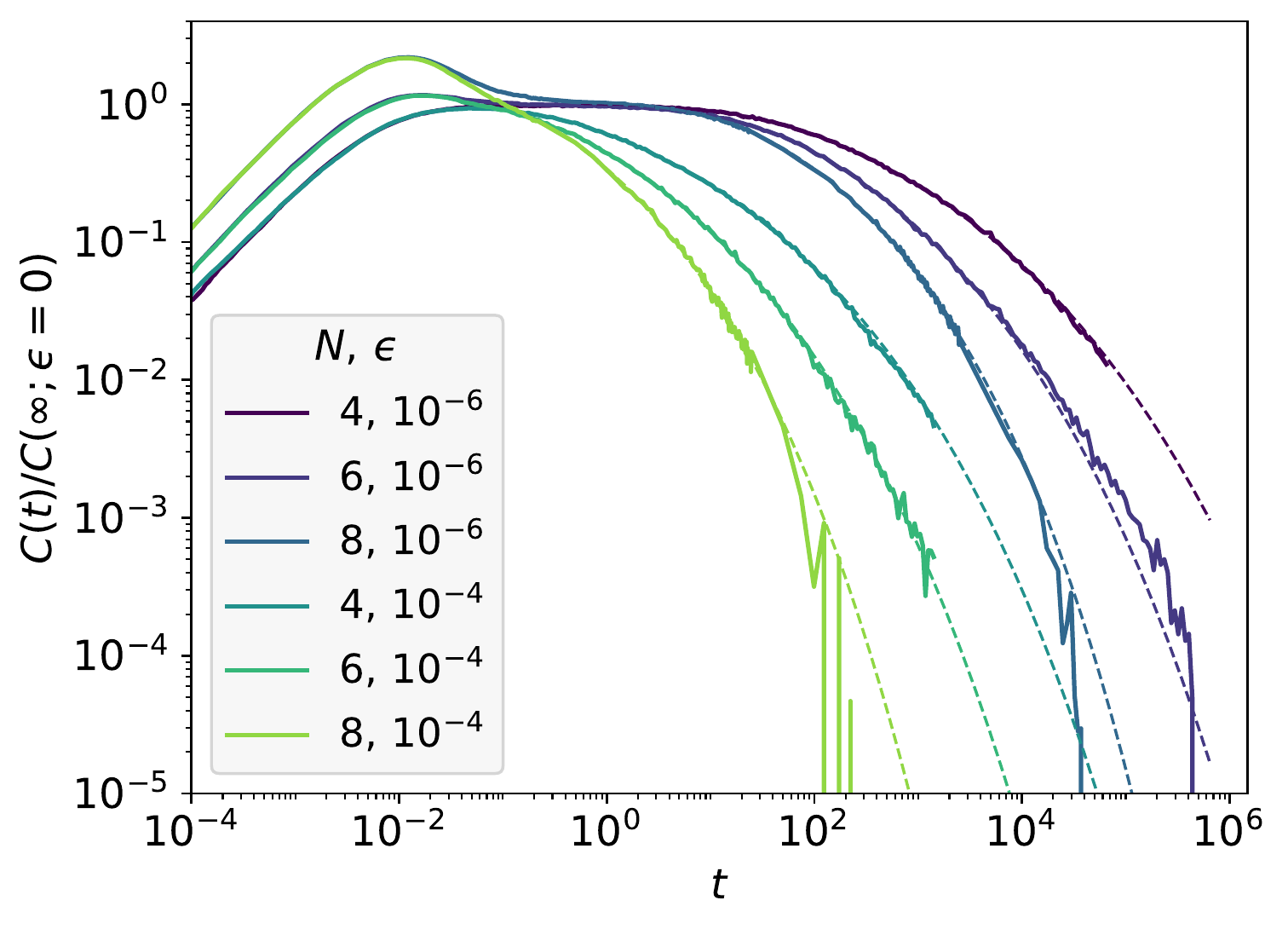}
    \caption{Stretched-exponential fit of the concurrence for $\epsilon=10^{-6}, 10^{-4}$, normalized to the plateau reached at $\epsilon=0$: $C(t; \epsilon)/C(\infty; \epsilon=0)$. Using as fitting function $\alpha \exp{\{-(\frac{t+t_0}{\tau})^\delta\}}$, we obtained $\delta\simeq0.2$ and $\tau=O(1)$. The plot shows the results for $\overline{\Delta}=0.1$, $\eta=10^5$, averaged over at least 1000 disorder realizations.} 
    \label{fig:dephasingStretchedExp}
\end{figure}

As can be seen from Fig.~\ref{fig:concAllCurves}, when $\epsilon$ is small enough the concurrence $C(t)$ reaches its maximum  at the same time as with unitary dynamics ($\epsilon=0$). Then, it decays from such peak and stabilizes around a finite value dependent on $N$ (cf. Sec.~\ref{sec:unitary_dynamics}), following the same behavior as in the case $\epsilon=0$. Ultimately, the dissipation forces $C(t)$ to vanish; $C(t)$ departs from the $\epsilon=0$ plateau, $C(\infty;\epsilon=0)$, with a stretched-exponential functional form (Fig.~\ref{fig:dephasingStretchedExp}). We can ascribe this feature to the interaction between TLSs and phonons: when $\epsilon \neq 0$, thanks to the dissipative terms in the GKSL equation \eqref{eq:our_Lindblad}, entanglement can spread among infinitely many phonons, preventing the concurrence from stabilizing around the plateau value.

\begin{figure}
\centering
\subfloat[]
{\includegraphics[width=\columnwidth]{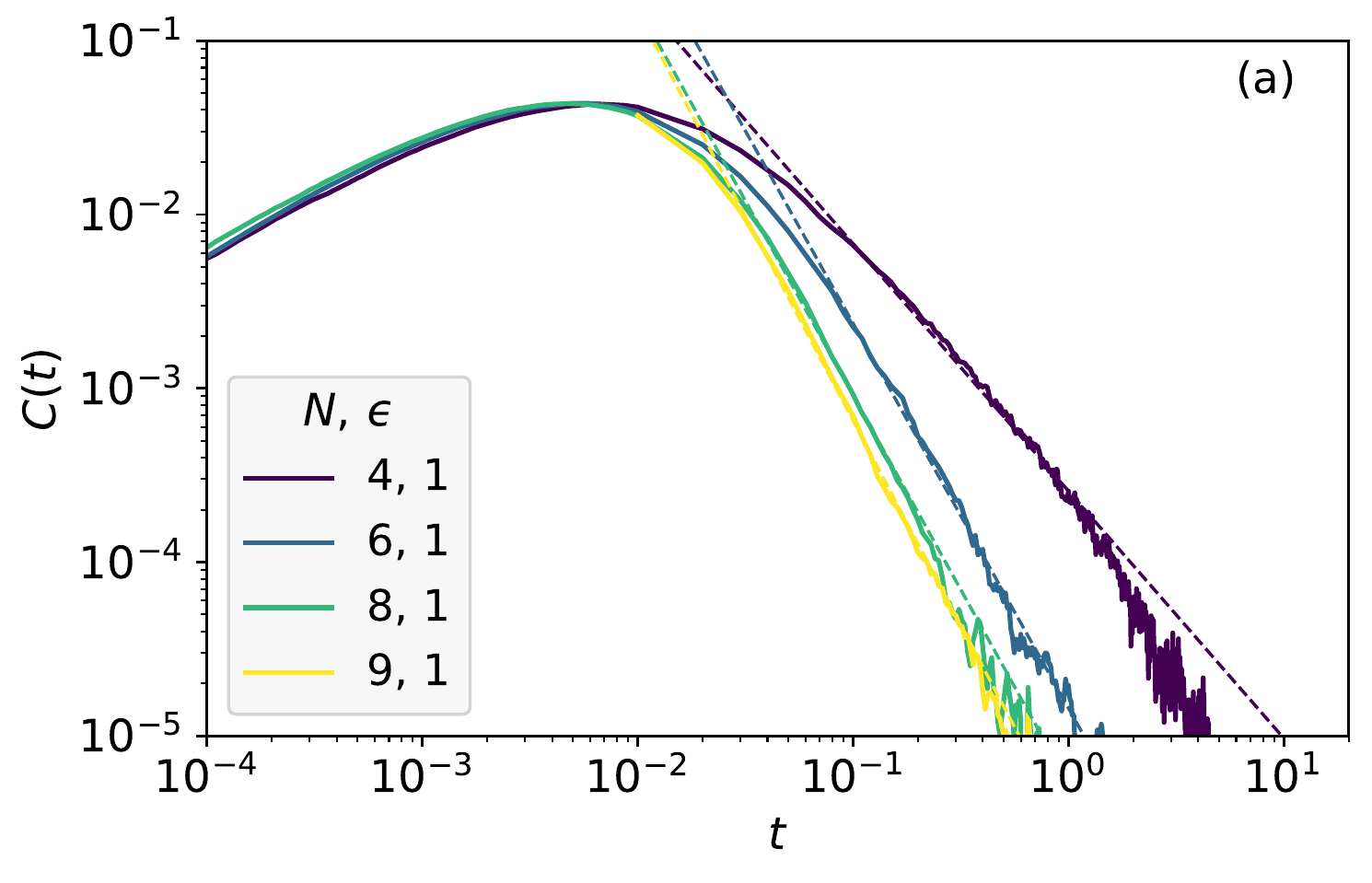}
\label{fig:dephasingLargeEps}}
\vspace{-6mm}
\subfloat[]
{\includegraphics[width=\columnwidth]{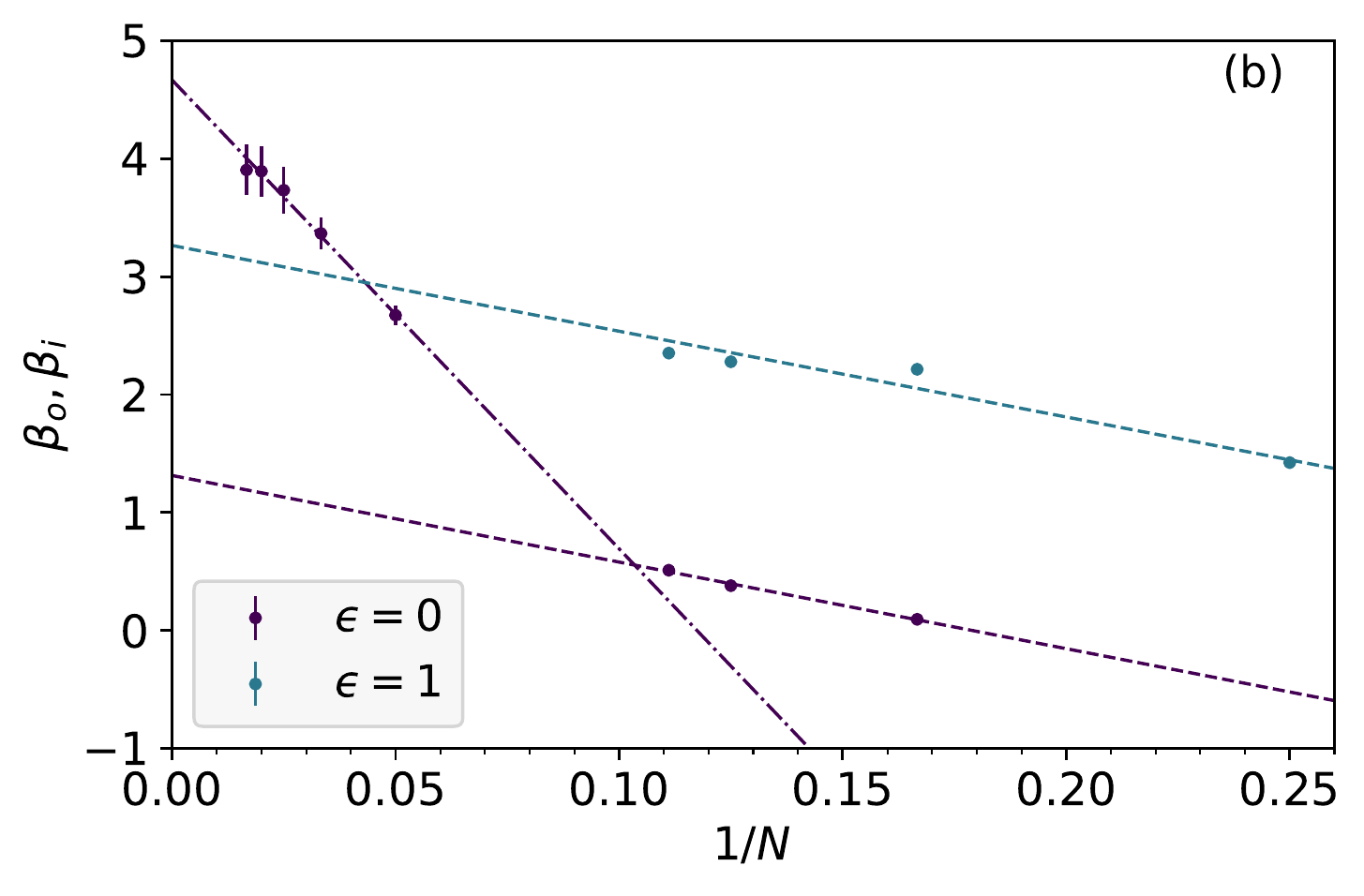}
\label{fig:dephasingLargeEps-beta}}
\caption{(a) Power-law fit of $C(t)$ at large times for $\epsilon=1$. (b) Power-law exponents $\beta_i$ ($\epsilon=0$; data from Fig.~\ref{fig:Ct-Unit}) and $\beta_o$ ($\epsilon=1$) as a function of $1/N$. We see that the concurrence decays  faster as $\epsilon$ increases (dashed lines). However, our data can capture the behavior of $C(t)$ in the presence of dissipation only at small $N$, i.e.\ in the pre-asymptotic region. We expect the large $N$ behavior to give a larger exponent $\beta_o$, as it happens for $\beta_i$ (dashed-dotted line). We set $\overline{\Delta}=0.1$, $\eta=10^5$, and averaged over at least 5000 disorder realizations. The errors are computed by using the statistical uncertainties of the concurrence values.}
\end{figure}

Furthermore, Fig.~\ref{fig:concAllCurves} shows that, increasing the dissipation strength ($\epsilon=1$), the concurrence maximum becomes smaller and is reached at earlier times. However, the decay from the maximum follows a power-law behavior as in the unitary case, albeit with a different exponent $\beta_o$, as reported in Fig.~\ref{fig:dephasingLargeEps}. This feature is very important since it shows that the signatures of localization are visible also in the presence of dissipation, if the latter is not too large. The reason at its origin might be linked to the specific (in particular, on-site) form of the dissipation operators in the GKSL equation \cite{levi2016robustness}. The power-law exponent $\beta_o$ depends on $\epsilon$ and $N$, as shown in Fig.~\ref{fig:dephasingLargeEps-beta}, and remains finite in the thermodynamic limit. Due to the small sizes accessible when integrating the full GKSL equation, we expect the extrapolated thermodynamic value of $\beta_o$ to be underestimated (see Fig.~\ref{fig:dephasingLargeEps-beta}, and the results on the unitary case $\epsilon=0$). 

Notice that the behavior of the concurrence is determined only by the ratio $\hbar Y_i/J_{ij}$. Remember that, in the unitary case, where the dissipation is absent, changing the typical strength of $J_{ij}$ through the parameter $\eta$ only shifts the timescale of $C(t)$, without modifying the shape of the curve (see Sec.~\ref{sec:unitary_dynamics}, Fig.~\ref{fig:rescaled-R}). Hence, in this Section, we employ the artificial parameter $\epsilon$ to investigate the behavior of pairwise entanglement in the different regions of the phase diagram (Fig.~\ref{fig:phaseDiag}) by (effectively) changing the ratio $\hbar Y_i/J_{ij}$.

\begin{figure}
    \centering
    \includegraphics[width=\columnwidth]{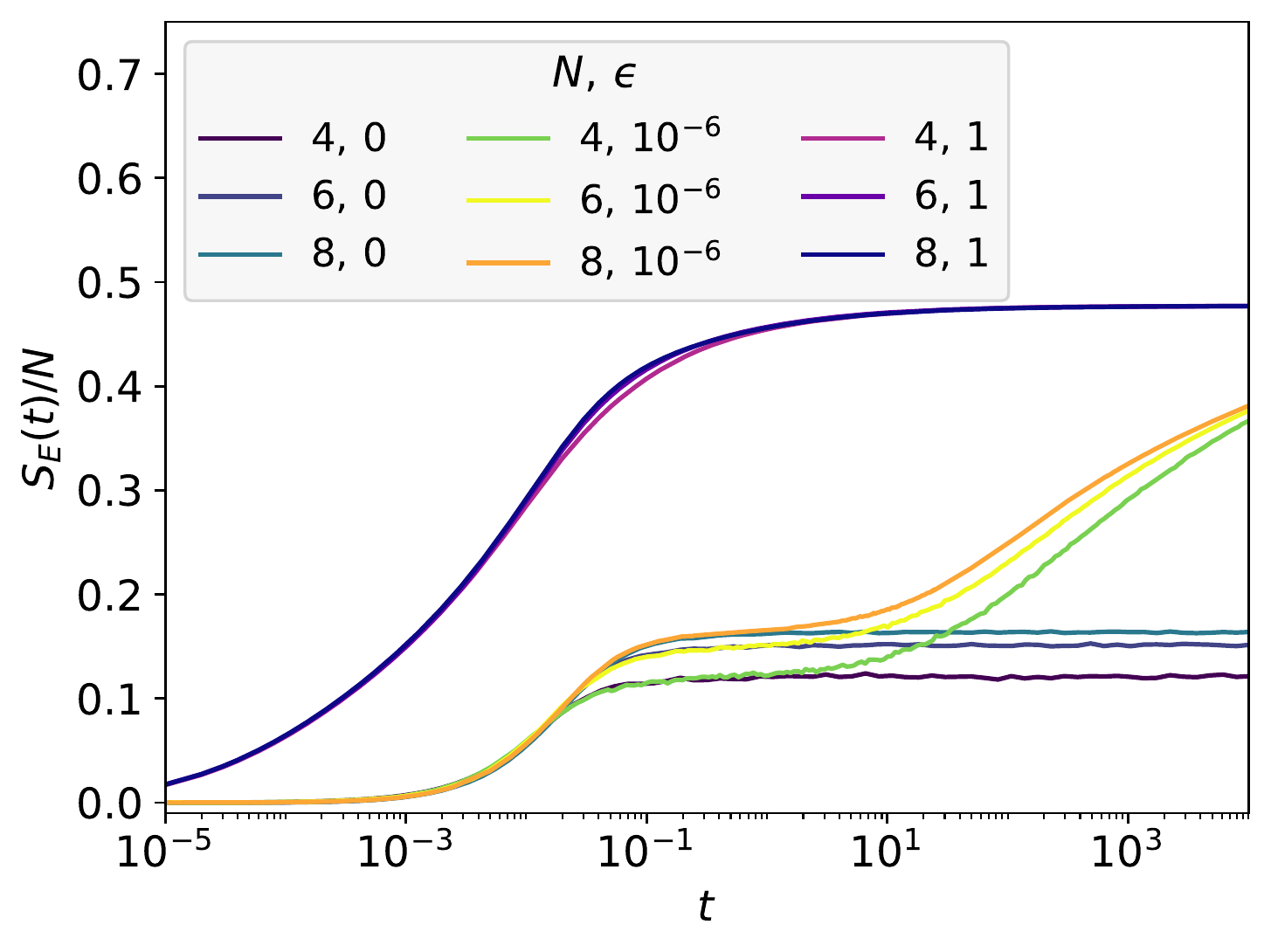}
    \caption{Half-system entanglement entropy $S_E(t)$, as defined in Sec.~\ref{sec:numerical_definitions}, per number of TLSs for various $N$ and $\epsilon$. The plot shows the results for $\overline{\Delta}=0.1$ and $\eta=10^5$, averaged over at least 1000 disorder realizations. For $\epsilon=10^{-6}$, we see that the entanglement spreading takes place in two steps: first, the TLSs become entangled with other TLSs and $S_{E}(t)/N$ reaches the plateau found in the case of unitary dynamics ($\epsilon=0$); then, HSEE grows further due to the spread of the entanglement among TLSs and phonons. For $\epsilon=1$, $S_E(t)/N$ is almost independent of $N$, indicating a volume law.}
    \label{fig:ent_entropy}
\end{figure}

Complementary to the concurrence is the half-system entanglement entropy (HSEE), $S_{E}(t)$, as defined in Sec.~\ref{sec:numerical_definitions}. Its behavior for various $N$ and $\epsilon$ is shown in Fig.~\ref{fig:ent_entropy}. As in the unitary case, HSEE starts to increase roughly when $C(t)$ reaches its maximum, i.e.\ when the TLSs start to evolve coherently. It keeps increasing at larger times when entanglement spreads inside the system. From the data at $\epsilon=10^{-6}$, it can be seen that the entanglement spreading takes place in two steps: first, the TLSs become entangled one with another and $S_{E}(t)$ reaches the plateau found with unitary dynamics ($\epsilon = 0$); then, the HSEE increases further due to the dissipative terms in the Lindbladian \eqref{eq:our_Lindblad}. Indeed, for $\epsilon \neq 0$ the TLSs entangle also with the thermal bath.

\section{Conclusions}
\label{sec:conclusions}

In this study, we investigated the well-known Two-Level System (TLS) model for glasses at low temperatures. We studied the quantum dynamics of tunnelling TLSs coupled to phonons. Within the framework of the Gorini–Kossakowski–Sudarshan–Lindblad (GKSL) master equation, and by means of a weak coupling approximation, we computed explicitly the phonon-mediated interactions among TLSs and the dissipation rates. 

We found that, as a consequence of disorder, the Hamiltonian responsible for the unitary part of the TLS dynamics, and accounting for TLS-TLS interactions, is completely expressed in terms of local integrals of motion, and is thus Many-Body Localized (MBL). Even though it differs from the effective l-bit Hamiltonian of standard MBL systems, in particular because the TLS-TLS interactions decay as a power law with distance, the TLS relaxation dynamics presents clear signatures of quantum many-body localization. Indeed, simulating the artificially isolated system with unitary dynamics governed by the TLS Hamiltonian, we found that the concurrence decays slowly in time as a power law, rather than exponentially fast as it would for an ergodic system. We also observed that the entanglement entropy grows slowly, as in standard MBL systems.

This picture is broken by the presence of dissipation, induced by real processes of TLSs and phonons, which destroy MBL and ultimately drive the system to a thermal state.

The competition between TLS-TLS interactions and dissipation determines the presence of two distinct regions in the dynamical phase diagram of the model: when interactions are comparable or stronger than dissipation, the system dynamics presents a transient \emph{bona fide} MBL region; in the opposite case, the system quickly thermalizes. Considering the typical disorder distribution parameters encompassed in the literature, it seems that real glassy materials sit in the bulk of the transient MBL region of the phase diagram. 

We explored numerically the dynamical phase diagram of the model, by tuning interaction and dissipation strengths. We found that, in  the MBL region of the phase diagram, even for dissipation strengths of the same order of magnitude of the interactions, the dynamics of the entanglement resembles the one in the absence of dissipation, showing clear signatures of localization: the concurrence decays as a power law as in the artificially isolated system. 

These findings suggest that the signatures of MBL might be experimentally accessible in real glassy samples at ultra-low temperatures, for instance using ultra-fast laser probes. The dynamics we have depicted in this paper should be robust from material to material and against the uncertainty in the characterization of the disorder distributions.

We are indebted to M.\ Fabrizio and T.\ Maimbourg for valuable discussions and the careful reading of the manuscript. We are very grateful to R.\ Fazio, G.\ Giudici, S.\ Pappalardi, A.\ Russomanno, and M.\ Secl\`i for useful discussions. A.S.\ is also indebted to A.\ Leggett and G.\ Parisi, who have (independently) suggested that MBL signatures may be found in the physics of TLSs at low temperatures. This work is within the activities of the TQT Institute in Trieste.

\bibliography{refs}

\begin{thebibliography}{83}%
\makeatletter
\providecommand \@ifxundefined [1]{%
 \@ifx{#1\undefined}
}%
\providecommand \@ifnum [1]{%
 \ifnum #1\expandafter \@firstoftwo
 \else \expandafter \@secondoftwo
 \fi
}%
\providecommand \@ifx [1]{%
 \ifx #1\expandafter \@firstoftwo
 \else \expandafter \@secondoftwo
 \fi
}%
\providecommand \natexlab [1]{#1}%
\providecommand \enquote  [1]{``#1''}%
\providecommand \bibnamefont  [1]{#1}%
\providecommand \bibfnamefont [1]{#1}%
\providecommand \citenamefont [1]{#1}%
\providecommand \href@noop [0]{\@secondoftwo}%
\providecommand \href [0]{\begingroup \@sanitize@url \@href}%
\providecommand \@href[1]{\@@startlink{#1}\@@href}%
\providecommand \@@href[1]{\endgroup#1\@@endlink}%
\providecommand \@sanitize@url [0]{\catcode `\\12\catcode `\$12\catcode
  `\&12\catcode `\#12\catcode `\^12\catcode `\_12\catcode `\%12\relax}%
\providecommand \@@startlink[1]{}%
\providecommand \@@endlink[0]{}%
\providecommand \url  [0]{\begingroup\@sanitize@url \@url }%
\providecommand \@url [1]{\endgroup\@href {#1}{\urlprefix }}%
\providecommand \urlprefix  [0]{URL }%
\providecommand \Eprint [0]{\href }%
\providecommand \doibase [0]{https://doi.org/}%
\providecommand \selectlanguage [0]{\@gobble}%
\providecommand \bibinfo  [0]{\@secondoftwo}%
\providecommand \bibfield  [0]{\@secondoftwo}%
\providecommand \translation [1]{[#1]}%
\providecommand \BibitemOpen [0]{}%
\providecommand \bibitemStop [0]{}%
\providecommand \bibitemNoStop [0]{.\EOS\space}%
\providecommand \EOS [0]{\spacefactor3000\relax}%
\providecommand \BibitemShut  [1]{\csname bibitem#1\endcsname}%
\let\auto@bib@innerbib\@empty
\bibitem [{\citenamefont {Deutsch}(1991)}]{deutsch1991quantum}%
  \BibitemOpen
  \bibfield  {author} {\bibinfo {author} {\bibfnamefont {J.~M.}\ \bibnamefont
  {Deutsch}},\ }\bibfield  {title} {\bibinfo {title} {Quantum statistical
  mechanics in a closed system},\ }\href
  {https://doi.org/10.1103/PhysRevA.43.2046} {\bibfield  {journal} {\bibinfo
  {journal} {Phys. Rev. A}\ }\textbf {\bibinfo {volume} {43}},\ \bibinfo
  {pages} {2046} (\bibinfo {year} {1991})}\BibitemShut {NoStop}%
\bibitem [{\citenamefont {Srednicki}(1994)}]{srednicki1994chaos}%
  \BibitemOpen
  \bibfield  {author} {\bibinfo {author} {\bibfnamefont {M.}~\bibnamefont
  {Srednicki}},\ }\bibfield  {title} {\bibinfo {title} {Chaos and quantum
  thermalization},\ }\href {https://doi.org/10.1103/PhysRevE.50.888} {\bibfield
   {journal} {\bibinfo  {journal} {Phys. Rev. E}\ }\textbf {\bibinfo {volume}
  {50}},\ \bibinfo {pages} {888} (\bibinfo {year} {1994})}\BibitemShut
  {NoStop}%
\bibitem [{\citenamefont {D'Alessio}\ \emph {et~al.}(2016)\citenamefont
  {D'Alessio}, \citenamefont {Kafri}, \citenamefont {Polkovnikov},\ and\
  \citenamefont {Rigol}}]{d2016quantum}%
  \BibitemOpen
  \bibfield  {author} {\bibinfo {author} {\bibfnamefont {L.}~\bibnamefont
  {D'Alessio}}, \bibinfo {author} {\bibfnamefont {Y.}~\bibnamefont {Kafri}},
  \bibinfo {author} {\bibfnamefont {A.}~\bibnamefont {Polkovnikov}},\ and\
  \bibinfo {author} {\bibfnamefont {M.}~\bibnamefont {Rigol}},\ }\bibfield
  {title} {\bibinfo {title} {From quantum chaos and eigenstate thermalization
  to statistical mechanics and thermodynamics},\ }\href
  {https://doi.org/10.1080/00018732.2016.1198134} {\bibfield  {journal}
  {\bibinfo  {journal} {Adv. Phys.}\ }\textbf {\bibinfo {volume} {65}},\
  \bibinfo {pages} {239} (\bibinfo {year} {2016})}\BibitemShut {NoStop}%
\bibitem [{\citenamefont {Rigol}\ and\ \citenamefont
  {Srednicki}(2012)}]{rigol2012alternatives}%
  \BibitemOpen
  \bibfield  {author} {\bibinfo {author} {\bibfnamefont {M.}~\bibnamefont
  {Rigol}}\ and\ \bibinfo {author} {\bibfnamefont {M.}~\bibnamefont
  {Srednicki}},\ }\bibfield  {title} {\bibinfo {title} {Alternatives to
  eigenstate thermalization},\ }\href
  {https://doi.org/10.1103/PhysRevLett.108.110601} {\bibfield  {journal}
  {\bibinfo  {journal} {Phys. Rev. Lett.}\ }\textbf {\bibinfo {volume} {108}},\
  \bibinfo {pages} {110601} (\bibinfo {year} {2012})}\BibitemShut {NoStop}%
\bibitem [{\citenamefont {Gornyi}\ \emph {et~al.}(2005)\citenamefont {Gornyi},
  \citenamefont {Mirlin},\ and\ \citenamefont
  {Polyakov}}]{gornyi2005interacting}%
  \BibitemOpen
  \bibfield  {author} {\bibinfo {author} {\bibfnamefont {I.~V.}\ \bibnamefont
  {Gornyi}}, \bibinfo {author} {\bibfnamefont {A.~D.}\ \bibnamefont {Mirlin}},\
  and\ \bibinfo {author} {\bibfnamefont {D.~G.}\ \bibnamefont {Polyakov}},\
  }\bibfield  {title} {\bibinfo {title} {Interacting electrons in disordered
  wires: Anderson localization and low-$t$ transport},\ }\href
  {https://doi.org/10.1103/PhysRevLett.95.206603} {\bibfield  {journal}
  {\bibinfo  {journal} {Phys. Rev. Lett.}\ }\textbf {\bibinfo {volume} {95}},\
  \bibinfo {pages} {206603} (\bibinfo {year} {2005})}\BibitemShut {NoStop}%
\bibitem [{\citenamefont {Basko}\ \emph {et~al.}(2006)\citenamefont {Basko},
  \citenamefont {Aleiner},\ and\ \citenamefont {Altshuler}}]{Basko:2006hh}%
  \BibitemOpen
  \bibfield  {author} {\bibinfo {author} {\bibfnamefont {D.~M.}\ \bibnamefont
  {Basko}}, \bibinfo {author} {\bibfnamefont {I.~L.}\ \bibnamefont {Aleiner}},\
  and\ \bibinfo {author} {\bibfnamefont {B.~L.}\ \bibnamefont {Altshuler}},\
  }\bibfield  {title} {\bibinfo {title} {{Metal--insulator transition in a
  weakly interacting many-electron system with localized single-particle
  states}},\ }\href {https://doi.org/https://doi.org/10.1016/j.aop.2005.11.014}
  {\bibfield  {journal} {\bibinfo  {journal} {Ann. Phys.}\ }\textbf {\bibinfo
  {volume} {321}},\ \bibinfo {pages} {1126} (\bibinfo {year}
  {2006})}\BibitemShut {NoStop}%
\bibitem [{\citenamefont {Luca}\ and\ \citenamefont
  {Scardicchio}(2013)}]{deLuca2013}%
  \BibitemOpen
  \bibfield  {author} {\bibinfo {author} {\bibfnamefont {A.~D.}\ \bibnamefont
  {Luca}}\ and\ \bibinfo {author} {\bibfnamefont {A.}~\bibnamefont
  {Scardicchio}},\ }\bibfield  {title} {\bibinfo {title} {Ergodicity breaking
  in a model showing many-body localization},\ }\href
  {https://doi.org/10.1209/0295-5075/101/37003} {\bibfield  {journal} {\bibinfo
   {journal} {{Europhys. Lett.}}\ }\textbf {\bibinfo {volume} {101}},\ \bibinfo
  {pages} {37003} (\bibinfo {year} {2013})}\BibitemShut {NoStop}%
\bibitem [{\citenamefont {Huse}\ \emph {et~al.}(2014)\citenamefont {Huse},
  \citenamefont {Nandkishore},\ and\ \citenamefont
  {Oganesyan}}]{huse2014phenomenology}%
  \BibitemOpen
  \bibfield  {author} {\bibinfo {author} {\bibfnamefont {D.~A.}\ \bibnamefont
  {Huse}}, \bibinfo {author} {\bibfnamefont {R.}~\bibnamefont {Nandkishore}},\
  and\ \bibinfo {author} {\bibfnamefont {V.}~\bibnamefont {Oganesyan}},\
  }\bibfield  {title} {\bibinfo {title} {Phenomenology of fully
  many-body-localized systems},\ }\href
  {https://doi.org/10.1103/PhysRevB.90.174202} {\bibfield  {journal} {\bibinfo
  {journal} {Phys. Rev. B}\ }\textbf {\bibinfo {volume} {90}},\ \bibinfo
  {pages} {174202} (\bibinfo {year} {2014})}\BibitemShut {NoStop}%
\bibitem [{\citenamefont {Luitz}\ \emph {et~al.}(2015)\citenamefont {Luitz},
  \citenamefont {Laflorencie},\ and\ \citenamefont {Alet}}]{luitz2015many}%
  \BibitemOpen
  \bibfield  {author} {\bibinfo {author} {\bibfnamefont {D.~J.}\ \bibnamefont
  {Luitz}}, \bibinfo {author} {\bibfnamefont {N.}~\bibnamefont {Laflorencie}},\
  and\ \bibinfo {author} {\bibfnamefont {F.}~\bibnamefont {Alet}},\ }\bibfield
  {title} {\bibinfo {title} {Many-body localization edge in the random-field
  heisenberg chain},\ }\href {https://doi.org/10.1103/PhysRevB.91.081103}
  {\bibfield  {journal} {\bibinfo  {journal} {Phys. Rev. B}\ }\textbf {\bibinfo
  {volume} {91}},\ \bibinfo {pages} {081103} (\bibinfo {year}
  {2015})}\BibitemShut {NoStop}%
\bibitem [{\citenamefont {Nandkishore}\ and\ \citenamefont
  {Huse}(2015)}]{huse2015review}%
  \BibitemOpen
  \bibfield  {author} {\bibinfo {author} {\bibfnamefont {R.}~\bibnamefont
  {Nandkishore}}\ and\ \bibinfo {author} {\bibfnamefont {D.~A.}\ \bibnamefont
  {Huse}},\ }\bibfield  {title} {\bibinfo {title} {{Many-Body Localization and
  Thermalization in Quantum Statistical Mechanics}},\ }\href
  {https://doi.org/https://doi.org/10.1146/annurev-conmatphys-031214-014726}
  {\bibfield  {journal} {\bibinfo  {journal} {Annu. Rev. Condens. Matter
  Phys.}\ }\textbf {\bibinfo {volume} {6}},\ \bibinfo {pages} {15} (\bibinfo
  {year} {2015})}\BibitemShut {NoStop}%
\bibitem [{\citenamefont {Abanin}\ and\ \citenamefont
  {Papi\'{c}}(2017)}]{abanin2017recent}%
  \BibitemOpen
  \bibfield  {author} {\bibinfo {author} {\bibfnamefont {D.~A.}\ \bibnamefont
  {Abanin}}\ and\ \bibinfo {author} {\bibfnamefont {Z.}~\bibnamefont
  {Papi\'{c}}},\ }\bibfield  {title} {\bibinfo {title} {Recent progress in
  many-body localization},\ }\href {https://doi.org/10.1002/andp.201700169}
  {\bibfield  {journal} {\bibinfo  {journal} {Ann. Phys.}\ }\textbf {\bibinfo
  {volume} {529}},\ \bibinfo {pages} {1700169} (\bibinfo {year}
  {2017})}\BibitemShut {NoStop}%
\bibitem [{\citenamefont {Abanin}\ \emph {et~al.}(2016)\citenamefont {Abanin},
  \citenamefont {{De Roeck}},\ and\ \citenamefont
  {Huveneers}}]{Abanin2016Theory}%
  \BibitemOpen
  \bibfield  {author} {\bibinfo {author} {\bibfnamefont {D.~A.}\ \bibnamefont
  {Abanin}}, \bibinfo {author} {\bibfnamefont {W.}~\bibnamefont {{De Roeck}}},\
  and\ \bibinfo {author} {\bibfnamefont {F.}~\bibnamefont {Huveneers}},\
  }\bibfield  {title} {\bibinfo {title} {Theory of many-body localization in
  periodically driven systems},\ }\href
  {https://doi.org/https://doi.org/10.1016/j.aop.2016.03.010} {\bibfield
  {journal} {\bibinfo  {journal} {Ann. Phys.}\ }\textbf {\bibinfo {volume}
  {372}},\ \bibinfo {pages} {1 } (\bibinfo {year} {2016})}\BibitemShut
  {NoStop}%
\bibitem [{\citenamefont {Zhang}\ \emph {et~al.}(2017)\citenamefont {Zhang},
  \citenamefont {Hess}, \citenamefont {Kyprianidis}, \citenamefont {Becker},
  \citenamefont {Lee}, \citenamefont {Smith}, \citenamefont {Pagano},
  \citenamefont {Potirniche}, \citenamefont {Potter}, \citenamefont
  {Vishwanath} \emph {et~al.}}]{zhang2017observation}%
  \BibitemOpen
  \bibfield  {author} {\bibinfo {author} {\bibfnamefont {J.}~\bibnamefont
  {Zhang}}, \bibinfo {author} {\bibfnamefont {P.}~\bibnamefont {Hess}},
  \bibinfo {author} {\bibfnamefont {A.}~\bibnamefont {Kyprianidis}}, \bibinfo
  {author} {\bibfnamefont {P.}~\bibnamefont {Becker}}, \bibinfo {author}
  {\bibfnamefont {A.}~\bibnamefont {Lee}}, \bibinfo {author} {\bibfnamefont
  {J.}~\bibnamefont {Smith}}, \bibinfo {author} {\bibfnamefont
  {G.}~\bibnamefont {Pagano}}, \bibinfo {author} {\bibfnamefont {I.-D.}\
  \bibnamefont {Potirniche}}, \bibinfo {author} {\bibfnamefont {A.~C.}\
  \bibnamefont {Potter}}, \bibinfo {author} {\bibfnamefont {A.}~\bibnamefont
  {Vishwanath}}, \emph {et~al.},\ }\bibfield  {title} {\bibinfo {title}
  {Observation of a discrete time crystal},\ }\href
  {https://doi.org/https://doi.org/10.1038/nature21413} {\bibfield  {journal}
  {\bibinfo  {journal} {Nature}\ }\textbf {\bibinfo {volume} {543}},\ \bibinfo
  {pages} {217} (\bibinfo {year} {2017})}\BibitemShut {NoStop}%
\bibitem [{\citenamefont {Sacha}\ and\ \citenamefont
  {Zakrzewski}(2017)}]{sacha2017time}%
  \BibitemOpen
  \bibfield  {author} {\bibinfo {author} {\bibfnamefont {K.}~\bibnamefont
  {Sacha}}\ and\ \bibinfo {author} {\bibfnamefont {J.}~\bibnamefont
  {Zakrzewski}},\ }\bibfield  {title} {\bibinfo {title} {Time crystals: a
  review},\ }\href {https://doi.org/10.1088/1361-6633/aa8b38} {\bibfield
  {journal} {\bibinfo  {journal} {Rep. Progr. Phys.}\ }\textbf {\bibinfo
  {volume} {81}},\ \bibinfo {pages} {016401} (\bibinfo {year}
  {2017})}\BibitemShut {NoStop}%
\bibitem [{\citenamefont {Schiulaz}\ and\ \citenamefont
  {M\"uller}(2014)}]{schiulaz2014glass}%
  \BibitemOpen
  \bibfield  {author} {\bibinfo {author} {\bibfnamefont {M.}~\bibnamefont
  {Schiulaz}}\ and\ \bibinfo {author} {\bibfnamefont {M.}~\bibnamefont
  {M\"uller}},\ }\bibfield  {title} {\bibinfo {title} {Ideal quantum glass
  transitions: Many-body localization without quenched disorder},\ }\href
  {https://doi.org/https://doi.org/10.1063/1.4893505} {\bibfield  {journal}
  {\bibinfo  {journal} {AIP Conf. Proc.}\ }\textbf {\bibinfo {volume} {1610}},\
  \bibinfo {pages} {11} (\bibinfo {year} {2014})}\BibitemShut {NoStop}%
\bibitem [{\citenamefont {Papi{\'c}}\ \emph {et~al.}(2015)\citenamefont
  {Papi{\'c}}, \citenamefont {Stoudenmire},\ and\ \citenamefont
  {Abanin}}]{Papic2015}%
  \BibitemOpen
  \bibfield  {author} {\bibinfo {author} {\bibfnamefont {Z.}~\bibnamefont
  {Papi{\'c}}}, \bibinfo {author} {\bibfnamefont {E.~M.}\ \bibnamefont
  {Stoudenmire}},\ and\ \bibinfo {author} {\bibfnamefont {D.~A.}\ \bibnamefont
  {Abanin}},\ }\bibfield  {title} {\bibinfo {title} {{Many-body localization in
  disorder-free systems: The importance of finite-size constraints}},\ }\href
  {https://doi.org/10.1016/j.aop.2015.08.024} {\bibfield  {journal} {\bibinfo
  {journal} {Ann. Phys.}\ }\textbf {\bibinfo {volume} {362}},\ \bibinfo {pages}
  {714} (\bibinfo {year} {2015})}\BibitemShut {NoStop}%
\bibitem [{\citenamefont {Pino}\ \emph {et~al.}(2015)\citenamefont {Pino},
  \citenamefont {Ioffe},\ and\ \citenamefont {Altshuler}}]{pino2015metallic}%
  \BibitemOpen
  \bibfield  {author} {\bibinfo {author} {\bibfnamefont {M.}~\bibnamefont
  {Pino}}, \bibinfo {author} {\bibfnamefont {L.~B.}\ \bibnamefont {Ioffe}},\
  and\ \bibinfo {author} {\bibfnamefont {B.~L.}\ \bibnamefont {Altshuler}},\
  }\bibfield  {title} {\bibinfo {title} {Nonergodic metallic and insulating
  phases of josephson junction chains},\ }\href
  {https://doi.org/https://doi.org/10.1073/pnas.1520033113} {\bibfield
  {journal} {\bibinfo  {journal} {Proc. Natl. Acad. Sci. U.S.A.}\ }\textbf
  {\bibinfo {volume} {113}},\ \bibinfo {pages} {536} (\bibinfo {year}
  {2015})}\BibitemShut {NoStop}%
\bibitem [{\citenamefont {Yao}\ \emph {et~al.}(2016)\citenamefont {Yao},
  \citenamefont {Laumann}, \citenamefont {Cirac}, \citenamefont {Lukin},\ and\
  \citenamefont {Moore}}]{yao2016quasi}%
  \BibitemOpen
  \bibfield  {author} {\bibinfo {author} {\bibfnamefont {N.~Y.}\ \bibnamefont
  {Yao}}, \bibinfo {author} {\bibfnamefont {C.~R.}\ \bibnamefont {Laumann}},
  \bibinfo {author} {\bibfnamefont {J.~I.}\ \bibnamefont {Cirac}}, \bibinfo
  {author} {\bibfnamefont {M.~D.}\ \bibnamefont {Lukin}},\ and\ \bibinfo
  {author} {\bibfnamefont {J.~E.}\ \bibnamefont {Moore}},\ }\bibfield  {title}
  {\bibinfo {title} {Quasi-many-body localization in translation-invariant
  systems},\ }\href {https://doi.org/10.1103/PhysRevLett.117.240601} {\bibfield
   {journal} {\bibinfo  {journal} {Phys. Rev. Lett.}\ }\textbf {\bibinfo
  {volume} {117}},\ \bibinfo {pages} {240601} (\bibinfo {year}
  {2016})}\BibitemShut {NoStop}%
\bibitem [{\citenamefont {Nandkishore}\ and\ \citenamefont
  {Sondhi}(2017)}]{Nandkishore2017LR}%
  \BibitemOpen
  \bibfield  {author} {\bibinfo {author} {\bibfnamefont {R.~M.}\ \bibnamefont
  {Nandkishore}}\ and\ \bibinfo {author} {\bibfnamefont {S.~L.}\ \bibnamefont
  {Sondhi}},\ }\bibfield  {title} {\bibinfo {title} {Many-body localization
  with long-range interactions},\ }\href
  {https://doi.org/10.1103/PhysRevX.7.041021} {\bibfield  {journal} {\bibinfo
  {journal} {Phys. Rev. X}\ }\textbf {\bibinfo {volume} {7}},\ \bibinfo {pages}
  {041021} (\bibinfo {year} {2017})}\BibitemShut {NoStop}%
\bibitem [{\citenamefont {Brenes}\ \emph {et~al.}(2018)\citenamefont {Brenes},
  \citenamefont {Dalmonte}, \citenamefont {Heyl},\ and\ \citenamefont
  {Scardicchio}}]{brenes2018many}%
  \BibitemOpen
  \bibfield  {author} {\bibinfo {author} {\bibfnamefont {M.}~\bibnamefont
  {Brenes}}, \bibinfo {author} {\bibfnamefont {M.}~\bibnamefont {Dalmonte}},
  \bibinfo {author} {\bibfnamefont {M.}~\bibnamefont {Heyl}},\ and\ \bibinfo
  {author} {\bibfnamefont {A.}~\bibnamefont {Scardicchio}},\ }\bibfield
  {title} {\bibinfo {title} {Many-body localization dynamics from gauge
  invariance},\ }\href {https://doi.org/10.1103/PhysRevLett.120.030601}
  {\bibfield  {journal} {\bibinfo  {journal} {Phys. Rev. Lett.}\ }\textbf
  {\bibinfo {volume} {120}},\ \bibinfo {pages} {030601} (\bibinfo {year}
  {2018})}\BibitemShut {NoStop}%
\bibitem [{\citenamefont {Schulz}\ \emph {et~al.}(2019)\citenamefont {Schulz},
  \citenamefont {Hooley}, \citenamefont {Moessner},\ and\ \citenamefont
  {Pollmann}}]{Schulz2019Stark}%
  \BibitemOpen
  \bibfield  {author} {\bibinfo {author} {\bibfnamefont {M.}~\bibnamefont
  {Schulz}}, \bibinfo {author} {\bibfnamefont {C.~A.}\ \bibnamefont {Hooley}},
  \bibinfo {author} {\bibfnamefont {R.}~\bibnamefont {Moessner}},\ and\
  \bibinfo {author} {\bibfnamefont {F.}~\bibnamefont {Pollmann}},\ }\bibfield
  {title} {\bibinfo {title} {Stark many-body localization},\ }\href
  {https://doi.org/10.1103/PhysRevLett.122.040606} {\bibfield  {journal}
  {\bibinfo  {journal} {Phys. Rev. Lett.}\ }\textbf {\bibinfo {volume} {122}},\
  \bibinfo {pages} {040606} (\bibinfo {year} {2019})}\BibitemShut {NoStop}%
\bibitem [{\citenamefont {van Nieuwenburg}\ \emph {et~al.}(2019)\citenamefont
  {van Nieuwenburg}, \citenamefont {Baum},\ and\ \citenamefont
  {Refael}}]{vanNieuwenburg2019From}%
  \BibitemOpen
  \bibfield  {author} {\bibinfo {author} {\bibfnamefont {E.}~\bibnamefont {van
  Nieuwenburg}}, \bibinfo {author} {\bibfnamefont {Y.}~\bibnamefont {Baum}},\
  and\ \bibinfo {author} {\bibfnamefont {G.}~\bibnamefont {Refael}},\
  }\bibfield  {title} {\bibinfo {title} {From bloch oscillations to many-body
  localization in clean interacting systems},\ }\href
  {https://doi.org/10.1073/pnas.1819316116} {\bibfield  {journal} {\bibinfo
  {journal} {Proc. Natl. Acad. Sci. U.S.A.}\ }\textbf {\bibinfo {volume}
  {116}},\ \bibinfo {pages} {9269} (\bibinfo {year} {2019})}\BibitemShut
  {NoStop}%
\bibitem [{\citenamefont {Giudici}\ \emph {et~al.}(2020)\citenamefont
  {Giudici}, \citenamefont {Surace}, \citenamefont {Ebot}, \citenamefont
  {Scardicchio},\ and\ \citenamefont {Dalmonte}}]{giudici2020breakdown}%
  \BibitemOpen
  \bibfield  {author} {\bibinfo {author} {\bibfnamefont {G.}~\bibnamefont
  {Giudici}}, \bibinfo {author} {\bibfnamefont {F.~M.}\ \bibnamefont {Surace}},
  \bibinfo {author} {\bibfnamefont {J.~E.}\ \bibnamefont {Ebot}}, \bibinfo
  {author} {\bibfnamefont {A.}~\bibnamefont {Scardicchio}},\ and\ \bibinfo
  {author} {\bibfnamefont {M.}~\bibnamefont {Dalmonte}},\ }\bibfield  {title}
  {\bibinfo {title} {Breakdown of ergodicity in disordered $u(1)$ lattice gauge
  theories},\ }\href {https://doi.org/10.1103/PhysRevResearch.2.032034}
  {\bibfield  {journal} {\bibinfo  {journal} {Phys. Rev. Res.}\ }\textbf
  {\bibinfo {volume} {2}},\ \bibinfo {pages} {032034} (\bibinfo {year}
  {2020})}\BibitemShut {NoStop}%
\bibitem [{\citenamefont {Zeller}\ and\ \citenamefont
  {Pohl}(1971)}]{ZellerPohl}%
  \BibitemOpen
  \bibfield  {author} {\bibinfo {author} {\bibfnamefont {R.~C.}\ \bibnamefont
  {Zeller}}\ and\ \bibinfo {author} {\bibfnamefont {R.~O.}\ \bibnamefont
  {Pohl}},\ }\bibfield  {title} {\bibinfo {title} {Thermal conductivity and
  specific heat of noncrystalline solids},\ }\href
  {https://doi.org/10.1103/PhysRevB.4.2029} {\bibfield  {journal} {\bibinfo
  {journal} {Phys. Rev. B}\ }\textbf {\bibinfo {volume} {4}},\ \bibinfo {pages}
  {2029} (\bibinfo {year} {1971})}\BibitemShut {NoStop}%
\bibitem [{\citenamefont {Stephens}(1973)}]{stephens1973low}%
  \BibitemOpen
  \bibfield  {author} {\bibinfo {author} {\bibfnamefont {R.~B.}\ \bibnamefont
  {Stephens}},\ }\bibfield  {title} {\bibinfo {title} {Low-temperature specific
  heat and thermal conductivity of noncrystalline dielectric solids},\ }\href
  {https://doi.org/10.1103/PhysRevB.8.2896} {\bibfield  {journal} {\bibinfo
  {journal} {Phys. Rev. B}\ }\textbf {\bibinfo {volume} {8}},\ \bibinfo {pages}
  {2896} (\bibinfo {year} {1973})}\BibitemShut {NoStop}%
\bibitem [{\citenamefont {Anderson}\ \emph {et~al.}(1972)\citenamefont
  {Anderson}, \citenamefont {Halperin},\ and\ \citenamefont
  {Varma}}]{anderson1972anomalous}%
  \BibitemOpen
  \bibfield  {author} {\bibinfo {author} {\bibfnamefont {P.~W.}\ \bibnamefont
  {Anderson}}, \bibinfo {author} {\bibfnamefont {B.~I.}\ \bibnamefont
  {Halperin}},\ and\ \bibinfo {author} {\bibfnamefont {C.~M.}\ \bibnamefont
  {Varma}},\ }\bibfield  {title} {\bibinfo {title} {Anomalous low-temperature
  thermal properties of glasses and spin glasses},\ }\href
  {https://doi.org/10.1080/14786437208229210} {\bibfield  {journal} {\bibinfo
  {journal} {Phil. Mag.}\ }\textbf {\bibinfo {volume} {25}},\ \bibinfo {pages}
  {1} (\bibinfo {year} {1972})}\BibitemShut {NoStop}%
\bibitem [{\citenamefont {Phillips}(1972)}]{phillips1972tunneling}%
  \BibitemOpen
  \bibfield  {author} {\bibinfo {author} {\bibfnamefont {W.}~\bibnamefont
  {Phillips}},\ }\bibfield  {title} {\bibinfo {title} {Tunneling states in
  amorphous solids},\ }\href@noop {} {\bibfield  {journal} {\bibinfo  {journal}
  {J. Low Temp. Phys.}\ }\textbf {\bibinfo {volume} {7}},\ \bibinfo {pages}
  {351} (\bibinfo {year} {1972})}\BibitemShut {NoStop}%
\bibitem [{\citenamefont {Phillips}(1987)}]{Phillips87}%
  \BibitemOpen
  \bibfield  {author} {\bibinfo {author} {\bibfnamefont {W.~A.}\ \bibnamefont
  {Phillips}},\ }\bibfield  {title} {\bibinfo {title} {Two-level states in
  glasses},\ }\href {https://doi.org/10.1088/0034-4885/50/12/003} {\bibfield
  {journal} {\bibinfo  {journal} {Rep. Progr. Phys.}\ }\textbf {\bibinfo
  {volume} {50}},\ \bibinfo {pages} {1657} (\bibinfo {year}
  {1987})}\BibitemShut {NoStop}%
\bibitem [{\citenamefont {Esquinazi}(2013)}]{Esquinazi2013}%
  \BibitemOpen
  \bibfield  {author} {\bibinfo {author} {\bibfnamefont {P.}~\bibnamefont
  {Esquinazi}},\ }\href {https://doi.org/10.1007/978-3-662-03695-2} {\emph
  {\bibinfo {title} {Tunneling systems in amorphous and crystalline solids}}}\
  (\bibinfo  {publisher} {Springer Science},\ \bibinfo {year}
  {2013})\BibitemShut {NoStop}%
\bibitem [{\citenamefont {Leggett}\ and\ \citenamefont {Yu}(1988)}]{YuLeggett}%
  \BibitemOpen
  \bibfield  {author} {\bibinfo {author} {\bibfnamefont {A.~J.}\ \bibnamefont
  {Leggett}}\ and\ \bibinfo {author} {\bibfnamefont {C.~C.}\ \bibnamefont
  {Yu}},\ }\bibfield  {title} {\bibinfo {title} {Low temperature properties of
  amorphous materials: Through a glass darkly},\ }\href@noop {} {\bibfield
  {journal} {\bibinfo  {journal} {Comments Condens. Matter Phys.}\ }\textbf
  {\bibinfo {volume} {14}},\ \bibinfo {pages} {231} (\bibinfo {year}
  {1988})}\BibitemShut {NoStop}%
\bibitem [{\citenamefont {Leggett}(1991)}]{Leggett91}%
  \BibitemOpen
  \bibfield  {author} {\bibinfo {author} {\bibfnamefont {A.~J.}\ \bibnamefont
  {Leggett}},\ }\bibfield  {title} {\bibinfo {title} {Amorphous materials at
  low temperatures: why are they so similar?},\ }\href
  {https://doi.org/https://doi.org/10.1016/0921-4526(91)90246-B} {\bibfield
  {journal} {\bibinfo  {journal} {Physica B}\ }\textbf {\bibinfo {volume}
  {169}},\ \bibinfo {pages} {322 } (\bibinfo {year} {1991})}\BibitemShut
  {NoStop}%
\bibitem [{\citenamefont {Leggett}\ and\ \citenamefont
  {Vural}(2013)}]{LeggettVural}%
  \BibitemOpen
  \bibfield  {author} {\bibinfo {author} {\bibfnamefont {A.~J.}\ \bibnamefont
  {Leggett}}\ and\ \bibinfo {author} {\bibfnamefont {D.~C.}\ \bibnamefont
  {Vural}},\ }\bibfield  {title} {\bibinfo {title} {``{Tunneling} two-level
  systems'' model of the low-temperature properties of glasses: Are
  ``smoking-gun'' tests possible?},\ }\href {https://doi.org/10.1021/jp402222g}
  {\bibfield  {journal} {\bibinfo  {journal} {J. Phys. Chem. B}\ }\textbf
  {\bibinfo {volume} {117}},\ \bibinfo {pages} {12966} (\bibinfo {year}
  {2013})}\BibitemShut {NoStop}%
\bibitem [{\citenamefont {Arnold}\ and\ \citenamefont
  {Hunklinger}(1975)}]{Arnold75}%
  \BibitemOpen
  \bibfield  {author} {\bibinfo {author} {\bibfnamefont {W.}~\bibnamefont
  {Arnold}}\ and\ \bibinfo {author} {\bibfnamefont {S.}~\bibnamefont
  {Hunklinger}},\ }\bibfield  {title} {\bibinfo {title} {Experimental evidence
  for the direct interaction between two-level systems in glasses at very low
  temperatures},\ }\href
  {https://doi.org/https://doi.org/10.1016/0038-1098(75)90743-7} {\bibfield
  {journal} {\bibinfo  {journal} {Solid State Comm.}\ }\textbf {\bibinfo
  {volume} {17}},\ \bibinfo {pages} {883 } (\bibinfo {year}
  {1975})}\BibitemShut {NoStop}%
\bibitem [{\citenamefont {Enss}\ and\ \citenamefont
  {Hunklinger}(1997)}]{Enss97}%
  \BibitemOpen
  \bibfield  {author} {\bibinfo {author} {\bibfnamefont {C.}~\bibnamefont
  {Enss}}\ and\ \bibinfo {author} {\bibfnamefont {S.}~\bibnamefont
  {Hunklinger}},\ }\bibfield  {title} {\bibinfo {title} {Incoherent tunneling
  in glasses at very low temperatures},\ }\href
  {https://doi.org/10.1103/PhysRevLett.79.2831} {\bibfield  {journal} {\bibinfo
   {journal} {Phys. Rev. Lett.}\ }\textbf {\bibinfo {volume} {79}},\ \bibinfo
  {pages} {2831} (\bibinfo {year} {1997})}\BibitemShut {NoStop}%
\bibitem [{\citenamefont {Strehlow}\ \emph {et~al.}(1998)\citenamefont
  {Strehlow}, \citenamefont {Enss},\ and\ \citenamefont
  {Hunklinger}}]{Strehlow98}%
  \BibitemOpen
  \bibfield  {author} {\bibinfo {author} {\bibfnamefont {P.}~\bibnamefont
  {Strehlow}}, \bibinfo {author} {\bibfnamefont {C.}~\bibnamefont {Enss}},\
  and\ \bibinfo {author} {\bibfnamefont {S.}~\bibnamefont {Hunklinger}},\
  }\bibfield  {title} {\bibinfo {title} {Evidence for a phase transition in
  glasses at very low temperature: A macroscopic quantum state of tunneling
  systems?},\ }\href {https://doi.org/10.1103/PhysRevLett.80.5361} {\bibfield
  {journal} {\bibinfo  {journal} {Phys. Rev. Lett.}\ }\textbf {\bibinfo
  {volume} {80}},\ \bibinfo {pages} {5361} (\bibinfo {year}
  {1998})}\BibitemShut {NoStop}%
\bibitem [{\citenamefont {Boiron}\ \emph {et~al.}(1999)\citenamefont {Boiron},
  \citenamefont {Tamarat}, \citenamefont {Lounis}, \citenamefont {Brown},\ and\
  \citenamefont {Orrit}}]{Boiron99}%
  \BibitemOpen
  \bibfield  {author} {\bibinfo {author} {\bibfnamefont {A.-M.}\ \bibnamefont
  {Boiron}}, \bibinfo {author} {\bibfnamefont {P.}~\bibnamefont {Tamarat}},
  \bibinfo {author} {\bibfnamefont {B.}~\bibnamefont {Lounis}}, \bibinfo
  {author} {\bibfnamefont {R.}~\bibnamefont {Brown}},\ and\ \bibinfo {author}
  {\bibfnamefont {M.}~\bibnamefont {Orrit}},\ }\bibfield  {title} {\bibinfo
  {title} {Are the spectral trails of single molecules consistent with the
  standard two-level system model of glasses at low temperatures?},\ }\href
  {https://doi.org/https://doi.org/10.1016/S0301-0104(99)00140-8} {\bibfield
  {journal} {\bibinfo  {journal} {Chem. Phys.}\ }\textbf {\bibinfo {volume}
  {247}},\ \bibinfo {pages} {119 } (\bibinfo {year} {1999})}\BibitemShut
  {NoStop}%
\bibitem [{\citenamefont {Classen}\ \emph {et~al.}(2000)\citenamefont
  {Classen}, \citenamefont {Burkert}, \citenamefont {Enss},\ and\ \citenamefont
  {Hunklinger}}]{Classen00}%
  \BibitemOpen
  \bibfield  {author} {\bibinfo {author} {\bibfnamefont {J.}~\bibnamefont
  {Classen}}, \bibinfo {author} {\bibfnamefont {T.}~\bibnamefont {Burkert}},
  \bibinfo {author} {\bibfnamefont {C.}~\bibnamefont {Enss}},\ and\ \bibinfo
  {author} {\bibfnamefont {S.}~\bibnamefont {Hunklinger}},\ }\bibfield  {title}
  {\bibinfo {title} {Anomalous frequency dependence of the internal friction of
  vitreous silica},\ }\href {https://doi.org/10.1103/PhysRevLett.84.2176}
  {\bibfield  {journal} {\bibinfo  {journal} {Phys. Rev. Lett.}\ }\textbf
  {\bibinfo {volume} {84}},\ \bibinfo {pages} {2176} (\bibinfo {year}
  {2000})}\BibitemShut {NoStop}%
\bibitem [{\citenamefont {Lisenfeld}\ \emph {et~al.}(2015)\citenamefont
  {Lisenfeld}, \citenamefont {Grabovskij}, \citenamefont {M{\"u}ller},
  \citenamefont {Cole}, \citenamefont {Weiss},\ and\ \citenamefont
  {Ustinov}}]{Lisenfeld2015}%
  \BibitemOpen
  \bibfield  {author} {\bibinfo {author} {\bibfnamefont {J.}~\bibnamefont
  {Lisenfeld}}, \bibinfo {author} {\bibfnamefont {G.~J.}\ \bibnamefont
  {Grabovskij}}, \bibinfo {author} {\bibfnamefont {C.}~\bibnamefont
  {M{\"u}ller}}, \bibinfo {author} {\bibfnamefont {J.~H.}\ \bibnamefont
  {Cole}}, \bibinfo {author} {\bibfnamefont {G.}~\bibnamefont {Weiss}},\ and\
  \bibinfo {author} {\bibfnamefont {A.~V.}\ \bibnamefont {Ustinov}},\
  }\bibfield  {title} {\bibinfo {title} {Observation of directly interacting
  coherent two-level systems in an amorphous material},\ }\href
  {https://doi.org/10.1038/ncomms7182} {\bibfield  {journal} {\bibinfo
  {journal} {Nature Comm.}\ }\textbf {\bibinfo {volume} {6}},\ \bibinfo {pages}
  {6182} (\bibinfo {year} {2015})}\BibitemShut {NoStop}%
\bibitem [{\citenamefont {Joffrin}\ and\ \citenamefont
  {Levelut}(1975)}]{JoffrinLevelut}%
  \BibitemOpen
  \bibfield  {author} {\bibinfo {author} {\bibfnamefont {J.}~\bibnamefont
  {Joffrin}}\ and\ \bibinfo {author} {\bibfnamefont {A.}~\bibnamefont
  {Levelut}},\ }\bibfield  {title} {\bibinfo {title} {{Virtual phonon exchange
  in glasses}},\ }\href {https://doi.org/10.1051/jphys:01975003609081100}
  {\bibfield  {journal} {\bibinfo  {journal} {{J. Phys.}}\ }\textbf {\bibinfo
  {volume} {36}},\ \bibinfo {pages} {811} (\bibinfo {year} {1975})}\BibitemShut
  {NoStop}%
\bibitem [{\citenamefont {Black}\ and\ \citenamefont
  {Halperin}(1977)}]{black1977spectral}%
  \BibitemOpen
  \bibfield  {author} {\bibinfo {author} {\bibfnamefont {J.~L.}\ \bibnamefont
  {Black}}\ and\ \bibinfo {author} {\bibfnamefont {B.~I.}\ \bibnamefont
  {Halperin}},\ }\bibfield  {title} {\bibinfo {title} {Spectral diffusion,
  phonon echoes, and saturation recovery in glasses at low temperatures},\
  }\href {https://doi.org/10.1103/PhysRevB.16.2879} {\bibfield  {journal}
  {\bibinfo  {journal} {Phys. Rev. B}\ }\textbf {\bibinfo {volume} {16}},\
  \bibinfo {pages} {2879} (\bibinfo {year} {1977})}\BibitemShut {NoStop}%
\bibitem [{\citenamefont {Kassner}\ and\ \citenamefont
  {Silbey}(1989)}]{Kassner-89}%
  \BibitemOpen
  \bibfield  {author} {\bibinfo {author} {\bibfnamefont {K.}~\bibnamefont
  {Kassner}}\ and\ \bibinfo {author} {\bibfnamefont {R.}~\bibnamefont
  {Silbey}},\ }\bibfield  {title} {\bibinfo {title} {Interactions of two-level
  systems in glasses},\ }\href {https://doi.org/10.1088/0953-8984/1/28/009}
  {\bibfield  {journal} {\bibinfo  {journal} {J. Phys. Condens. Matter}\
  }\textbf {\bibinfo {volume} {1}},\ \bibinfo {pages} {4599} (\bibinfo {year}
  {1989})}\BibitemShut {NoStop}%
\bibitem [{\citenamefont {Burin}\ \emph {et~al.}(1996)\citenamefont {Burin},
  \citenamefont {Maksimov},\ and\ \citenamefont {Polishchuk}}]{Burin96}%
  \BibitemOpen
  \bibfield  {author} {\bibinfo {author} {\bibfnamefont {A.~L.}\ \bibnamefont
  {Burin}}, \bibinfo {author} {\bibfnamefont {L.~A.}\ \bibnamefont
  {Maksimov}},\ and\ \bibinfo {author} {\bibfnamefont {I.~Y.}\ \bibnamefont
  {Polishchuk}},\ }\bibfield  {title} {\bibinfo {title} {The dephasing rate in
  glasses at ultra low temperatures},\ }\href
  {https://doi.org/10.1007/BF02571127} {\bibfield  {journal} {\bibinfo
  {journal} {Czech. J. Phys.}\ }\textbf {\bibinfo {volume} {46}},\ \bibinfo
  {pages} {2271} (\bibinfo {year} {1996})}\BibitemShut {NoStop}%
\bibitem [{\citenamefont {Asban}\ \emph {et~al.}(2017)\citenamefont {Asban},
  \citenamefont {Amir}, \citenamefont {Imry},\ and\ \citenamefont
  {Schechter}}]{Schechter-PRB17}%
  \BibitemOpen
  \bibfield  {author} {\bibinfo {author} {\bibfnamefont {O.}~\bibnamefont
  {Asban}}, \bibinfo {author} {\bibfnamefont {A.}~\bibnamefont {Amir}},
  \bibinfo {author} {\bibfnamefont {Y.}~\bibnamefont {Imry}},\ and\ \bibinfo
  {author} {\bibfnamefont {M.}~\bibnamefont {Schechter}},\ }\bibfield  {title}
  {\bibinfo {title} {Effect of interactions and disorder on the relaxation of
  two-level systems in amorphous solids},\ }\href
  {https://doi.org/10.1103/PhysRevB.95.144207} {\bibfield  {journal} {\bibinfo
  {journal} {Phys. Rev. B}\ }\textbf {\bibinfo {volume} {95}},\ \bibinfo
  {pages} {144207} (\bibinfo {year} {2017})}\BibitemShut {NoStop}%
\bibitem [{\citenamefont {Nandkishore}\ \emph {et~al.}(2014)\citenamefont
  {Nandkishore}, \citenamefont {Gopalakrishnan},\ and\ \citenamefont
  {Huse}}]{nandkishore2014spectral}%
  \BibitemOpen
  \bibfield  {author} {\bibinfo {author} {\bibfnamefont {R.}~\bibnamefont
  {Nandkishore}}, \bibinfo {author} {\bibfnamefont {S.}~\bibnamefont
  {Gopalakrishnan}},\ and\ \bibinfo {author} {\bibfnamefont {D.~A.}\
  \bibnamefont {Huse}},\ }\bibfield  {title} {\bibinfo {title} {Spectral
  features of a many-body-localized system weakly coupled to a bath},\ }\href
  {https://doi.org/10.1103/PhysRevB.90.064203} {\bibfield  {journal} {\bibinfo
  {journal} {Phys. Rev. B}\ }\textbf {\bibinfo {volume} {90}},\ \bibinfo
  {pages} {064203} (\bibinfo {year} {2014})}\BibitemShut {NoStop}%
\bibitem [{\citenamefont {Levi}\ \emph {et~al.}(2016)\citenamefont {Levi},
  \citenamefont {Heyl}, \citenamefont {Lesanovsky},\ and\ \citenamefont
  {Garrahan}}]{levi2016robustness}%
  \BibitemOpen
  \bibfield  {author} {\bibinfo {author} {\bibfnamefont {E.}~\bibnamefont
  {Levi}}, \bibinfo {author} {\bibfnamefont {M.}~\bibnamefont {Heyl}}, \bibinfo
  {author} {\bibfnamefont {I.}~\bibnamefont {Lesanovsky}},\ and\ \bibinfo
  {author} {\bibfnamefont {J.~P.}\ \bibnamefont {Garrahan}},\ }\bibfield
  {title} {\bibinfo {title} {Robustness of many-body localization in the
  presence of dissipation},\ }\href
  {https://doi.org/10.1103/PhysRevLett.116.237203} {\bibfield  {journal}
  {\bibinfo  {journal} {Phys. Rev. Lett.}\ }\textbf {\bibinfo {volume} {116}},\
  \bibinfo {pages} {237203} (\bibinfo {year} {2016})}\BibitemShut {NoStop}%
\bibitem [{\citenamefont {Fischer}\ \emph {et~al.}(2016)\citenamefont
  {Fischer}, \citenamefont {Maksymenko},\ and\ \citenamefont
  {Altman}}]{fischer2016dynamics}%
  \BibitemOpen
  \bibfield  {author} {\bibinfo {author} {\bibfnamefont {M.~H.}\ \bibnamefont
  {Fischer}}, \bibinfo {author} {\bibfnamefont {M.}~\bibnamefont
  {Maksymenko}},\ and\ \bibinfo {author} {\bibfnamefont {E.}~\bibnamefont
  {Altman}},\ }\bibfield  {title} {\bibinfo {title} {Dynamics of a
  many-body-localized system coupled to a bath},\ }\href
  {https://doi.org/10.1103/PhysRevLett.116.160401} {\bibfield  {journal}
  {\bibinfo  {journal} {Phys. Rev. Lett.}\ }\textbf {\bibinfo {volume} {116}},\
  \bibinfo {pages} {160401} (\bibinfo {year} {2016})}\BibitemShut {NoStop}%
\bibitem [{\citenamefont {Medvedyeva}\ \emph {et~al.}(2016)\citenamefont
  {Medvedyeva}, \citenamefont {Prosen},\ and\ \citenamefont {\ifmmode
  \check{Z}\else \v{Z}\fi{}nidari\ifmmode~\check{c}\else
  \v{c}\fi{}}}]{medvedyeva2016influence}%
  \BibitemOpen
  \bibfield  {author} {\bibinfo {author} {\bibfnamefont {M.~V.}\ \bibnamefont
  {Medvedyeva}}, \bibinfo {author} {\bibfnamefont {T.~c.~v.}\ \bibnamefont
  {Prosen}},\ and\ \bibinfo {author} {\bibfnamefont {M.}~\bibnamefont {\ifmmode
  \check{Z}\else \v{Z}\fi{}nidari\ifmmode~\check{c}\else \v{c}\fi{}}},\
  }\bibfield  {title} {\bibinfo {title} {Influence of dephasing on many-body
  localization},\ }\href {https://doi.org/10.1103/PhysRevB.93.094205}
  {\bibfield  {journal} {\bibinfo  {journal} {Phys. Rev. B}\ }\textbf {\bibinfo
  {volume} {93}},\ \bibinfo {pages} {094205} (\bibinfo {year}
  {2016})}\BibitemShut {NoStop}%
\bibitem [{\citenamefont {Everest}\ \emph {et~al.}(2017)\citenamefont
  {Everest}, \citenamefont {Lesanovsky}, \citenamefont {Garrahan},\ and\
  \citenamefont {Levi}}]{everest2017role}%
  \BibitemOpen
  \bibfield  {author} {\bibinfo {author} {\bibfnamefont {B.}~\bibnamefont
  {Everest}}, \bibinfo {author} {\bibfnamefont {I.}~\bibnamefont {Lesanovsky}},
  \bibinfo {author} {\bibfnamefont {J.~P.}\ \bibnamefont {Garrahan}},\ and\
  \bibinfo {author} {\bibfnamefont {E.}~\bibnamefont {Levi}},\ }\bibfield
  {title} {\bibinfo {title} {Role of interactions in a dissipative many-body
  localized system},\ }\href {https://doi.org/10.1103/PhysRevB.95.024310}
  {\bibfield  {journal} {\bibinfo  {journal} {Phys. Rev. B}\ }\textbf {\bibinfo
  {volume} {95}},\ \bibinfo {pages} {024310} (\bibinfo {year}
  {2017})}\BibitemShut {NoStop}%
\bibitem [{\citenamefont {Nandkishore}\ and\ \citenamefont
  {Gopalakrishnan}(2017)}]{nandkishore2017many}%
  \BibitemOpen
  \bibfield  {author} {\bibinfo {author} {\bibfnamefont {R.}~\bibnamefont
  {Nandkishore}}\ and\ \bibinfo {author} {\bibfnamefont {S.}~\bibnamefont
  {Gopalakrishnan}},\ }\bibfield  {title} {\bibinfo {title} {Many body
  localized systems weakly coupled to baths},\ }\href
  {https://doi.org/10.1002/andp.201600181} {\bibfield  {journal} {\bibinfo
  {journal} {Annalen der Physik}\ }\textbf {\bibinfo {volume} {529}},\ \bibinfo
  {pages} {1600181} (\bibinfo {year} {2017})}\BibitemShut {NoStop}%
\bibitem [{\citenamefont {Vakulchyk}\ \emph {et~al.}(2018)\citenamefont
  {Vakulchyk}, \citenamefont {Yusipov}, \citenamefont {Ivanchenko},
  \citenamefont {Flach},\ and\ \citenamefont
  {Denisov}}]{vakulchyk2018signatures}%
  \BibitemOpen
  \bibfield  {author} {\bibinfo {author} {\bibfnamefont {I.}~\bibnamefont
  {Vakulchyk}}, \bibinfo {author} {\bibfnamefont {I.}~\bibnamefont {Yusipov}},
  \bibinfo {author} {\bibfnamefont {M.}~\bibnamefont {Ivanchenko}}, \bibinfo
  {author} {\bibfnamefont {S.}~\bibnamefont {Flach}},\ and\ \bibinfo {author}
  {\bibfnamefont {S.}~\bibnamefont {Denisov}},\ }\bibfield  {title} {\bibinfo
  {title} {Signatures of many-body localization in steady states of open
  quantum systems},\ }\href {https://doi.org/10.1103/PhysRevB.98.020202}
  {\bibfield  {journal} {\bibinfo  {journal} {Phys. Rev. B}\ }\textbf {\bibinfo
  {volume} {98}},\ \bibinfo {pages} {020202} (\bibinfo {year}
  {2018})}\BibitemShut {NoStop}%
\bibitem [{\citenamefont {Gopalakrishnan}\ and\ \citenamefont
  {Parameswaran}(2020)}]{Gopalakrishnan20}%
  \BibitemOpen
  \bibfield  {author} {\bibinfo {author} {\bibfnamefont {S.}~\bibnamefont
  {Gopalakrishnan}}\ and\ \bibinfo {author} {\bibfnamefont {S.}~\bibnamefont
  {Parameswaran}},\ }\bibfield  {title} {\bibinfo {title} {Dynamics and
  transport at the threshold of many-body localization},\ }\href
  {https://doi.org/https://doi.org/10.1016/j.physrep.2020.03.003} {\bibfield
  {journal} {\bibinfo  {journal} {Physics Reports}\ }\textbf {\bibinfo {volume}
  {862}},\ \bibinfo {pages} {1 } (\bibinfo {year} {2020})}\BibitemShut
  {NoStop}%
\bibitem [{\citenamefont {Wybo}\ \emph {et~al.}(2020)\citenamefont {Wybo},
  \citenamefont {Knap},\ and\ \citenamefont {Pollmann}}]{wybo2020entanglement}%
  \BibitemOpen
  \bibfield  {author} {\bibinfo {author} {\bibfnamefont {E.}~\bibnamefont
  {Wybo}}, \bibinfo {author} {\bibfnamefont {M.}~\bibnamefont {Knap}},\ and\
  \bibinfo {author} {\bibfnamefont {F.}~\bibnamefont {Pollmann}},\ }\bibfield
  {title} {\bibinfo {title} {Entanglement dynamics of a many-body localized
  system coupled to a bath},\ }\href
  {https://doi.org/10.1103/PhysRevB.102.064304} {\bibfield  {journal} {\bibinfo
   {journal} {Phys. Rev. B}\ }\textbf {\bibinfo {volume} {102}},\ \bibinfo
  {pages} {064304} (\bibinfo {year} {2020})}\BibitemShut {NoStop}%
\bibitem [{\citenamefont {Schreiber}\ \emph {et~al.}(2015)\citenamefont
  {Schreiber}, \citenamefont {Hodgman}, \citenamefont {Bordia}, \citenamefont
  {L{\"u}schen}, \citenamefont {Fischer}, \citenamefont {Vosk}, \citenamefont
  {Altman}, \citenamefont {Schneider},\ and\ \citenamefont
  {Bloch}}]{schreiber2015observation}%
  \BibitemOpen
  \bibfield  {author} {\bibinfo {author} {\bibfnamefont {M.}~\bibnamefont
  {Schreiber}}, \bibinfo {author} {\bibfnamefont {S.~S.}\ \bibnamefont
  {Hodgman}}, \bibinfo {author} {\bibfnamefont {P.}~\bibnamefont {Bordia}},
  \bibinfo {author} {\bibfnamefont {H.~P.}\ \bibnamefont {L{\"u}schen}},
  \bibinfo {author} {\bibfnamefont {M.~H.}\ \bibnamefont {Fischer}}, \bibinfo
  {author} {\bibfnamefont {R.}~\bibnamefont {Vosk}}, \bibinfo {author}
  {\bibfnamefont {E.}~\bibnamefont {Altman}}, \bibinfo {author} {\bibfnamefont
  {U.}~\bibnamefont {Schneider}},\ and\ \bibinfo {author} {\bibfnamefont
  {I.}~\bibnamefont {Bloch}},\ }\bibfield  {title} {\bibinfo {title}
  {Observation of many-body localization of interacting fermions in a
  quasirandom optical lattice},\ }\href
  {https://doi.org/10.1126/science.aaa7432} {\bibfield  {journal} {\bibinfo
  {journal} {Science}\ }\textbf {\bibinfo {volume} {349}},\ \bibinfo {pages}
  {842} (\bibinfo {year} {2015})}\BibitemShut {NoStop}%
\bibitem [{\citenamefont {Bordia}\ \emph {et~al.}(2016)\citenamefont {Bordia},
  \citenamefont {L{\"u}schen}, \citenamefont {Hodgman}, \citenamefont
  {Schreiber}, \citenamefont {Bloch},\ and\ \citenamefont
  {Schneider}}]{bordia2016coupling}%
  \BibitemOpen
  \bibfield  {author} {\bibinfo {author} {\bibfnamefont {P.}~\bibnamefont
  {Bordia}}, \bibinfo {author} {\bibfnamefont {H.~P.}\ \bibnamefont
  {L{\"u}schen}}, \bibinfo {author} {\bibfnamefont {S.~S.}\ \bibnamefont
  {Hodgman}}, \bibinfo {author} {\bibfnamefont {M.}~\bibnamefont {Schreiber}},
  \bibinfo {author} {\bibfnamefont {I.}~\bibnamefont {Bloch}},\ and\ \bibinfo
  {author} {\bibfnamefont {U.}~\bibnamefont {Schneider}},\ }\bibfield  {title}
  {\bibinfo {title} {Coupling identical one-dimensional many-body localized
  systems},\ }\href {https://doi.org/10.1103/PhysRevLett.116.140401} {\bibfield
   {journal} {\bibinfo  {journal} {Phys. Rev. Lett.}\ }\textbf {\bibinfo
  {volume} {116}},\ \bibinfo {pages} {140401} (\bibinfo {year}
  {2016})}\BibitemShut {NoStop}%
\bibitem [{\citenamefont {Smith}\ \emph {et~al.}(2016)\citenamefont {Smith},
  \citenamefont {Lee}, \citenamefont {Richerme}, \citenamefont {Neyenhuis},
  \citenamefont {Hess}, \citenamefont {Hauke}, \citenamefont {Heyl},
  \citenamefont {Huse},\ and\ \citenamefont {Monroe}}]{smith2016many}%
  \BibitemOpen
  \bibfield  {author} {\bibinfo {author} {\bibfnamefont {J.}~\bibnamefont
  {Smith}}, \bibinfo {author} {\bibfnamefont {A.}~\bibnamefont {Lee}}, \bibinfo
  {author} {\bibfnamefont {P.}~\bibnamefont {Richerme}}, \bibinfo {author}
  {\bibfnamefont {B.}~\bibnamefont {Neyenhuis}}, \bibinfo {author}
  {\bibfnamefont {P.~W.}\ \bibnamefont {Hess}}, \bibinfo {author}
  {\bibfnamefont {P.}~\bibnamefont {Hauke}}, \bibinfo {author} {\bibfnamefont
  {M.}~\bibnamefont {Heyl}}, \bibinfo {author} {\bibfnamefont {D.~A.}\
  \bibnamefont {Huse}},\ and\ \bibinfo {author} {\bibfnamefont
  {C.}~\bibnamefont {Monroe}},\ }\bibfield  {title} {\bibinfo {title}
  {Many-body localization in a quantum simulator with programmable random
  disorder},\ }\href {https://doi.org/10.1038/NPHYS3783} {\bibfield  {journal}
  {\bibinfo  {journal} {Nature Phys.}\ }\textbf {\bibinfo {volume} {12}},\
  \bibinfo {pages} {907} (\bibinfo {year} {2016})}\BibitemShut {NoStop}%
\bibitem [{\citenamefont {L{\"u}schen}\ \emph {et~al.}(2017)\citenamefont
  {L{\"u}schen}, \citenamefont {Bordia}, \citenamefont {Hodgman}, \citenamefont
  {Schreiber}, \citenamefont {Sarkar}, \citenamefont {Daley}, \citenamefont
  {Fischer}, \citenamefont {Altman}, \citenamefont {Bloch},\ and\ \citenamefont
  {Schneider}}]{luschen2017signatures}%
  \BibitemOpen
  \bibfield  {author} {\bibinfo {author} {\bibfnamefont {H.~P.}\ \bibnamefont
  {L{\"u}schen}}, \bibinfo {author} {\bibfnamefont {P.}~\bibnamefont {Bordia}},
  \bibinfo {author} {\bibfnamefont {S.~S.}\ \bibnamefont {Hodgman}}, \bibinfo
  {author} {\bibfnamefont {M.}~\bibnamefont {Schreiber}}, \bibinfo {author}
  {\bibfnamefont {S.}~\bibnamefont {Sarkar}}, \bibinfo {author} {\bibfnamefont
  {A.~J.}\ \bibnamefont {Daley}}, \bibinfo {author} {\bibfnamefont {M.~H.}\
  \bibnamefont {Fischer}}, \bibinfo {author} {\bibfnamefont {E.}~\bibnamefont
  {Altman}}, \bibinfo {author} {\bibfnamefont {I.}~\bibnamefont {Bloch}},\ and\
  \bibinfo {author} {\bibfnamefont {U.}~\bibnamefont {Schneider}},\ }\bibfield
  {title} {\bibinfo {title} {Signatures of many-body localization in a
  controlled open quantum system},\ }\href
  {https://doi.org/10.1103/PhysRevX.7.011034} {\bibfield  {journal} {\bibinfo
  {journal} {Physical Review X}\ }\textbf {\bibinfo {volume} {7}},\ \bibinfo
  {pages} {011034} (\bibinfo {year} {2017})}\BibitemShut {NoStop}%
\bibitem [{\citenamefont {Hill}\ and\ \citenamefont
  {Wootters}(1997)}]{hill1997entanglement}%
  \BibitemOpen
  \bibfield  {author} {\bibinfo {author} {\bibfnamefont {S.}~\bibnamefont
  {Hill}}\ and\ \bibinfo {author} {\bibfnamefont {W.~K.}\ \bibnamefont
  {Wootters}},\ }\bibfield  {title} {\bibinfo {title} {Entanglement of a pair
  of quantum bits},\ }\href {https://doi.org/10.1103/PhysRevLett.78.5022}
  {\bibfield  {journal} {\bibinfo  {journal} {Phys. Rev. Lett.}\ }\textbf
  {\bibinfo {volume} {78}},\ \bibinfo {pages} {5022} (\bibinfo {year}
  {1997})}\BibitemShut {NoStop}%
\bibitem [{\citenamefont {Wootters}(1998)}]{Wootters98}%
  \BibitemOpen
  \bibfield  {author} {\bibinfo {author} {\bibfnamefont {W.~K.}\ \bibnamefont
  {Wootters}},\ }\bibfield  {title} {\bibinfo {title} {Entanglement of
  formation of an arbitrary state of two qubits},\ }\href
  {https://doi.org/10.1103/PhysRevLett.80.2245} {\bibfield  {journal} {\bibinfo
   {journal} {Phys. Rev. Lett.}\ }\textbf {\bibinfo {volume} {80}},\ \bibinfo
  {pages} {2245} (\bibinfo {year} {1998})}\BibitemShut {NoStop}%
\bibitem [{\citenamefont {Amico}\ \emph {et~al.}(2008)\citenamefont {Amico},
  \citenamefont {Fazio}, \citenamefont {Osterloh},\ and\ \citenamefont
  {Vedral}}]{Amico08}%
  \BibitemOpen
  \bibfield  {author} {\bibinfo {author} {\bibfnamefont {L.}~\bibnamefont
  {Amico}}, \bibinfo {author} {\bibfnamefont {R.}~\bibnamefont {Fazio}},
  \bibinfo {author} {\bibfnamefont {A.}~\bibnamefont {Osterloh}},\ and\
  \bibinfo {author} {\bibfnamefont {V.}~\bibnamefont {Vedral}},\ }\bibfield
  {title} {\bibinfo {title} {Entanglement in many-body systems},\ }\href
  {https://doi.org/10.1103/RevModPhys.80.517} {\bibfield  {journal} {\bibinfo
  {journal} {Rev. Mod. Phys.}\ }\textbf {\bibinfo {volume} {80}},\ \bibinfo
  {pages} {517} (\bibinfo {year} {2008})}\BibitemShut {NoStop}%
\bibitem [{\citenamefont {Carruzzo}\ and\ \citenamefont
  {Yu}(2020)}]{Carruzzo-Yu}%
  \BibitemOpen
  \bibfield  {author} {\bibinfo {author} {\bibfnamefont {H.~M.}\ \bibnamefont
  {Carruzzo}}\ and\ \bibinfo {author} {\bibfnamefont {C.~C.}\ \bibnamefont
  {Yu}},\ }\bibfield  {title} {\bibinfo {title} {Why phonon scattering in
  glasses is universally small at low temperatures},\ }\href
  {https://doi.org/10.1103/PhysRevLett.124.075902} {\bibfield  {journal}
  {\bibinfo  {journal} {Phys. Rev. Lett.}\ }\textbf {\bibinfo {volume} {124}},\
  \bibinfo {pages} {075902} (\bibinfo {year} {2020})}\BibitemShut {NoStop}%
\bibitem [{\citenamefont {Ruzicka}\ \emph {et~al.}(2004)\citenamefont
  {Ruzicka}, \citenamefont {Scopigno}, \citenamefont {Caponi}, \citenamefont
  {Fontana}, \citenamefont {Pilla}, \citenamefont {Giura}, \citenamefont
  {Monaco}, \citenamefont {Pontecorvo}, \citenamefont {Ruocco},\ and\
  \citenamefont {Sette}}]{Ruzicka-04}%
  \BibitemOpen
  \bibfield  {author} {\bibinfo {author} {\bibfnamefont {B.}~\bibnamefont
  {Ruzicka}}, \bibinfo {author} {\bibfnamefont {T.}~\bibnamefont {Scopigno}},
  \bibinfo {author} {\bibfnamefont {S.}~\bibnamefont {Caponi}}, \bibinfo
  {author} {\bibfnamefont {A.}~\bibnamefont {Fontana}}, \bibinfo {author}
  {\bibfnamefont {O.}~\bibnamefont {Pilla}}, \bibinfo {author} {\bibfnamefont
  {P.}~\bibnamefont {Giura}}, \bibinfo {author} {\bibfnamefont
  {G.}~\bibnamefont {Monaco}}, \bibinfo {author} {\bibfnamefont
  {E.}~\bibnamefont {Pontecorvo}}, \bibinfo {author} {\bibfnamefont
  {G.}~\bibnamefont {Ruocco}},\ and\ \bibinfo {author} {\bibfnamefont
  {F.}~\bibnamefont {Sette}},\ }\bibfield  {title} {\bibinfo {title} {Evidence
  of anomalous dispersion of the generalized sound velocity in glasses},\
  }\href {https://doi.org/10.1103/PhysRevB.69.100201} {\bibfield  {journal}
  {\bibinfo  {journal} {Phys. Rev. B}\ }\textbf {\bibinfo {volume} {69}},\
  \bibinfo {pages} {100201} (\bibinfo {year} {2004})}\BibitemShut {NoStop}%
\bibitem [{Note1()}]{Note1}%
  \BibitemOpen
  \bibinfo {note} {Isotropy is due to structural disorder and holds up to short
  scales. A different dispersion is encountered in ultra-stable vapor-deposited
  glasses \cite {Ediger17}, which are essentially two-dimensional.}\BibitemShut
  {Stop}%
\bibitem [{\citenamefont {Berret}\ and\ \citenamefont
  {Mei{\ss}ner}(1988)}]{Berret1988How}%
  \BibitemOpen
  \bibfield  {author} {\bibinfo {author} {\bibfnamefont {J.~F.}\ \bibnamefont
  {Berret}}\ and\ \bibinfo {author} {\bibfnamefont {M.}~\bibnamefont
  {Mei{\ss}ner}},\ }\bibfield  {title} {\bibinfo {title} {How universal are the
  low temperature acoustic properties of glasses?},\ }\href
  {https://doi.org/10.1007/BF01320540} {\bibfield  {journal} {\bibinfo
  {journal} {Z. Phys. B}\ }\textbf {\bibinfo {volume} {70}},\ \bibinfo {pages}
  {65} (\bibinfo {year} {1988})}\BibitemShut {NoStop}%
\bibitem [{\citenamefont {Hunklinger}\ and\ \citenamefont
  {Raychaudhuri}(1986)}]{Hunklinger1986Thermal}%
  \BibitemOpen
  \bibfield  {author} {\bibinfo {author} {\bibfnamefont {S.}~\bibnamefont
  {Hunklinger}}\ and\ \bibinfo {author} {\bibfnamefont {A.}~\bibnamefont
  {Raychaudhuri}},\ }\bibfield  {title} {\bibinfo {title} {Chapter 3: Thermal
  and elastic anomalies in glasses at low temperatures}\ }(\bibinfo
  {publisher} {Elsevier},\ \bibinfo {year} {1986})\ pp.\ \bibinfo {pages}
  {265--344}\BibitemShut {NoStop}%
\bibitem [{\citenamefont {Breuer}\ and\ \citenamefont
  {Petruccione}(2007)}]{BreuerPetruccione}%
  \BibitemOpen
  \bibfield  {author} {\bibinfo {author} {\bibfnamefont {H.-P.}\ \bibnamefont
  {Breuer}}\ and\ \bibinfo {author} {\bibfnamefont {F.}~\bibnamefont
  {Petruccione}},\ }\href
  {https://doi.org/10.1093/acprof:oso/9780199213900.001.0001} {\emph {\bibinfo
  {title} {The Theory of Open Quantum Systems}}}\ (\bibinfo  {publisher}
  {Oxford University Press},\ \bibinfo {address} {Oxford},\ \bibinfo {year}
  {2007})\ p.\ \bibinfo {pages} {656}\BibitemShut {NoStop}%
\bibitem [{\citenamefont {Manzano}(2020)}]{Manzano2020Lindblad}%
  \BibitemOpen
  \bibfield  {author} {\bibinfo {author} {\bibfnamefont {D.}~\bibnamefont
  {Manzano}},\ }\bibfield  {title} {\bibinfo {title} {A short introduction to
  the lindblad master equation},\ }\href {https://doi.org/10.1063/1.5115323}
  {\bibfield  {journal} {\bibinfo  {journal} {AIP Adv.}\ }\textbf {\bibinfo
  {volume} {10}},\ \bibinfo {pages} {025106} (\bibinfo {year}
  {2020})}\BibitemShut {NoStop}%
\bibitem [{Note2()}]{Note2}%
  \BibitemOpen
  \bibinfo {note} {Even accounting for rare interactions in the $S^x$--$S^x$
  channel, the picture is not modified. Indeed, terms of the form $K_{ij} S_i^x
  S_j^x$ will still decay with the distance $r_{ij}$: the probability of having
  a resonant $ij$ couple that is \protect \emph {also close in real space} is
  vanishingly small. Therefore, the MBL-breaking effect of weak $S^x_i S^x_j$
  terms \cite {Yao14dipolar,Burin15MBL,deng2020anisotropymediated} is
  negligible in comparison to the Lindblad dissipator.}\BibitemShut {Stop}%
\bibitem [{\citenamefont {O'Brien}\ \emph {et~al.}(2016)\citenamefont
  {O'Brien}, \citenamefont {Abanin}, \citenamefont {Vidal},\ and\ \citenamefont
  {Papi\'{c}}}]{Abanin2016Explicit}%
  \BibitemOpen
  \bibfield  {author} {\bibinfo {author} {\bibfnamefont {T.~E.}\ \bibnamefont
  {O'Brien}}, \bibinfo {author} {\bibfnamefont {D.~A.}\ \bibnamefont {Abanin}},
  \bibinfo {author} {\bibfnamefont {G.}~\bibnamefont {Vidal}},\ and\ \bibinfo
  {author} {\bibfnamefont {Z.}~\bibnamefont {Papi\'{c}}},\ }\bibfield  {title}
  {\bibinfo {title} {Explicit construction of local conserved operators in
  disordered many-body systems},\ }\href
  {https://doi.org/10.1103/PhysRevB.94.144208} {\bibfield  {journal} {\bibinfo
  {journal} {Phys. Rev. B}\ }\textbf {\bibinfo {volume} {94}},\ \bibinfo
  {pages} {144208} (\bibinfo {year} {2016})}\BibitemShut {NoStop}%
\bibitem [{\citenamefont {Ros}\ \emph {et~al.}(2015)\citenamefont {Ros},
  \citenamefont {Müller},\ and\ \citenamefont
  {Scardicchio}}]{ros2015integrals}%
  \BibitemOpen
  \bibfield  {author} {\bibinfo {author} {\bibfnamefont {V.}~\bibnamefont
  {Ros}}, \bibinfo {author} {\bibfnamefont {M.}~\bibnamefont {Müller}},\ and\
  \bibinfo {author} {\bibfnamefont {A.}~\bibnamefont {Scardicchio}},\
  }\bibfield  {title} {\bibinfo {title} {Integrals of motion in the many-body
  localized phase},\ }\href
  {https://doi.org/https://doi.org/10.1016/j.nuclphysb.2014.12.014} {\bibfield
  {journal} {\bibinfo  {journal} {Nucl. Phys. B}\ }\textbf {\bibinfo {volume}
  {891}},\ \bibinfo {pages} {420 } (\bibinfo {year} {2015})}\BibitemShut
  {NoStop}%
\bibitem [{\citenamefont {{Imbrie}}(2016)}]{imbrie2014many}%
  \BibitemOpen
  \bibfield  {author} {\bibinfo {author} {\bibfnamefont {J.~Z.}\ \bibnamefont
  {{Imbrie}}},\ }\bibfield  {title} {\bibinfo {title} {{On Many-Body
  Localization for Quantum Spin Chains}},\ }\href
  {https://doi.org/https://doi.org/10.1007/s10955-016-1508-x} {\bibfield
  {journal} {\bibinfo  {journal} {J. Stat. Phys.}\ }\textbf {\bibinfo {volume}
  {163}},\ \bibinfo {pages} {998} (\bibinfo {year} {2016})}\BibitemShut
  {NoStop}%
\bibitem [{\citenamefont {Imbrie}\ \emph {et~al.}(2017)\citenamefont {Imbrie},
  \citenamefont {Ros},\ and\ \citenamefont {Scardicchio}}]{imbrie2016review}%
  \BibitemOpen
  \bibfield  {author} {\bibinfo {author} {\bibfnamefont {J.~Z.}\ \bibnamefont
  {Imbrie}}, \bibinfo {author} {\bibfnamefont {V.}~\bibnamefont {Ros}},\ and\
  \bibinfo {author} {\bibfnamefont {A.}~\bibnamefont {Scardicchio}},\
  }\bibfield  {title} {\bibinfo {title} {Local integrals of motion in many-body
  localized systems},\ }\href {https://doi.org/10.1002/andp.201600278}
  {\bibfield  {journal} {\bibinfo  {journal} {Ann. Phys.}\ }\textbf {\bibinfo
  {volume} {529}},\ \bibinfo {pages} {1600278} (\bibinfo {year}
  {2017})}\BibitemShut {NoStop}%
\bibitem [{\citenamefont {Serbyn}\ \emph {et~al.}(2014)\citenamefont {Serbyn},
  \citenamefont {Papi\ifmmode~\acute{c}\else \'{c}\fi{}},\ and\ \citenamefont
  {Abanin}}]{serbyn2014quantum}%
  \BibitemOpen
  \bibfield  {author} {\bibinfo {author} {\bibfnamefont {M.}~\bibnamefont
  {Serbyn}}, \bibinfo {author} {\bibfnamefont {Z.}~\bibnamefont
  {Papi\ifmmode~\acute{c}\else \'{c}\fi{}}},\ and\ \bibinfo {author}
  {\bibfnamefont {D.~A.}\ \bibnamefont {Abanin}},\ }\bibfield  {title}
  {\bibinfo {title} {Quantum quenches in the many-body localized phase},\
  }\href {https://doi.org/10.1103/PhysRevB.90.174302} {\bibfield  {journal}
  {\bibinfo  {journal} {Phys. Rev. B}\ }\textbf {\bibinfo {volume} {90}},\
  \bibinfo {pages} {174302} (\bibinfo {year} {2014})}\BibitemShut {NoStop}%
\bibitem [{\citenamefont {Iemini}\ \emph {et~al.}(2016)\citenamefont {Iemini},
  \citenamefont {Russomanno}, \citenamefont {Rossini}, \citenamefont
  {Scardicchio},\ and\ \citenamefont {Fazio}}]{iemini2016signatures}%
  \BibitemOpen
  \bibfield  {author} {\bibinfo {author} {\bibfnamefont {F.}~\bibnamefont
  {Iemini}}, \bibinfo {author} {\bibfnamefont {A.}~\bibnamefont {Russomanno}},
  \bibinfo {author} {\bibfnamefont {D.}~\bibnamefont {Rossini}}, \bibinfo
  {author} {\bibfnamefont {A.}~\bibnamefont {Scardicchio}},\ and\ \bibinfo
  {author} {\bibfnamefont {R.}~\bibnamefont {Fazio}},\ }\bibfield  {title}
  {\bibinfo {title} {Signatures of many-body localization in the dynamics of
  two-site entanglement},\ }\href {https://doi.org/10.1103/PhysRevB.94.214206}
  {\bibfield  {journal} {\bibinfo  {journal} {Phys. Rev. B}\ }\textbf {\bibinfo
  {volume} {94}},\ \bibinfo {pages} {214206} (\bibinfo {year}
  {2016})}\BibitemShut {NoStop}%
\bibitem [{\citenamefont {\v{Z}nidari\v{c}}(2018)}]{Znidaric18}%
  \BibitemOpen
  \bibfield  {author} {\bibinfo {author} {\bibfnamefont {M.}~\bibnamefont
  {\v{Z}nidari\v{c}}},\ }\bibfield  {title} {\bibinfo {title} {Entanglement in
  a dephasing model and many-body localization},\ }\href
  {https://doi.org/10.1103/PhysRevB.97.214202} {\bibfield  {journal} {\bibinfo
  {journal} {Phys. Rev. B}\ }\textbf {\bibinfo {volume} {97}},\ \bibinfo
  {pages} {214202} (\bibinfo {year} {2018})}\BibitemShut {NoStop}%
\bibitem [{\citenamefont {Pino}(2014)}]{pino2014entanglement}%
  \BibitemOpen
  \bibfield  {author} {\bibinfo {author} {\bibfnamefont {M.}~\bibnamefont
  {Pino}},\ }\bibfield  {title} {\bibinfo {title} {Entanglement growth in
  many-body localized systems with long-range interactions},\ }\href
  {https://doi.org/10.1103/PhysRevB.90.174204} {\bibfield  {journal} {\bibinfo
  {journal} {Phys. Rev. B}\ }\textbf {\bibinfo {volume} {90}},\ \bibinfo
  {pages} {174204} (\bibinfo {year} {2014})}\BibitemShut {NoStop}%
\bibitem [{\citenamefont {Bardarson}\ \emph {et~al.}(2012)\citenamefont
  {Bardarson}, \citenamefont {Pollmann},\ and\ \citenamefont
  {Moore}}]{bardarson2012unbounded}%
  \BibitemOpen
  \bibfield  {author} {\bibinfo {author} {\bibfnamefont {J.~H.}\ \bibnamefont
  {Bardarson}}, \bibinfo {author} {\bibfnamefont {F.}~\bibnamefont
  {Pollmann}},\ and\ \bibinfo {author} {\bibfnamefont {J.~E.}\ \bibnamefont
  {Moore}},\ }\bibfield  {title} {\bibinfo {title} {Unbounded growth of
  entanglement in models of many-body localization},\ }\href
  {https://doi.org/10.1103/PhysRevLett.109.017202} {\bibfield  {journal}
  {\bibinfo  {journal} {Physical review letters}\ }\textbf {\bibinfo {volume}
  {109}},\ \bibinfo {pages} {017202} (\bibinfo {year} {2012})}\BibitemShut
  {NoStop}%
\bibitem [{\citenamefont {Serbyn}\ \emph {et~al.}(2013)\citenamefont {Serbyn},
  \citenamefont {Papi\'{c}},\ and\ \citenamefont
  {Abanin}}]{serbyn2013universal}%
  \BibitemOpen
  \bibfield  {author} {\bibinfo {author} {\bibfnamefont {M.}~\bibnamefont
  {Serbyn}}, \bibinfo {author} {\bibfnamefont {Z.}~\bibnamefont {Papi\'{c}}},\
  and\ \bibinfo {author} {\bibfnamefont {D.~A.}\ \bibnamefont {Abanin}},\
  }\bibfield  {title} {\bibinfo {title} {Universal slow growth of entanglement
  in interacting strongly disordered systems},\ }\href
  {https://doi.org/10.1103/PhysRevLett.110.260601} {\bibfield  {journal}
  {\bibinfo  {journal} {Phys. Rev. Lett.}\ }\textbf {\bibinfo {volume} {110}},\
  \bibinfo {pages} {260601} (\bibinfo {year} {2013})}\BibitemShut {NoStop}%
\bibitem [{\citenamefont {Ediger}(2017)}]{Ediger17}%
  \BibitemOpen
  \bibfield  {author} {\bibinfo {author} {\bibfnamefont {M.~D.}\ \bibnamefont
  {Ediger}},\ }\bibfield  {title} {\bibinfo {title} {Perspective: Highly stable
  vapor-deposited glasses},\ }\href {https://doi.org/10.1063/1.5006265}
  {\bibfield  {journal} {\bibinfo  {journal} {J. Chem. Phys.}\ }\textbf
  {\bibinfo {volume} {147}},\ \bibinfo {pages} {210901} (\bibinfo {year}
  {2017})}\BibitemShut {NoStop}%
\bibitem [{\citenamefont {Yao}\ \emph {et~al.}(2014)\citenamefont {Yao},
  \citenamefont {Laumann}, \citenamefont {Gopalakrishnan}, \citenamefont
  {Knap}, \citenamefont {M\"uller}, \citenamefont {Demler},\ and\ \citenamefont
  {Lukin}}]{Yao14dipolar}%
  \BibitemOpen
  \bibfield  {author} {\bibinfo {author} {\bibfnamefont {N.~Y.}\ \bibnamefont
  {Yao}}, \bibinfo {author} {\bibfnamefont {C.~R.}\ \bibnamefont {Laumann}},
  \bibinfo {author} {\bibfnamefont {S.}~\bibnamefont {Gopalakrishnan}},
  \bibinfo {author} {\bibfnamefont {M.}~\bibnamefont {Knap}}, \bibinfo {author}
  {\bibfnamefont {M.}~\bibnamefont {M\"uller}}, \bibinfo {author}
  {\bibfnamefont {E.~A.}\ \bibnamefont {Demler}},\ and\ \bibinfo {author}
  {\bibfnamefont {M.~D.}\ \bibnamefont {Lukin}},\ }\bibfield  {title} {\bibinfo
  {title} {Many-body localization in dipolar systems},\ }\href
  {https://doi.org/10.1103/PhysRevLett.113.243002} {\bibfield  {journal}
  {\bibinfo  {journal} {Phys. Rev. Lett.}\ }\textbf {\bibinfo {volume} {113}},\
  \bibinfo {pages} {243002} (\bibinfo {year} {2014})}\BibitemShut {NoStop}%
\bibitem [{\citenamefont {Burin}(2015)}]{Burin15MBL}%
  \BibitemOpen
  \bibfield  {author} {\bibinfo {author} {\bibfnamefont {A.~L.}\ \bibnamefont
  {Burin}},\ }\bibfield  {title} {\bibinfo {title} {Many-body delocalization in
  a strongly disordered system with long-range interactions: Finite-size
  scaling},\ }\href {https://doi.org/10.1103/PhysRevB.91.094202} {\bibfield
  {journal} {\bibinfo  {journal} {Phys. Rev. B}\ }\textbf {\bibinfo {volume}
  {91}},\ \bibinfo {pages} {094202} (\bibinfo {year} {2015})}\BibitemShut
  {NoStop}%
\bibitem [{\citenamefont {Deng}\ \emph {et~al.}(2020)\citenamefont {Deng},
  \citenamefont {Burin},\ and\ \citenamefont
  {Khaymovich}}]{deng2020anisotropymediated}%
  \BibitemOpen
  \bibfield  {author} {\bibinfo {author} {\bibfnamefont {X.}~\bibnamefont
  {Deng}}, \bibinfo {author} {\bibfnamefont {A.~L.}\ \bibnamefont {Burin}},\
  and\ \bibinfo {author} {\bibfnamefont {I.~M.}\ \bibnamefont {Khaymovich}},\
  }\bibfield  {title} {\bibinfo {title} {Anisotropy-mediated reentrant
  localization},\ }\href@noop {} {\bibfield  {journal} {\bibinfo  {journal}
  {arXiv:2002.00013}\ } (\bibinfo {year} {2020})}\BibitemShut {NoStop}%
\bibitem [{\citenamefont {Lloyd}\ and\ \citenamefont {Pagels}(1988)}]{Lloyd88}%
  \BibitemOpen
  \bibfield  {author} {\bibinfo {author} {\bibfnamefont {S.}~\bibnamefont
  {Lloyd}}\ and\ \bibinfo {author} {\bibfnamefont {H.}~\bibnamefont {Pagels}},\
  }\bibfield  {title} {\bibinfo {title} {Complexity as thermodynamic depth},\
  }\href {https://doi.org/https://doi.org/10.1016/0003-4916(88)90094-2}
  {\bibfield  {journal} {\bibinfo  {journal} {Ann. Phys.}\ }\textbf {\bibinfo
  {volume} {188}},\ \bibinfo {pages} {186 } (\bibinfo {year}
  {1988})}\BibitemShut {NoStop}%
\bibitem [{\citenamefont {Page}(1993)}]{DonPage93}%
  \BibitemOpen
  \bibfield  {author} {\bibinfo {author} {\bibfnamefont {D.~N.}\ \bibnamefont
  {Page}},\ }\bibfield  {title} {\bibinfo {title} {Average entropy of a
  subsystem},\ }\href {https://doi.org/10.1103/PhysRevLett.71.1291} {\bibfield
  {journal} {\bibinfo  {journal} {Phys. Rev. Lett.}\ }\textbf {\bibinfo
  {volume} {71}},\ \bibinfo {pages} {1291} (\bibinfo {year}
  {1993})}\BibitemShut {NoStop}%
\end{thebibliography}%

\appendix

\begin{widetext}
\section{Explicit form of interactions and dissipator}
\label{app:sec:computation_Lindblad}

We want to compute explicitly $Y_i$ and $J_{ij}$ defined in  Eqs.~\eqref{eq:def_Yi} and \eqref{eq:def_Jij}, respectively. To do so, we need $\Gamma_{ij}^\omega$ defined in Eq.~\eqref{eq:def_Gamma}, that we reproduce here:
\begin{equation}
    \Gamma_{ij}^\omega := \frac{1}{\hbar^2} \int_0^\infty ds \, e^{i \omega s} \, \Tr_B \left[ \rho_B^T \, \hat E_i^\dagger(t) \, \hat E_j(t-s) \right].
\end{equation}
Therefore, as a first thing we need to evolve the operators $E_i$ in the interaction picture. Recalling that (Eqs.~\eqref{eq:def_xi_vectors} and \eqref{eq:def_Ei})
\begin{equation}
    E_j := \xi_{j k} \psi_k + \mathrm{h.c.} = - i \sqrt{\frac{\hbar}{2 V \rho \, \omega_k}} \gamma_j D_j^{ab} e^{ab}_k e^{i \mathbf{q} \cdot \mathbf{r}_j} \psi_k + \mathrm{h.c.}, 
\end{equation}
it holds
\begin{align}
    \notag
    \hat E_i(t) &= e^{i H_{\mathit{ph}} t /\hbar} E_i e^{-i H_{\mathit{ph}} t /\hbar} \\
    &= \sum_k \big( \xi_{ik} e^{-i \omega_k t} \psi_k + \xi_{ik}^* e^{i \omega_k t} \psi_k^\dagger \big).
\end{align}
Thus, it follows
\begin{align}
    \notag
    \hbar^2 \Gamma_{ij}^\omega 
    &= \int_0^\infty ds \, e^{i\omega s} \, \Tr_B \bigg\{ 
    \rho_B^T \sum_{kl}  \big( \xi_{ik} e^{-i \omega_k t} \psi_k + {\rm h.c.} \big) \big( \xi_{jl} e^{-i \omega_l (t-s)} \psi_l + {\rm h.c.} \big) \bigg\} \\ \notag
    &= \int_0^\infty ds \, e^{i\omega s} \, \sum_k \left\{ \xi_{ik} \xi_{jk}^* \, e^{-i \omega_k s} \, \Tr_B \big[\rho_B^T \psi_k^{\phantom{\dagger}} \psi_k^\dagger \big]
    + \xi_{ik}^* \xi_{jk} \, e^{i \omega_k s} \, \Tr_B \big[\rho_B^T \psi_k^\dagger \psi_k^{\phantom{\dagger}} \big] \right\} \\
    & = \int_0^\infty ds \, e^{i\omega s} \, \sum_k \left[ \xi_{ik} \xi_{jk}^* \, e^{-i \omega_k s} \, (f_T(\hbar \omega_k) + 1)
    + \xi_{ik}^* \xi_{jk} \, e^{i \omega_k s} \, f_T(\hbar \omega_k) \right],
\end{align}
where we recall $f_T$ is the Bose-Einstein distribution function at temperature $T$. We perform the time integral using the identity
\begin{equation}
    \int_0^\infty ds \, e^{i \zeta s} = i \, \PV \frac{1}{\zeta} + \pi \delta(\zeta).
\end{equation}
Plugging in the explicit expression of $\xi_{ik}$ from Eqs.~\eqref{eq:def_xi_vectors}, we arrive at
\begin{multline}
    \label{eq:our_Gamma}
    \Gamma_{ij}^\omega = \frac{\gamma_i \gamma_j}{2 \rho} \sum_{abcd} D_i^{ab} D_j^{cd}
    \sum_\alpha \int \frac{d^3 q}{(2\pi)^3} \, \frac{1}{\hbar\omega_{\mathbf{q}\alpha}} \, e^{ab}_{\mathbf{q}\alpha} e^{cd}_{\mathbf{q}\alpha} \left[ (f_T(\hbar \omega_{\mathbf{q}\alpha})+1) \left( i \PV \frac{1}{\omega-\omega_{\mathbf{q}\alpha}} + \pi \delta(\omega-\omega_{\mathbf{q}\alpha}) \right) e^{i \mathbf{q} \cdot (\mathbf{r}_i - \mathbf{r}_j)} \right. \\
    + \left. f_T(\hbar \omega_{\mathbf{q}\alpha}) \left( i \PV \frac{1}{\omega+\omega_{\mathbf{q}\alpha}} + \pi \delta(\omega+\omega_{\mathbf{q}\alpha}) \right) e^{-i \mathbf{q} \cdot (\mathbf{r}_i - \mathbf{r}_j)} \right].
\end{multline}

\subsection{The dissipation rates}

The dissipation rates $Y_i$ can be computed from Eq.~\eqref{eq:our_Gamma} by taking the real part of $\Gamma^{\omega}_{ii}$ (see Eq.~\eqref{eq:def_Yi}):
\begin{equation}
    \label{eq:def_Yi_alternative}
    Y_i = \left(\frac{\Delta_i}{\hbar \nu_i} \right)^2 2 \, \Re \Gamma_{ii}^{\nu_i} \big|_{T=0}.
\end{equation}
Hence, we need to compute (see Eq.~\eqref{eq:our_Gamma})
\begin{equation}
    \Re \Gamma_{ii}^\omega = \frac{\pi \gamma_i^2}{2 \rho} \sum_{abcd} D_i^{ab} D_i^{cd} \sum_\alpha \int \frac{d^3 q}{(2\pi)^3} \, \frac{1}{\hbar\omega_{\mathbf{q}\alpha}} \, e^{ab}_{\mathbf{q}\alpha}e^{cd}_{\mathbf{q}\alpha}
    \left[ (f_T(\hbar \omega_{\mathbf{q}\alpha})+1) \delta(\omega-\omega_{\mathbf{q}\alpha}) + f_T(\hbar \omega_{\mathbf{q}\alpha}) \delta(\omega+\omega_{\mathbf{q}\alpha}) \right].
\end{equation}
One could in principle consider the longitudinal and transverse polarizations separately, however it is convenient to employ an isotropic Debye model with sound velocity
\begin{equation}
    \label{eq:average_v}
    \frac{1}{v^3} := \frac{1}{3} \sum_\alpha \frac{1}{v_\alpha^3}. 
\end{equation}
Within this assumption, it is convenient to compute the angular averages summing over all polarizations as
\begin{equation}
    \label{eq:TrD2}
    \frac{1}{4\pi} \sum_{abcd} \sum_\alpha \int d\Omega \, e^{ab}_{\mathbf{q} \alpha} e^{cd}_{\mathbf{q} \alpha} D_i^{ab} D_i^{cd} = \frac{1}{3}\, \Tr(D_i^2) \, q^2,
\end{equation}
and therefore,
\begin{equation}
    \label{eq:ReGamma}
    \Re \Gamma_{ii}^\omega
    = \frac{\gamma_i^2 \Tr(D_i^2)}{12 \pi \rho \hbar v}
    \int_0^\infty dq \, q^3 \big[ (f_T(\hbar v q)+1) \delta(\omega-v q) + f_T(\hbar v q) \delta(\omega + vq) \big].
\end{equation}

We know from the theory of the GKSL equation \cite{BreuerPetruccione,Manzano2020Lindblad} that dissipation and dephasing rates are obtained by setting respectively $\omega = \pm \nu_i, 0$ in Eq.~\eqref{eq:ReGamma}. However, as argued in the main text we are effectively at zero temperature: $f_{T=0} = 0$, and we are left with only
\begin{equation}
    \Re \Gamma_{ii}^{\nu_{i}}\big|_{T=0} = \frac{\gamma_i^2 \nu_i^3 \Tr(D_i^2) }{12 \pi \rho \hbar v^5}.
\end{equation}
Notice in particular that $\Gamma_{ii}^0 = 0$ since the phonons have zero density of states at $\omega = 0$. Using Eq.~\eqref{eq:def_Yi_alternative}, we finally arrive at 
\begin{equation}
    Y_i = \frac{\Delta_i^2 \gamma_i^2 \nu_i \Tr (D^2_i) }{12 \pi\rho \hbar^3 v^5},
\end{equation}
that is exactly Eq.~\eqref{eq:Y_i}.

\subsection{The interaction strengths}

From the general considerations reported in the main text (see Sec.~\ref{sec:coupling_phonons} and Fig.~\ref{fig:processes}), we know that interactions take place mostly in the $S^z$--$S^z$ channel. What we need to compute is the coefficient $J_{ij}$ in front, that comes from the imaginary part of $\Gamma_{ij}^0$ of Eq.~\eqref{eq:our_Gamma}. With hindsight, we note that the temperature-dependent terms will not contribute; therefore, we just need to compute the following quantity:
\begin{equation}
    \Pi_{ij} := -i \big[ \Gamma_{ij}^0 - \big( \Gamma_{ji}^0 \big)^* \big]
    = \frac{\gamma_i \gamma_j}{4 \rho} \sum_{abcd} D_i^{ab} D_j^{cd} \sum_\alpha \PV \! \int \frac{d^3 q}{(2\pi)^3} \, \frac{1}{\hbar\omega_{\mathbf{q}\alpha}} \, (q^a \hat e^b_{\mathbf{q}\alpha} + q^b \hat e^a_{\mathbf{q}\alpha}) (q^c \hat e^d_{\mathbf{q}\alpha} + q^d \hat e^c_{\mathbf{q}\alpha}) \frac{e^{i \mathbf{q} \cdot (\mathbf{r}_i-\mathbf{r}_j)} }{-\omega_{\mathbf{q}\alpha}} .
\end{equation}
The interactions $J_{ij}$ are then given by (see Eq.~\eqref{eq:def_Jij})
\begin{equation}
    \label{eq:J_Pi}
    J_{ij} = \frac{2\veps_i}{\hbar \nu_i} \frac{2\veps_j}{\hbar \nu_j} \,\frac{\hbar}{2} \, \Pi_{ij}.
\end{equation}
We can proceed as follows: we split the different polarization contributions, then evaluate the angular integrals, and, finally, the $|\mathbf{q}|$ integral. Treating separately the different polarization here is crucial: as will be clear from Eq.~\eqref{eq:def_Dij}, there is a contribution that vanishes if $v_L = v_T$.  

Let us define some quantities that will soon appear in the computation: 
\begin{align}
    \label{eq:Iabcd}
    I^{abcd}(\zeta) &:= \frac{1}{4\pi} \int d\Omega \, \hat q^a \hat q^b \hat q^c \hat q^d e^{i\zeta \cos\theta}, \\
    \label{eq:Iab}
    I^{ab}(\zeta) &:= \frac{1}{4\pi} \int d\Omega \, \hat q^a \hat q^b e^{i\zeta \cos\theta}.
\end{align}
Explicitly, they read:
\begin{equation}
\begin{aligned}
    I^{xxyy}(\zeta) &= \frac{1}{3} I^{xxxx}(\zeta) = -\frac{3\zeta\cos\zeta + (\zeta^2-3)\sin\zeta}{\zeta^5}, \\
    I^{xxzz}(\zeta) &= -\frac{\zeta(\zeta^2-12)\cos\zeta - (5\zeta^2-12)\sin\zeta}{\zeta^5}, \\
    I^{zzzz}(\zeta) &= \frac{4\zeta(\zeta^2-6)\cos\zeta + (\zeta^4 -12\zeta^2 +24)\sin\zeta}{\zeta^5}, \\
    I^{xx}(\zeta) &= \frac{-\zeta\cos\zeta + \sin\zeta}{\zeta^3}, \\
    I^{zz}(\zeta) &= \frac{2\zeta\cos\zeta + (\zeta^2-2) \sin\zeta}{\zeta^3}.
\end{aligned}
\end{equation}
Similar ones are obtained exchanging $x$ and $y$ and permuting the indices; all the others are zero. We can parametrize them as
\begin{equation}
\begin{aligned}
    I^{abcd}(\zeta) &= \frac{1}{\zeta^5} \sum_{l=0}^4 C^{abcd}_l \, \zeta^l \, s_l(\zeta), \\
    I^{ab}(\zeta) &= \frac{1}{\zeta^3} \sum_{l=0}^2 C^{ab}_l \, \zeta^l \, s_l(\zeta)
\end{aligned}
\end{equation}
where
\begin{equation}
    s_l(\zeta) := 
    \begin{cases}
        \sin \zeta      &l \text{ even} \\
        \cos \zeta      &l \text{ odd.}
    \end{cases}
\end{equation} 

Let us start considering the longitudinally polarized modes. Since $\mathbf{\hat e}_{\mathbf{q} L} = \mathbf{\hat q}$, we find
\begin{equation}
    \label{eq:Pi_L}
    \big[ \Pi_{ij} \big]_L = - \frac{\gamma_i \gamma_j}{\rho} \sum_{abcd} D_i^{ab} D_j^{cd} \, \PV \! \int \frac{d^3 q}{(2\pi)^3} \, \frac{1}{\hbar v_L^2 q^2} \,q^2 \hat q^a \hat q^b \hat q^c \hat q^d e^{i \mathbf{q} \cdot (\mathbf{r}_i-\mathbf{r}_j)}  .
\end{equation}
Setting the $\hat z$ axis along $(\mathbf{r}_i - \mathbf{r}_j)$, defining the modulus distance $|\mathbf{r}_i-\mathbf{r}_j|=r_{ij}$ and $\zeta=qr_{ij}$, and using the definition of $I^{abcd}$ in Eq.~\eqref{eq:Iabcd} above, we find 
\begin{align}
    \notag
    \big[ \Pi_{ij} \big]_L &= - \frac{\gamma_i \gamma_j}{2 \pi^2 \rho \hbar v_L^2} \sum_{abcd} D_i^{ab} D_j^{cd} \, \PV \! \int dq \, q^2 I^{abcd}(q r_{ij}) \\
    &= - \frac{\gamma_i \gamma_j}{2 \pi^2 \rho \hbar v_L^2 r_{ij}^3 } \sum_{l=0}^4 \left[ \sum_{abcd} C^{abcd}_l D_i^{ab} D_j^{cd} \right] \, \PV \! \int d\zeta \, \zeta^{l-3} s_l(\zeta) .
\end{align}
One can check that all the IR divergences cancel out (since $C_{0}^{abcd} = - C_1^{abcd}$), while the UV divergences are harmless thanks to the oscillating functions $s_l(\zeta)$. We find
\begin{align}
    \nonumber
    \big[ \Pi_{ij} \big]_L &= - \frac{\gamma_i \gamma_j}{2 \pi^2 \rho \hbar v_L^2 r_{ij}^3 } \sum_{abcd} \left( \frac{\pi}{4} C^{abcd}_0 + \frac{\pi}{2} C^{abcd}_2 \right) D_i^{ab} D_j^{cd} \\ 
    \label{eq:Pi_L_final}
    &= - \frac{\gamma_i \gamma_j}{8 \pi \rho \hbar v_L^2 r_{ij}^3 } \sum_{abcd} ( C^{abcd}_0 + 2 C^{abcd}_2) D_i^{ab} D_j^{cd} .
\end{align} 

Now we perform a similar computation for the transversely polarized modes. Using the relation,
\begin{equation}
    \sum_{\alpha \; \text{trans.}} \hat e_{\mathbf{q} \alpha}^a \hat e_{\mathbf{q} \alpha}^b = \delta^{ab}-\hat q^a \hat q^b,
\end{equation}
we see that there are terms involving $I^{ab}$ (coming from $\delta^{ab}$) and terms involving $I^{abcd}$ (coming form $\hat q^a \hat q^b$). It is easy to check that the result is
\begin{equation}
    \label{eq:Pi_T_final}
    \big[ \Pi_{ij} \big]_T = - \frac{\gamma_i \gamma_j}{8 \pi \rho \hbar v_T^2 r_{ij}^3 }  \sum_{abcd} (2 C_0^{ac} \delta^{bd} - C^{abcd}_0 - 2 C^{abcd}_2) D_i^{ab} D_j^{cd}.
\end{equation}

Summing the longitudinal and transverse contributions in Eqs.~\eqref{eq:Pi_L_final} and \eqref{eq:Pi_T_final}, we finally obtain
\begin{equation}
    \Pi_{ij} = \frac{\gamma_i \gamma_j \mathbb{D}_{ij} }{8 \pi \rho \hbar v^2 r_{ij}^3 } 
\end{equation}
having defined
\begin{equation}
    \label{eq:def_Dij}
    \mathbb{D}_{ij} := v^2 \sum_{abcd} \left[ - \frac{2 C_0^{ac} \delta^{bd}}{v_T^2} + (C^{abcd}_0 + 2 C^{abcd}_2) \left(\frac{1}{v_T^2} - \frac{1}{v_L^2} \right) \right] D_i^{ab} D_j^{cd},
\end{equation}
where $v$ is the average velocity defined in Eq.~\eqref{eq:average_v}. Despite the cumbersome appearance, $\mathbb{D}_{ij}$ are dimensionless random variables with zero average and standard deviation of order 1. Finally, by means of Eq.~\eqref{eq:J_Pi}:
\begin{equation}
    J_{ij} = \frac{\gamma_i \veps_i}{\hbar \nu_i} \frac{\gamma_j \veps_j}{\hbar \nu_j} \, \frac{\mathbb{D}_{ij} }{4 \pi \rho v^2 r_{ij}^3 },
\end{equation}
that is Eq.~\eqref{eq:J_ij} of the main text.


\section{Two-site observables within the diagonal unitary evolution}
\label{app:sec:numerics_unitary_dynamics}

In this Section, we show how to compute with $O(N)$ steps the two-site density matrix $\rho_{ij}$, and therefore any two-site observable, for the Hamiltonian \eqref{eq:lbit_hamiltonian}. Call the initial density matrix 
\begin{equation}
    \rho_0 = \bigotimes_{i=1}^N \rho_{0,i} = \bigotimes_{i=1}^N \sum_{s_i, s_i'} \rho_{0,i}^{s_i s_i'} \ket{s_i} \bra{s_i'},
\end{equation}
and recall that the Hamiltonian \eqref{eq:lbit_hamiltonian} reads explicitly
\begin{equation*}
    H_{\it TLS} + H_{\it LS} = - \frac{1}{2} \sum_i \hbar \nu_i S_i^z + \sum_{ij} J_{ij} S_i^z S_j^z.
\end{equation*}
Time evolving the density matrix according to the von Neumann equation and rearranging the sum, one finds
\begin{equation}
    \rho(t) = \sum_{s,s'} \prod_i \rho_{0,i}^{s_i s_i'} \ket{s} \bra{s'} e^{-i (H[s] - H[s'])t/\hbar}
\end{equation}
with $H[s] = -\frac 1 2 \sum_{i} \hbar \nu_i s_i + \sum_{ij} J_{ij} s_i s_j$, where $s_i=\pm 1$ is the projection of the spin-1/2 on the $z$ axis. Without loss of generality, we can trace out all the spins but the first two. The matrix elements of the two-site reduced density matrix read

\begin{equation}
\begin{aligned}
    \bra{s_1 s_2} \rho_{12}(t) \ket{s_1' s_2'} 
    &= \bra{s_1 s_2} \Tr_{3\cdots N} \, \rho(t) \ket{s_1' s_2'} \\
    &= \sum_{s_3 \cdots s_N} \rho_{0,1}^{s_1,s_1'} \rho_{0,2}^{s_2,s_2'} \rho_{0,3}^{s_3,s_3} \cdots \rho_{0,N}^{s_N,s_N} e^{-i (H[s_1 s_2 s_3 \cdots s_N] - H[s'_1 s'_2 s_3 \cdots s_N])t/\hbar} \\
    &= \rho_{0,1}^{s_1,s_1'} \rho_{0,2}^{s_2,s_2'} e^{-i\Delta H_{12}[s] t / \hbar} \prod_{j=3}^N \Big[ \rho_{0,j}^{\uparrow,\uparrow} e^{-i\Delta H_{12j}[s] t / \hbar} + \rho_{0,j}^{\downarrow,\downarrow} e^{i\Delta H_{12j}[s] t / \hbar}  \Big],
\end{aligned}
\end{equation}
having defined
\begin{equation}
\begin{aligned}
    \Delta H_{12}[s] &:= 2 J_{12} (s_1s_2 - s'_1 s'_2) - \frac{\hbar \nu_1}{2} (s_1-s'_1) -  \frac{\hbar \nu_2}{2} (s_2 - s'_2), \\
    \Delta H_{12j}[s] &:= 2 J_{1j} (s_1-s'_1) + 2 J_{2j} (s_2-s'_2).
\end{aligned}
\end{equation}
From the knowledge of $\rho_{ij}$, the concurrence follows by using Eq.~\eqref{eq:def_concurrence}.

Notice that an analogue procedure gives the $k$-site reduced density matrix with $O(k^2 N)$ steps. Thus, this computation allows to access few-sites observables for large system sizes.

\section{Concurrence in a random state}
\label{app:sec:random_state}

Let us consider a system of $N$ spin-1/2. A random, uniformly distributed state is $\ket{\psi} = U \ket{\psi_0}$, $U$ being a Haar-random unitary, and $\ket{\psi_0}$ a reference state. Equivalently, a random state is $\ket{\psi} = \sum_{\{s\}} A_{\{s\}} \ket{\{s\}}$, with the coefficients $A_{\{s\}}$ being uniformly distributed over $\mathbb{C}P^{M-1}$, with $M=2^N$. 

The concurrence of two spins, say sites 1 and 2 wlog.,\ follows from the knowledge of the square roots of the eigenvalues of the matrix $R_{12} = \rho_{12} (\sigma_y \otimes \sigma_y) \rho^*_{12} (\sigma_y \otimes \sigma_y)$. The exact determination of such eigenvalues has evaded our analytical attempts, but we can give an heuristic argument that captures the scaling with $N$. Consider, instead of the square roots of the eigenvalues of $R_{12}$, directly the eigenvalues $\lambda_a$ of $\rho_{12}$. Classical works \cite{Lloyd88,DonPage93} give us their probability density function:
\begin{equation}
    p(\lambda_1, \lambda_2, \lambda_3, \lambda_4)
    \propto \delta \Big( 1 - \sum_{a=1}^4 \lambda_a \Big) \prod_{a=1}^4 \lambda_a^{M-4} \prod_{a < b} (\lambda_a - \lambda_b)^2 
\end{equation}
with the constraint $\lambda_a > 0$, $a=1,\dots,4$. With hindsight, we perform the change of variables 
\begin{align}
    \rho_{12} \equiv \frac 1 4 \mathrm{Id} + \frac{1}{4 \sqrt{M-4}} \tau_{12}, \qquad
    \lambda_a \equiv \frac 1 4 + \frac{\mu_a}{4 \sqrt{M-4}},
\end{align}
so that
\begin{align}
    \notag
    p(\mu_1, \mu_2, \mu_3, \mu_4) 
    & \propto \delta \Big( \sum_{a=1}^4 \mu_a \Big) \prod_{a=1}^4 \left( 1 + \frac{\mu_a}{\sqrt{M-4}} \right)^{M-4} \, \prod_{a < b} (\mu_a - \mu_b)^2  \\
    & \propto \delta \Big( \sum_{a=1}^4 \mu_a \Big) \exp \left[ - \frac{1}{2} \sum_a \mu_a^2 + O\left(\frac{1}{\sqrt{M-4}} \right) \right] \prod_{a < b} (\mu_a - \mu_b)^2 .
\end{align}
We see that, at this order, we can let $\mu_a$ range from $- \infty$ to $+ \infty$ if $N$ is big enough. 

At this point we note that not only the eigenvalues of $\tau_{12}$, but every entry of the matrix is at most of order 1 because of our rescaling. This enables us to expand 
\begin{align}
    \notag
    \sqrt{R_{12}} &= \left[\frac{1}{16} \mathrm{Id} + \frac{1}{16 \sqrt{M-4}} \big[ \tau_{12} +(\sigma_y \otimes \sigma_y) \tau_{12}^* (\sigma_y \otimes \sigma_y) \big] + O\left(\frac{1}{M}\right) \right]^{1/2} \\ 
    &= \frac{1}{4} \mathrm{Id} + \frac{1}{8 \sqrt{M-4}} \big[ \tau_{12} +(\sigma_y \otimes \sigma_y) \tau_{12}^* (\sigma_y \otimes \sigma_y) \big] + O\left(\frac{1}{M}\right)
\end{align}
The matrix $\frac{1}{2} [\tau_{12} +(\sigma_y \otimes \sigma_y) \tau_{12}^* (\sigma_y \otimes \sigma_y)]$ is traceless and very roughly its eigenvalues will have a joint probability density function very similar to that of $\tau_{12}$. For this reason, we can approximate the average concurrence with
\begin{equation}
    \langle C \rangle \approx \int d\vec{\mu} \, p(\vec{\mu})\, \max\left\{0, \frac{2\mu_1 - 1}{4 \sqrt{M-4}} - \frac{1}{2} \right\} ,
\end{equation}
where we have used the $\delta$-function constraint and called $\mu_1$ the largest eigenvalue. Integrating only on $\mu_1$, and forgetting the presence of $\mu_2$, $\mu_3$, $\mu_4$ (otherwise the integration becomes rather cumbersome), we find 
\begin{equation}
    \langle C \rangle \approx \frac{e^{-(M + \sqrt{M-4})/2 }}{2 \sqrt{2 \pi} (M-4)^{3/2}},
\end{equation}
from which
\begin{equation}
    \label{eq:rand_conc_approx}
    \log_2 \big( - \log \langle C \rangle \big) \approx \log(a) + b N + \cdots
\end{equation}
with $a = 1/2$ and $b=1$. As can be seen from Figure \ref{fig:rand_conc}, this scaling is correct, but the numerical factor $a$ is different.

\begin{figure}[t]
    \centering
    \includegraphics[width=0.35\columnwidth]{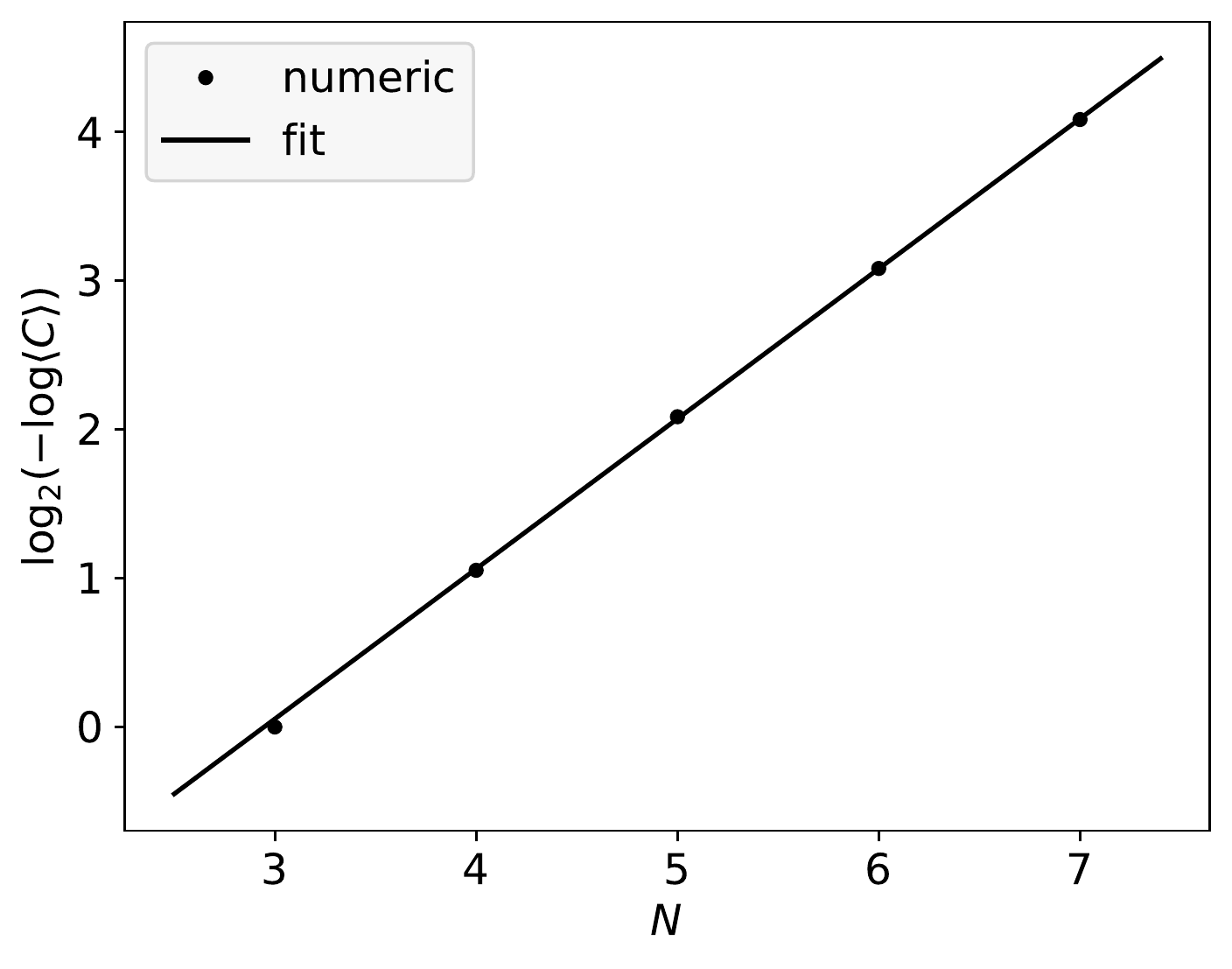}
    \caption{The average concurrence in a random state follows the scaling $\langle C \rangle \sim e^{-a 2^{bN}}$. The dots show the concurrence averaged over $10^7$ randomly generated states, and over every couple of spins for each state. A linear fit is shown for comparison: $b=1.009(6)$, but $a = 0.127(3)$, differing from $a = 1/2$ found analytically (Eq.~\eqref{eq:rand_conc_approx}).}
    \label{fig:rand_conc}
\end{figure}

\section{Integration of the GKSL master equation}
\label{app:sec:full_lindblad}

The density matrix of the system can be parametrized as
\begin{equation}
    \rho(t) = \sum_{\mu_1 \cdots \mu_N} C_{\mu_1 \cdots \mu_N}(t) S_1^{\mu_1} \otimes \cdots \otimes S_N^{\mu_N} ,
\end{equation}
where $S_i^{\mu_i} \in \{\mathrm{Id}_i, S^+_i, S^-_i, S^z_i\}$. Writing explicitly the GKSL equation (see Eqs.~\eqref{eq:our_Lindblad} and \eqref{eq:lbit_hamiltonian} in the main text), we get
\begin{multline}
    \partial_t \rho(t) = - \frac{i}{\hbar} \bigg[ - \sum_i \frac{\hbar \nu_i}{2} S_i^z + \sum_{ij} J_{ij} S_i^z S_j^z, \; \rho(t) \bigg] \\
    + \sum_{i} Y_i f_T(\hbar \nu_i) \left[ S_i^+ \rho(t) S_i^- +  S_i^- \rho(t) S_i^+  - 4 \rho(t) \right]
    + \sum_{i} Y_i \left[ S_i^+ \rho(t) S_i^- + \left\{ \rho(t),S^z_i \right\} - 2 \rho(t) \right].
\end{multline}

In the absence of the interactions (i.e.\ ignoring the term $\sum_{ij}J_{ij}S^z_iS^z_j$), the evolution can be easily computed, and the density matrix evolves as
\begin{equation}
    \partial_t C_{\mu_1 \cdots \mu_N} =  \Big[ \sum_i \lambda_i^{\mu_i} \Big] C_{\mu_1 \cdots \mu_N} \\
    + \sum_i 4 Y_i\ \delta^{\mu_i z} C_{\mu_1 \cdots 0_i \cdots \mu_N},
\end{equation}
where the $\delta^{\mu_iz}$ are Kronecker deltas, the $\lambda_i^{\mu_i}$'s are given by
\begin{equation}
    \lambda_i^{z} = - 4 Y_i (1 + 2f_T), \qquad 
    \lambda^\pm_i = \frac{1}{2} \lambda^z_i \pm i \nu_i,
\end{equation}
and $\lambda_i^{0} = 0$. When interactions are suppressed, the TLSs evolve independently one from the other and any factorized initial state will remain such at all times. One has
\begin{equation}
    \rho(t) = \bigotimes_{i=1}^N \sum_{\mu_i} P_i^{\mu_i}(t) S_i^{\mu_i} \\ \implies
    C_{\mu_1 \cdots \mu_N}(t) = P_1^{\mu_1}(t) \cdots P_N^{\mu_N}(t) \quad \forall t.
\end{equation}

The interactions among TLSs make the evolution more complicated. Computing the commutator
\begin{multline}
    \left[ S_i^z S_j^z, S_i^{\mu_i} S_j^{\mu_j} \right] = S_i^z  S_i^{\mu_i} \left[  S_j^z, S_j^{\mu_j} \right] +  \left[  S_i^z, S_i^{\mu_i} \right] S_j^{\mu_j} S_j^z \\ 
    = 2 \sum_{\mu'_i \mu'_j} \big[ (\delta^{\mu_i 0} \delta^{\mu'_i z} + \delta^{\mu_i z} \delta^{\mu'_i 0} + \delta^{\mu_i +} \delta^{\mu'_i +} - \delta^{\mu_i -} \delta^{\mu'_i -}) (\delta^{\mu_j +} \delta^{\mu'_j +} - \delta^{\mu_j -} \delta^{\mu'_j -})\\
    +(\delta^{\mu_j 0} \delta^{\mu'_j z} + \delta^{\mu_j z} \delta^{\mu'_j 0} - \delta^{\mu_j +} \delta^{\mu'_j +} + \delta^{\mu_j -} \delta^{\mu'_j -}) (\delta^{\mu_i +} \delta^{\mu'_i +} - \delta^{\mu_i -} \delta^{\mu'_i -}) \big] S_i^{\mu'_i} S_j^{\mu'_j},
\end{multline}
and defining
\begin{align}
    \zeta^{\mu \mu'} := \delta^{\mu 0} \delta^{\mu' 3} + \delta^{\mu 3} \delta^{\mu' 0}, \qquad
    \kappa^{\mu \mu'} := 2 \delta^{\mu +} \delta^{\mu' +} - 2\delta^{\mu -} \delta^{\mu' -},
\end{align}
one arrives at
\begin{equation}
    \sum_{i \neq j} J_{ij} \left[ S_i^z S_j^z, S_i^{\mu_i} S_j^{\mu_j} \right]
    = 2 \sum_{i < j} J_{ij} \sum_{\mu'_i \mu'_j} \big[ \zeta^{\mu_i \mu'_i} \kappa^{\mu_j \mu'_j} + (i \leftrightarrow j) \big] S_i^{\mu'_i} S_j^{\mu'_j}.
\end{equation}
The full evolution of the density matrix is given by
\begin{equation}
    \partial_t C_{\mu_1 \cdots \mu_N} =
    \sum_i \lambda_i^{\mu_i} C_{\mu_1 \cdots \mu_N} + \sum_i 4Y_i\ \delta^{\mu_i z} C_{\mu_1 \cdots 0_i \cdots \mu_N}
    - \frac{2i}{\hbar} \sum_{i < j} J_{ij} \sum_{\mu'_i \mu'_j} \left( \zeta^{\mu_i \mu'_i} \kappa^{\mu_j \mu'_j} + \kappa^{\mu_i \mu'_i} \zeta^{\mu_j \mu'_j} \right) C_{\mu_1 \cdots \mu'_i \cdots \mu'_j \cdots \mu_N}.
\end{equation}
This is a systems of $4^N$ partial differential equations. We solved it by matrix exponentiation, using the library for linear algebra with sparse matrices contained in \texttt{SciPy(Python)}.
\end{widetext}

\end{document}